\pdfoutput=1

\documentclass[longauth]{aa}  

\usepackage{comment}
\usepackage{graphicx}
\usepackage{multirow}
\usepackage{booktabs}
\usepackage{textgreek}
\usepackage{xargs}
\usepackage[dvipsnames]{xcolor}
\usepackage{xspace}
\usepackage{caption}
\usepackage{subcaption}
\usepackage{float}
\usepackage{enumitem}

\usepackage{txfonts}
\usepackage{hyperref}
\hypersetup{
    colorlinks=true,
    linkcolor=blue,
    citecolor=blue,
    filecolor=magenta,      
    urlcolor=cyan,
    }
\usepackage{subcaption}
\usepackage{adjustbox}

\graphicspath{{./}}
\usepackage{etoolbox}
\makeatletter
\newcommand\sendemail[3]{
\edef\@tempa{mailto:#1?subject=#2 }%
\edef\@tempb{\expandafter\html@spaces\@tempa\@empty}%
\href{\@tempb}{#3}}

\catcode\%=11
\def\html@spaces#1 #2{#1
\catcode\%=14
\makeatother

\newcommand{\citationneeded}{\textcolor{ForestGreen}{$^{\rm citation\;needed}$}}
\let\oldtextsigma\textsigma
\renewcommand{\textsigma}{\oldtextsigma\xspace}
\let\oldtextalpha\textalpha
\renewcommand{\textalpha}{\oldtextalpha\xspace}
\let\oldAA\AA
\renewcommand{\AA}{\text{\oldAA}\xspace}
\let\oldtextdegree\textdegree
\renewcommand{\textdegree}{\oldtextdegree\xspace}

\newcommand{\orcid}[2]{\href{http://orcid.org/#2}{#1{\includegraphics[height=10pt]{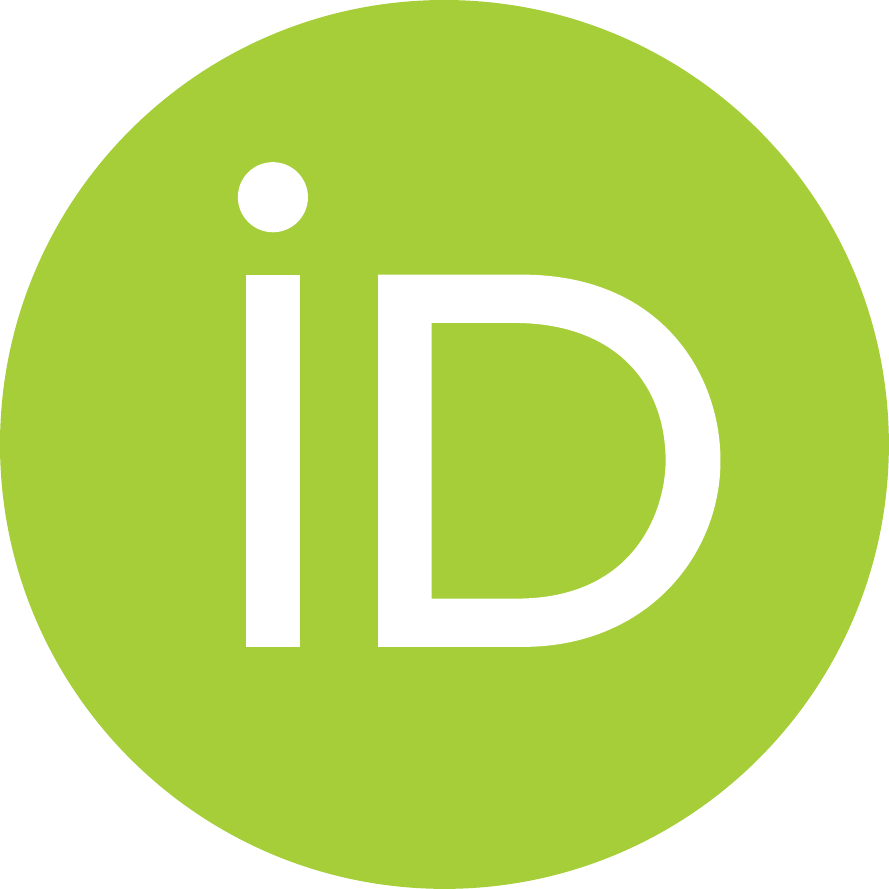}}}}

\newcommand{\kms}{\ensuremath{\mathrm{km\,s^{-1}}}\xspace}
\newcommand{\MSun}{\ensuremath{{\rm M}_\odot}\xspace}
\newcommand{\yr}{\ensuremath{{\rm yr}}\xspace}
\newcommand{\Myr}{\ensuremath{{\rm Myr}}\xspace}
\newcommand{\Gyr}{\ensuremath{{\rm Gyr}}\xspace}
\newcommand{\peryr}{\ensuremath{{\rm yr^{-1}}}\xspace}
\newcommand{\Lsun}{\hbox{\,${\rm L}_\odot$}}
\newcommand{\mum}{\text{\textmu m}\xspace}
\newcommand{\dex}{\text{dex}\xspace}
\newcommand{\kpc}{\text{kpc}\xspace}
\newcommand{\ZH}{\text{[Z/H]}\xspace}
\newcommand{\CO}{\text{[C/O]}\xspace}
\newcommand{\NO}{\text{[N/O]}\xspace}
\newcommand{\FeH}{\text{[Fe/H]}\xspace}
\newcommand{\percm}[1]{\ensuremath{\rm cm^{#1}}\xspace}

\newcommand{\eps}{\ensuremath{\epsilon}\xspace}
\newcommand{\mstar}{\ensuremath{M_\star}\xspace}
\newcommand{\mgas}{\ensuremath{M_\mathrm{gas}}\xspace}
\newcommand{\re}{\ensuremath{R_\mathrm{e}}\xspace}
\newcommand{\NH}{\ensuremath{N_\mathrm{H}}\xspace}
\newcommand{\tauv}{\ensuremath{\tau_\mathrm{V}}\xspace}
\newcommand{\AV}{\ensuremath{A_\mathrm{V}}\xspace}
\newcommand{\xid}{\ensuremath{\xi_\mathrm{d}}\xspace}
\newcommand{\logoh}{\ensuremath{12 + \log\,(\mathrm{O/H})}\xspace}

\newcommand{\nelec}{\ensuremath{n_\mathrm{e}}\xspace}
\newcommandx{\Mout}[2][1=,2=]{\ensuremath{M_{\mathrm{out}{#2}}^{#1}}\xspace}
\newcommandx{\Mdotout}[2][1=,2=]{\ensuremath{\dot{M}_{\mathrm{out}{#2}}^{#1}}\xspace}

\newcommandx{\fluxdcgs}[1][1=-20]{\ensuremath{\mathrm{10^{#1}~erg~s^{-1}~cm^{-2}~\AA^{-1}}}\xspace}
\newcommandx{\fluxcgs}[1][1=-20]{\ensuremath{\mathrm{10^{#1}~erg~s^{-1}~cm^{-2}}}\xspace}
\newcommandx{\powercgs}[1][1=44]{$\times 10^{#1}$~erg~s$^{-1}$\xspace}
\newcommand{\Av}{\ensuremath{A_V}\xspace}

\newcommand{\Te}{\ensuremath{T_\text{e}}\xspace}
\newcommand{\Tiii}{\Te[\ion{O}{iii}]\xspace}
\newcommand{\Tii}{\Te[\ion{O}{ii}]\xspace}
\newcommand{\TSii}{\Te[\ion{S}{ii}]\xspace}
\newcommand{\TSiii}{\Te[\ion{S}{iii}]\xspace}

\newcommand{\Ne}{n$_{\text{e}}$\xspace}

\newcommand{\EWr}{EW$_{\text{0}}$\xspace}

\newcommand{\jwst}{\textit{JWST}\xspace}
\newcommand{\hst}{\textit{HST}\xspace}
\newcommand{\ppxf}{{\sc ppxf}\xspace}
\newcommand{\beagle}{{\sc beagle}\xspace}
\newcommand{\forcepho}{{\sc forcepho}\xspace}
\newcommand{\prospector}{{\sc prospector}\xspace}

\newcommand{\Lyalpha}{\text{Ly\textalpha}\xspace}
\newcommand{\Halpha}{\text{H\textalpha}\xspace}
\newcommand{\Hbeta}{\text{H\textbeta}\xspace}
\newcommand{\Hgamma}{\text{H\textgamma}\xspace}
\newcommand{\Hdelta}{\text{H\textdelta}\xspace}
\newcommand{\Pabeta}{\text{Pa\textbeta}\xspace}
\newcommand{\Hepsilon}{\text{H\textepsilon}\xspace}

\newcommandx{\permittedEL}[6][1=O,2=III,3=,4=,5=,6=]{\text{{#1}\,{\sc {#2}}{#3}{#4}{#5}{#6}}\xspace}
\newcommandx{\semiforbiddenEL}[6][1=O,2=III,3=,4=,5=,6=]{\text{{#1}\,{\sc{#2}}]{#3}{#4}{#5}{#6}}\xspace}
\newcommandx{\forbiddenEL}[6][1=O,2=III,3=,4=,5=,6=]{\text{[{#1}\,{\sc{#2}}]{#3}{#4}{#5}{#6}}\xspace}

\newcommand{\EW}[1]{\text{EW(#1)}\xspace}

\newcommand{\hii}{\permittedEL[H][ii]}

\newcommand{\NV}{\permittedEL[N][v]}
\newcommandx{\NVL}[1][1=1243]{\permittedEL[N][v][\textlambda][#1]}
\newcommandx{\NVall}{\permittedEL[N][v][\textlambda][\textlambda][1239,][1243]}
\newcommand{\NII}{\forbiddenEL[N][ii]}

\newcommand{\NIV}{\semiforbiddenEL[N][iv]}
\newcommandx{\NIVL}[1][1=1486]{\semiforbiddenEL[N][iv][\textlambda][#1]}

\newcommand{\CIV}{\permittedEL[C][iv]}
\newcommandx{\CIVL}[1][1=1550]{\permittedEL[C][iv][\textlambda][#1]}
\newcommand{\CIVall}{\permittedEL[C][iv][\textlambda][\textlambda][1549,][1551]}

\newcommand{\HeII}{\permittedEL[He][ii]}
\newcommandx{\HeIIL}[1][1=1640]{\permittedEL[He][ii][\textlambda][#1]}

\newcommand{\OIII}{\semiforbiddenEL[O][iii]}
\newcommandx{\OIIIL}[1][1=1666]{\semiforbiddenEL[O][iii][\textlambda][#1]}
\newcommand{\OIIIall}{\semiforbiddenEL[O][iii][\textlambda][\textlambda][1661,][1666]}

\newcommand{\OIIIopt}{\forbiddenEL[O][iii]}
\newcommandx{\OIIIoptL}[1][1=5007]{\forbiddenEL[O][iii][\textlambda][#1]}
\newcommand{\OIIIoptall}{\forbiddenEL[O][iii][\textlambda][\textlambda][4959,][5007]}

\newcommand{\NIII}{\semiforbiddenEL[N][iii]}
\newcommandx{\NIIIL}[1][1=1750]{\semiforbiddenEL[N][iii][\textlambda][#1]}
\newcommand{\NIIIall}{\semiforbiddenEL[N][iii][\textlambda][\textlambda][1747--][1754]}

\newcommandx{\CIII}{\semiforbiddenEL[C][iii]}
\newcommandx{\CIIIL}[1][1=1909]{\semiforbiddenEL[C][iii][\textlambda][#1]}
\newcommand{\CIIIall}{\semiforbiddenEL[C][iii][\textlambda][\textlambda][1907,][1909]}

\newcommand{\CIIIp}{\permittedEL[C][iii]}
\newcommand{\NIIIp}{\permittedEL[N][iii]}

\newcommand{\NeIV}{\forbiddenEL[Ne][iv]}
\newcommandx{\NeIVL}[1][1=2424]{\forbiddenEL[Ne][iv][\textlambda][#1]}
\newcommand{\NeIVall}{\forbiddenEL[Ne][iv][\textlambda][\textlambda][2422,][2424]}

\newcommand{\MgII}{\permittedEL[Mg][ii]}
\newcommandx{\MgIIL}[1][1=2803]{\permittedEL[Mg][ii][\textlambda][#1]}
\newcommand{\MgIIall}{\permittedEL[Mg][ii][\textlambda][\textlambda][2796,][2803]}

\newcommand{\NeV}{\forbiddenEL[Ne][v]}
\newcommandx{\NeVL}[1][1=3426]{\forbiddenEL[Ne][v][\textlambda][#1]}
\newcommand{\NeVall}{\forbiddenEL[Ne][v][\textlambda][\textlambda][3346,][3426]}

\newcommand{\OII}{\forbiddenEL[O][ii]}
\newcommandx{\OIIL}[1][1=3727]{\forbiddenEL[O][ii][\textlambda][#1]}
\newcommand{\OIIall}{\forbiddenEL[O][ii][\textlambda][\textlambda][3726,][3729]}

\newcommand{\OIIaur}{\forbiddenEL[O][ii][\textlambda][\textlambda][7320,][7330]}

\newcommand{\SII}{\forbiddenEL[S][ii]}
\newcommandx{\SIIL}[1][1=6725]{\forbiddenEL[S][ii][\textlambda][#1]}
\newcommand{\SIIall}{\semiforbiddenEL[S][ii][\textlambda][\textlambda][6718,][6732]}

\newcommand{\SIII}{\forbiddenEL[S][iii]}
\newcommandx{\SIIIL}[1][1=9068]{\forbiddenEL[S][iii][\textlambda][#1]}

\newcommand{\SIIIall}{\semiforbiddenEL[S][iii][\textlambda][\textlambda][9068,][9532]}

\newcommand{\NeIII}{\forbiddenEL[Ne][iii]}
\newcommandx{\NeIIIL}[1][1=3869]{\forbiddenEL[Ne][iii][\textlambda][#1]}
\newcommand{\NeIIIall}{\forbiddenEL[Ne][iii][\textlambda][\textlambda][3869,][39xx]}

\newcommand{\FeII}{\forbiddenEL[Fe][ii]
[\textlambda][4359]}

\newcommand{\ArIII}{\forbiddenEL[Ar][iii]}
\newcommand{\ArIIIL}{\forbiddenEL[Ar][iii]
[\textlambda][7135]}

\newcommand{\OI}{\forbiddenEL[O][i]}
\newcommand{\OIL}{\forbiddenEL[O][i]
[\textlambda][6300]}

\newcommand{\hda}{\ensuremath{\mathrm{H\text{\textdelta}_A}}\xspace}
\newcommand{\hga}{\ensuremath{\mathrm{H\text{\textgamma}_A}}\xspace}

\begin{document}

\title{MARTA: Temperature-temperature relationships and strong-line metallicity calibrations from multiple auroral-line detections at cosmic noon}


\titlerunning{Multiple auroral lines at cosmic noon}

\author{
E. Cataldi\inst{\ref{UNIFI}, \ref{arcetri}, \ref{eso}} \and
F. Belfiore\inst{\ref{arcetri}} \and
M. Curti\inst{\ref{eso}} \and
B. Moreschini\inst{\ref{UNIFI},\ref{arcetri}} \and
F. Mannucci\inst{\ref{arcetri}} \and
Q. D'Amato\inst{\ref{arcetri}} \and
G. Cresci\inst{\ref{arcetri}} \and
A. Feltre\inst{\ref{arcetri}} \and
M. Ginolfi\inst{\ref{UNIFI},\ref{arcetri}} \and
A. Marconi\inst{\ref{UNIFI},\ref{arcetri}} \and
A. Amiri\inst{\ref{arkansas}} \and
M. Arnaboldi\inst{\ref{eso}} \and
E. Bertola\inst{\ref{arcetri}}\and
C. Bracci\inst{\ref{UNIFI},\ref{arcetri}}\and
S. Carniani\inst{\ref{SNS}} \and
M. Ceci\inst{\ref{UNIFI},\ref{arcetri}}\and
A. Chakraborty\inst{\ref{arcetri}}\and
M. Cirasuolo\inst{\ref{eso}}\and
F. Cullen\inst{\ref{ROE}} \and
C. Kobayashi\inst{\ref{herts}}\and
N. Kumari\inst{\ref{aura}} \and
R. Maiolino\inst{\ref{cavensish}, \ref{kicc}} \and
C. Marconcini\inst{\ref{UNIFI},\ref{arcetri}}\and
M. Scialpi\inst{\ref{trento},\ref{arcetri},\ref{UNIFI}}\and
L. Ulivi\inst{\ref{trento},\ref{arcetri},\ref{UNIFI}}
}

\institute{
\label{UNIFI} Università di Firenze, Dipartimento di Fisica e Astronomia, via G. Sansone 1, 50019 Sesto Fiorentino, Florence, Italy \and
\label{arcetri} INAF -- Arcetri Astrophysical Observatory, Largo E. Fermi 5, I-50125, Florence, Italy \and
\label{eso} European Southern Observatory, Karl-Schwarzschild Straße 2, D-85748 Garching bei München, Germany \and
\label{arkansas} Department of Physics, University of Arkansas, 226 Physics Building, 825 West Dickson Street, Fayetteville, AR 72701, USA \and
\label{SNS} Scuola Normale Superiore, Piazza dei Cavalieri 7, I-56126 Pisa, Italy \and
\label{ROE} Institute for Astronomy, University of Edinburgh, Royal Observatory, Edinburgh, EH9 3HJ, UK \and
\label{herts} Centre for Astrophysics Research, Department of Physics, Astronomy and Mathematics, University of Hertfordshire, Hatfield AL10 9AB, UK \and
\label{aura} AURA for European Space Agency, Space Telescope Science Institute, 3700 San Martin Drive. Baltimore, MD, 21210 \and
\label{cavensish} Cavendish Laboratory, University of Cambridge, 19 JJ Thomson Avenue, Cambridge, CB3 0HE, UK \and
\label{kicc} Kavli Institute for Cosmology, University of Cambridge, Madingley Road, Cambridge, CB3 0HA, UK \and 
\label{trento} University of Trento, Via Sommarive 14, I-38123 Trento, Italy
}

\authorrunning{E. Cataldi et al.}
\date{}


\abstract{We present the first results from MARTA (Measuring Abundances at high Redshift with the T$_e$ Approach), a programme leveraging ultra-deep, medium-resolution JWST/NIRSpec spectroscopy to probe the interstellar medium (ISM) of star-forming galaxies at $z \sim 2 - 3$. We report detections of one or more auroral lines, including \OIIIopt$\lambda4363$, \OII$\lambda\lambda7320,7330$, \SII $\lambda4068$, and \SIII $\lambda6312$ for 16 galaxies in the sample, providing measurements of multiple ionic temperatures. We tested the validity of the T\OII-T\OIIIopt relation at high redshift considering a total sample of 21 objects including literature data, and obtained a shallower slope than in the low-$z$ literature. However, such a slope is consistent with low-redshift data when ultra-low metallicity objects are considered.
We assessed the correlation of the T\OII-T\OIIIopt relationship and its scatter on different physical parameters, finding a mild correlation with the ionisation parameter and radiation field hardness, while no significant correlation with gas density.
The location of high-redshift data is also consistent with the low-$z$ literature in the T\OII-T\SII, and T\SIII-T\OIIIopt relations, although this conclusion is limited with low-number statistics.
Finally, we leveraged our sample together with a comprehensive compilation of galaxies with \OIIIopt$\lambda4363$ detections from the literature to recalibrate classical strong-line diagnostics at high redshift. MARTA represents a key addition in this space because it provides direct metallicities at moderately high oxygen abundances (12+log(O/H) $\sim8.0-8.4$).
}


   \maketitle
%

\section{Introduction}
\label{sec:intro}

Nebular emission from star-forming regions is a fundamental window into the physical conditions and ionizing sources within galaxies across cosmic time. For instance, hydrogen recombination lines provide measurements of star formation rate (SFR) and dust attenuation \citep{kennicutt_star_1998, osterbrock_astrophysics_2006, kennicutt_star_2012},

while ratios of collisionally excited lines from elements like oxygen, nitrogen, and sulphur are sensitive to electron temperature ($T_e$), electron density ($n_e$), and the ionization state of the gas, enabling the derivation of key physical properties of the ionized gas, including its metal content \citep{osterbrock_astrophysics_2006}.

The measurement of chemical abundances in galaxies is a cornerstone of modern astrophysics, offering crucial insights into nucleosynthesis, the history of star formation, and  the exchange of gas between galaxies and their surrounding environment \citep{maiolino_re_2019, kewley_understanding_2019}.
In extragalactic astrophysics, the `gold standard' approach to determine the metallicity of the interstellar medium (ISM) involves deriving electron temperatures from faint auroral lines \citep{peimbert_temperature_1967, stasinska_abundance_2002}. In particular, ratios of auroral to nebular collisionally excited lines (CELs) are exponentially sensitive to the electron temperature as auroral lines originate from higher energy levels compared to nebular ones \citep{peimbert_temperature_1967, osterbrock_astrophysics_2006, 2017PASP..129h2001P}.

For the purposes of inferring abundances, \hii regions are generally modelled using a two-zone model consisting of a low- and high-ionization zone. 
Among optical lines, the auroral \OIIIopt$\lambda$4363 line has been used as a workhorse to infer oxygen electron temperature of the high-ionization zone and consequently the abundance of O$^{++}$. Measurements of the total oxygen abundance, however, require the detection of multiple auroral lines from both low- and high-ionization species to accurately constrain electron temperatures throughout the nebula \citep[e.g.,][]{pilyugin_electron_2007, andrews_mass-metallicity_2013, berg_chaos_2015, curti_new_2017}. Alternatively, a relation between the temperatures of the different ionization zones (temperature-temperature, or $T-T$ relation) is adopted \citep[e.g.][]{campbell_stellar_1986, garnett_electron_1992, pilyugin_electron_2009, mendez_delgado_density_biases_2023}.

A fundamental limitation to measuring chemical abundances using auroral lines is that they are often significantly 
fainter than the corresponding strong nebular lines (from tens to even hundreds of times, especially at high metallicity) since their excitation requires higher-energy electrons \citep{kennicutt_composition_2003, esteban_reappraisal_2004, berg_chaos_2020}. Empirical metallicity calibrators, based on strong line ratios such as \NII$\lambda$6584/H$\alpha$, $\OIIIopt~\lambda$5007/H$\beta$, $\OIIIopt~\lambda$5007/$\OII~\lambda\lambda$3727,3729 or $\OIIIopt~\lambda$5007/$\NII~\lambda$6584, offer an alternative approach by utilizing only strong nebular emission lines (e.g., \citealt{pettini_oiiinii_2004, curti_new_2017}, see \citealt{maiolino_re_2019} for a review). However, such calibrators are subject to significant limitations as strong-line ratios depend not only on metallicity but also on other parameters, such as gas density, ionization parameter, hardness of the ionizing spectrum and relative element abundances \citep{2002astro.ph.11148G, 10.1111/j.1365-2966.2006.10892.x, perez-montero_impact_2009, kewley_understanding_2019}. 

Moreover, empirical calibrations established for local galaxies are not directly applicable to high-redshift galaxies. Extensive observational evidence has demonstrated that galaxies at $z\sim2$ exhibit harder ionizing spectra—likely driven by $\alpha$-enhancement in high-redshift star-forming galaxies \citep{steidel_reconciling_2016, topping_mosdef-lris_2020_i, Cullen_NIRVANDELS_alpha_enhance_2021, 2024MNRAS.532.3102S}—, and higher electron densities compared to their $z\sim 0$ counterparts  
\citep{erb_h_2006,hainline_rest-frame_2009, 2010ApJ...725.1877B, 2012PASJ...64...60Y, 2013ApJ...768...74T, kewley_cosmic_2013, masters_physical_2014, nakajima_ionization_2014, sanders_mosdef_2016,2019ApJ...881L..35S, 2025MNRAS.541.1707T}. 
These differences are reflected in diagnostic diagrams such as the N2-BPT (Baldwin, Phillips \& Terlevich, \citealt{baldwin_classification_1981}) diagram, log(\OIIIopt/\Hbeta) versus log(\NII/\Halpha)), where high-redshift galaxies are systematically offset from their local counterparts \citep{steidel_strong_2014, masters_physical_2014, shapley_mosdef_2015, masters_tight_2016, strom_measuring_2017, topping_mosdef-lris_2020_i}.
Strong-line ratios like \NII/\Halpha and \OIIIopt/\Hbeta become increasingly influenced by other physical parameters at higher redshifts, such as the shape of the ionizing spectrum and the ionization parameter. As a result, their reliability as metallicity tracers is compromised, leading to potential biases when applied to high-redshift galaxies \citep[][]{bian_ldquodirectrdquo_2018, sanders_mosdef_mzr_2021}.

The ideal solution is to measure metallicities directly using auroral lines sensitive to electron temperature, bypassing the uncertainties inherent to empirical calibrators. In cases where only a single auroral line is available, locally calibrated relations (e.g., between the temperature of O$^{+}$ and O$^{++}$, known as $T_2$–$T_3$ relations) must be assumed, a situation which introduces additional uncertainties. Detecting these faint lines in individual high-redshift galaxies was extremely challenging prior to JWST, with only a few, low-significance detections of \OIIIopt~$\lambda4363$ reported at \textit{z}~$>$~2 \citep[e.g.,][]{christensen_gravitationally_2012, james_testing_2014, sanders_mosdef_2020}, see also the compilation by \citealt{patricio_testing_2018}; and rare cases of auroral \OII~$\lambda\lambda7320,7330$ detections, such as those presented in \citet{Sanders_2023} for two galaxies at $z>2$. This scarcity hindered the accurate determination of metallicities in distant galaxies, necessitating the use of empirical metallicity indicators calibrated on lower-redshift systems.

The JWST Near Infrared Spectrograph (NIRSpec, \citealt{jakobsen_nirspec_2022}), with its unparalleled sensitivity and wide spectroscopic wavelength coverage of the near-infrared regime, has revolutionized this field, enabling the detection of auroral lines in galaxies both at Cosmic Noon ($z\sim2-3$, e.g. \citealt{Strom2023CECILIA:,Welch2024TEMPLATES:,welch2024sunburst, 2024ApJ...964L..12R}) as well as at much higher redshifts \citep{curti_smacs_2023, sanders_calibrations_2023, laseter_auroral_jades_2023, Yang2023Analytical, Morishita2024Diverse, shapley2024aurora, 2025MNRAS.540.2991A, 2025arXiv250111099C, Stanton_excels_2025, 2025arXiv250210499S}. These efforts delivered some of the most accurate determinations of gas-phase abundances in early galaxies.
Interestingly, the higher sensitivity of JWST/NIRSpec toward the redder part of the spectrum, combined with the metal deficiency, higher ionization conditions, and compact nature of high-$z$ systems, as well as with the tendency of most JWST programs to focus on the high-$z$ Universe,
has made auroral line detections more likely at very high-$z$ than at $z\sim2-3$, where galaxies are typically more massive, extended, and metal-rich than at $z\gtrsim 5$. Moreover, very few observations have enabled detections of multiple auroral lines, necessary to probe the temperature and ionisation structure of high-redshift galaxies.

While this paper was under review, an independent analysis by \citet{2025arXiv250810099S} presented results from the AURORA survey, combining new JWST/NIRSpec observations with literature data to investigate temperature–temperature relations and to calibrate strong-line metallicity diagnostics at $z\sim$2–10. They found that local $T_2$–$T_3$ relations remain valid at high redshift, while empirical strong-line calibrations show clear offsets with respect to the local ones, consistent with evolving ISM conditions; in particular, diagnostics based on $\alpha$-elements provide the most reliable metallicity estimates, whereas nitrogen-based indicators exhibit significant scatter likely due to variations in N/O at fixed O/H.


In this paper we present novel observations of multiple auroral lines in spectra of individual galaxies at Cosmic Noon from the MARTA survey (Measuring Abundances at high Redshift with the T$_{\rm e}$ Approach; PID 1879, PI Curti), which targeted 127 galaxies in the COSMOS field at redshifts $z\sim2-3$ by means of very deep NIRSpec Micro Shutter Array (MSA) observations.
In Curti et al. (in prep), we discuss the multiple science cases addressed by such a program by focusing on the ISM properties and chemical abundances of one of the galaxies in the sample.

Here, we leverage a sample of 16 galaxies with (multiple) detections of auroral lines, mainly \OIIIopt$\lambda$4363 and \OII$\lambda$7320,7330, but including also \SII$\lambda$4069 and \SIII$\lambda$6312. This sample extends the parameter space probed by Cosmic Noon galaxies with auroral line detections towards higher metallicity than most existing studies.
We focus on the relationships between temperatures inferred from different ionic species and their possible redshift evolution. We discuss the implications for the applicability of the $T_{\rm e}$-method for measuring metallicity in high-$z$ galaxies and investigate the redshift evolution of some of the most widely adopted strong-line metallicity diagnostics.

The paper is structured as follows. Section \ref{sec:survey} outlines the survey design, including the target selection, observational strategy, and data reduction process. In Section \ref{sec:fitting}, we detail the data analysis methodology and the spectral fitting procedure. Section \ref{sec:MARTA_sample} defines the two subsamples of the MARTA survey, golden and silver, which are used for all subsequent analyses. In Sec. \ref{sec:dust_attenuation} we describe the dust correction procedure applied.
Section \ref{sec:tt_relation} focuses on the oxygen temperature-temperature relation,  the analysis of their dependency on other parameters such as density and ionization parameter, and the comparison with sulphur temperatures. Section \ref{sec:strong_lines} presents the results on strong-line calibrations. Finally, in Section \ref{sec:summary} we provide a summary of our conclusions and prospects for future research.

Throughout the rest of this work, we use the following notation to refer to the strong-line diagnostics:
\begin{itemize}
    \item R3 = \(\log{ ( \,  \text{\OIIIopt} \, \lambda 5007 / \text{H}\beta \, ) } \)
    \item R2 = \(\log ( \, \text{\OII} \, \lambda 3726,3729 / \text{H}\beta \, )\)
    \item O32 = \( \text{R3} - \text{R2} \)
    \item R23 = \(\log \left[ \, ( \, \text{\OII} \, \lambda 3726,3729 + \text{\OIIIopt} \, \lambda 4959,5007 \, ) / \text{H}\beta \, \right]\)
    \item $\tilde{R}$ = \( 0.46 \, \text{R2} + 0.88 \, \text{R3} \)
    \item{ Ne3O2 = \( \log{ ( \, \text{[NeIII]} \, \lambda 3869 / \text{\OII} \, \lambda 3726,29 } \, ) \) }
    \item{ N2 = \(  \log \left( \,  \text{\NII} \, \lambda 6585 / \text{H}\alpha \, \right) \) }
    \item{ O3N2 = \( \text{R3} - \text{N2} \) }
    \item{ S2 = \( \log \left[ \, ( \, \text{\SII} \, \lambda 6716 + \text{\SII} \, \lambda 6730 \, ) / \text{H}\alpha  \, \right]\)  }
    \item{ O3S2 = \( \text{R3} - \text{S2} \)}

\end{itemize}

\section{Observations and data reduction}
\label{sec:survey}

\subsection{Target selection, priority classes, and observing strategy}
\label{sec:selection}

The MARTA programme focuses on the redshift range around the peak of cosmic star-formation activity (Cosmic Noon), with the primary aim to detect the auroral lines of \OIIIopt$\lambda$4363 in the G140M (R $\sim$ 1000) grating and, for a subsample, the \OII$\lambda$7320,7330 lines in the G235M (R $\sim$ 1000) grating. Considering the need to cover \OII$\lambda\lambda$3727,29 to measure electron temperatures and derive metallicities, this constrains the primary redshift range of our targets to $1.7<z<3.1$. However, a few higher redshift sources were also observed, according to selection and prioritisation criteria described below.

The MARTA parent catalogue was generated from the COSMOS2020 \citep{Weaver_COSMOS2020_2022} and the 3D-HST catalogues \citep{skelton_3d-hst_2014, momcheva_3d-hst_2016}   
based on a multi-step process. The COSMOS2020 catalogue provided photometric redshifts
while the 3D-HST catalog provided grism spectroscopic redshift data, which were complemented by spectroscopic redshifts from higher spectral resolution, ground-based observations available at the time of MSA mask design (including zCOSMOS-bright, \citealt{2009ApJS..184..218L}; FMOS, \citealt{2015ApJS..220...12S}; MOSDEF, \citealt{kriek_mosfire_2015}; zFIRE, \citealt{2016ApJ...828...21N}; VUDS, \citealt{2017A&A...600A.110T}; hCOSMOS, \citealt{Damjanov_2018}; DEIMOS 10k, \citealt{2018ApJ...858...77H}; LEGA-C, \citealt{2018ApJS..239...27S}; KMOS3D, \citealt{schreiber_kmos_2019}; KLEVER, \citealt{curti_klever_2020}; and data from \citealt{2010ApJ...709..572K, 2014MNRAS.443.2679B, 2015ApJ...806L..35K, 2015yCat..35750040C, 2017yCat..17550169M, 2018yCat..36180085S}). 
When ground-based spectroscopic redshifts were not available, lower resolution, grism-based redshifts were adopted. In the absence of both, photometric redshifts were used, provided they were consistent between \textsc{LePhare} \citep{Ilbert_photz_2006} and \textsc{eazy} \citep{Brammer_EAZY_2008}, within a defined tolerance ($\Delta(z)/z < 0.1$, considering the redshift estimates from \textsc{eazy}).

\begin{figure*}[!t]
    \centering
    \includegraphics[width=0.85\linewidth]{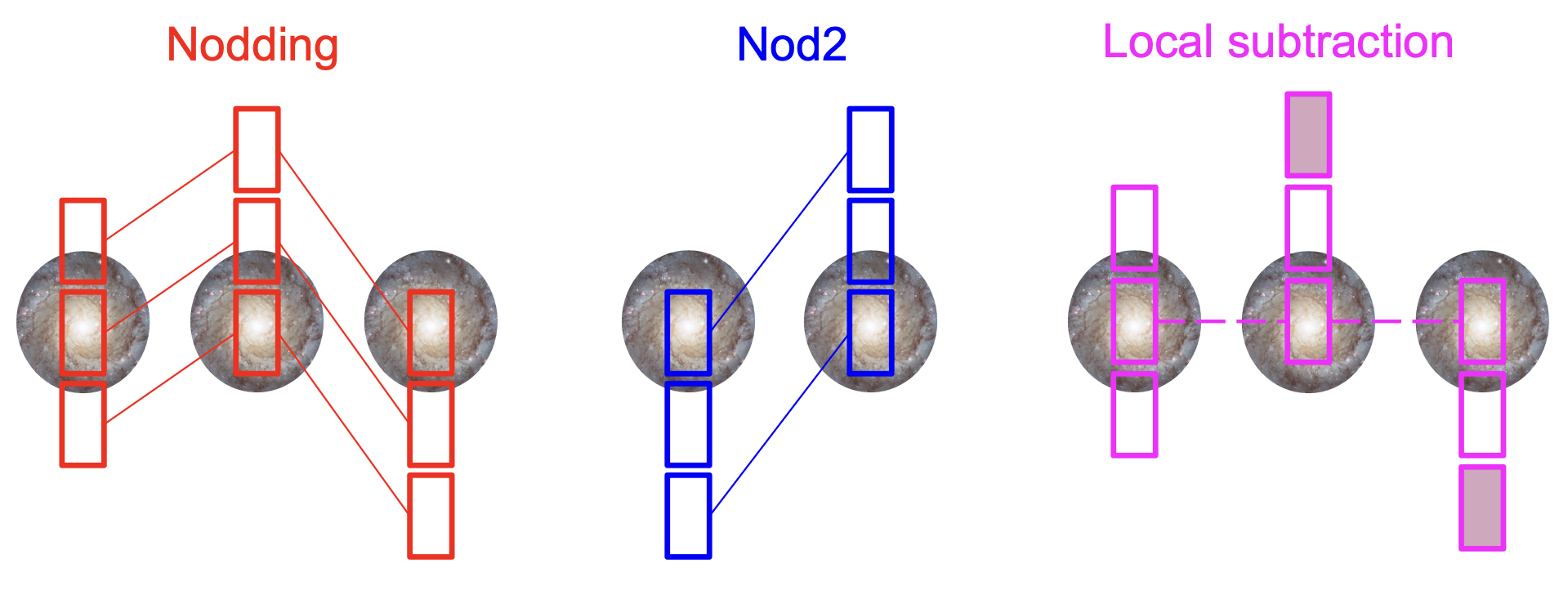}
    \caption{Comparison of strategies for background subtraction in NIRSpec/MSA observations. 
    Left (red): The standard three-point nodding strategy, in which the galaxy is observed in three positions within the shutter, and the background for each position is estimated using the corresponding shutter exposures from the other two nods. 
    Center (blue): The nod2 strategy, a two-position scheme where only two out of the three available positions are used for the nodding procedure, in order to use as background the shutters furthest from the galaxy. 
    Right (magenta): The local subtraction approach, which uses the outer shutters from the two off-source positions (indicated by the filled-in shutters in the figure) to estimate the sky background. In this method, the object signal is taken from the sum of all three nod positions (connected by dashed magenta lines), but no background is subtracted from the same shutters. As such, this method can be more sensitive to local systematics or background gradients.} 
    \label{fig:observational_strategies}
\end{figure*}

For each galaxy in the MARTA parent catalogue, we estimated the expected flux of the \OIIIoptL[4363] emission line based on the available information on stellar masses ($M_\star)$, SFR, and metallicity, as described in the following.
The stellar masses, SFR, and dust attenuation (A$_V$) were adopted from spectral energy distribution (SED) fitting of the photometry extracted by COSMOS2020 \citep{Weaver_COSMOS2020_2022}.
When ground-based (either slit or IFU) spectroscopy in COSMOS covering \Halpha was available, as provided e.g. by KMOS3D \citep{wisnioski_kmos_2019}, KLEVER \citep{curti_klever_2020}, MOSDEF \citep{kriek_mosfire_2015} surveys, the SFR was estimated from the extinction-corrected \Halpha flux using the relation from \cite{kennicutt_star_2012} and the A$_V$ derived from either photometry or from the Balmer decrements. 
As of gas-phase metallicity, we assumed the value for oxygen abundance derived from strong-line diagnostics and the \cite{bian_ldquodirectrdquo_2018} calibrations, specifically those based on local analogs of high-redshift galaxies, for galaxies with available spectroscopy and detection of at least three nebular emission lines, whereas we estimated the expected metallicity for each source with no previously available spectra assuming that the relationship between metallicity, mass, and SFR (so-called fundamental metallicity relation, FMR, \citealt{mannucci_fundamental_2010}) does not evolve with redshift up to $z\sim2$ \citep{cresci_fundamental_2018, sanders_mosdef_mzr_2021}, adopting specifically the parametrization of \cite{curti_massmetallicity_2020}. 

Finally, we based our indirect estimate of the \OIIIoptL[4363] flux on  the relationship between the \OIIIoptL[4363]/\Halpha ratio and metallicity, the latter calibrated on a combined sample of local \hii\ regions and low-metallicity galaxies from \cite{guseva_vlt_2011, berg_chaos_2020}, and stacks from SDSS galaxies \citep{curti_new_2017} with `direct', T$_e$-based oxygen abundances. 
Such intrinsic \OIIIoptL[4363] flux was then attenuated by the galaxy $A_V$, assuming the attenuation curve of \cite{cardelli_relationship_1989} and an $R_V=3.1$, to derive our final estimate of the expected \OIIIoptL[4363] flux in each source.

Galaxies in the MARTA parent catalogue were assigned to four different priority classes (each one with different weights) to drive the MSA target allocation. Such classes were established on the basis of the  expected emission line fluxes (estimated as detailed above), 
while additional factors such as dust attenuation and redshift reliability were also considered to fine-tune the weights assignment and optimize the final MSA mask. 
The detailed list of the criteria assumed in target prioritisation is reported in Table~\ref{tab:target_priorities}.
A total of 126 sources, including 18 `priority one', 22 `priority two', 12 `priority three', 21 `priority four' targets and 53 `filler' sources, were observed using a single MSA configuration, though only galaxies belonging to the top two classes were used to drive the choice of the pointing coordinates.

Each galaxy was observed through a 3-shutter slitlet, and observations were taken adopting a standard nodding strategy with three nod positions along the slitlet. 
To maximize the number of targets observable within a single pointing, we allowed for overlap of spectral traces on the detector, as the probability of overlapping emission lines from two different sources was estimated to be minimal. However, due to the depth of our observations, significant continuum emission was detected in many cases. The impact of this observing choice on background subtraction is discussed in Sec \ref{data_reduction}. 

Observations were obtained totalling an integration time of 31.9 hours in G140M, 7.4 hours in G235M, and 2.9 hours in G235H. The G235H is intended for studies of emission-line kinematics and is therefore not used in this work, where we instead focus on the G140M and G235M data.

\begin{table*}
    \centering
    \caption{Target prioritization criteria.}
    \label{tab:target_priorities}
    \begin{tabular}{l c c c}
        \hline \hline
        Priority Class & Expected \OIIIopt$\lambda$4363 Flux & Expected \Halpha Flux & Additional criteria for prioritization \\
        & $\rm erg\, s^{-1} \rm cm^{-2}$ & $\rm erg\, s^{-1} \rm cm^{-2}$ & \\
        \hline
        Priority one & $> 8 \times 10^{-19}$ & - & Lower extinction, reliable spectroscopic redshift \\
        Priority two & $[1.5 - 8] \times 10^{-19}$ & $> 1.5 \times 10^{-17}$ & Lower extinction, reliable spectroscopic redshift \\
        Priority three & $[1 - 1.5] \times 10^{-19}$ & $> 1.5 \times 10^{-17}$ & Includes $3.1 < z < 5$ if G235M covers \OIIIopt$\lambda$4363 \\
        Priority four & - & $> 7.5 \times 10^{-18}$ & Higher weights for \OIIIopt$\lambda$4363 $> 3 \times 10^{-20}$ \\
        Filler & - & $> 1 \times 10^{-20}$ & Higher weights for larger \Halpha fluxes \\
        \hline
    \end{tabular}
    \label{tab:priority_classes}
\end{table*}

\subsection{Data reduction}
\label{data_reduction}

The data reduction of our observations was performed using the JWST calibration pipeline, which is divided into three stages. We applied the pipeline version `2023\_2a' and CRDS context files `jwst\_1183.pmap'.  
Stage 1 applies detector-level corrections, including group-by-group corrections and ramp fitting, producing a count rate image for each exposure. 
Stage 2 focuses on instrument-specific and observing-mode calibrations, resulting in fully calibrated individual exposures. At this stage, the standard nodding background subtraction is performed, and path-loss corrections estimated for point-like sources are applied. 

Stage 3 combines multiple exposures, such as those from nodding and dither patterns, into a single 2D spectral product, while applying outlier rejection and additional corrections as needed. Although the pipeline also provides a final extracted 1D spectrum, we performed a bespoke extraction of 1D spectra for each target as described later in this section.
Other parameters in the data reduction pipeline were kept to their default values.

Due to our observing strategy (allowing for spectral overlap) and the spatially extended nature of our sources, in several cases a dedicated background subtraction is required.
In particular, our data suffers from spectral contamination from sources located in different MSA slits, whose continuum emission could contaminate both the central shutter of the slitlet and/or the flanking `background' shutters. Moreover, in some cases, the flux from the target object is extended or not well-centered, and extends into the micro-shutters adjacent to the central one. As a result, applying the standard nodding background subtraction, which subtracts the background pixel-per-pixel using the same shutter after offsetting (Fig. \ref{fig:observational_strategies}, left), can lead to either over-subtraction of the target’s flux and/or to a poor modeling of the background itself.

To mitigate oversubtraction in case of extended sources, we calculated the expected contamination of the flanking shutters by the source flux using NIRCAM imaging data from the PRIMER survey 
(PID:1837, \citealt{2021jwst.prop.1837D}) in bands F115W, F150W, F200W, F277W (broadly corresponding to the wavelength range probed by our NIRSpec observations in G140M and G235M). The reduced and calibrated NIRCam imaging data products were retrieved from the Dawn JWST Archive (DJA\footnote{\url{https://dawn-cph.github.io/dja/}}). 

For galaxies where PRIMER imaging and 2D spectra indicate that the emission is visibly extended beyond the central shutter, we adopt a so-called `nod2 subtraction' strategy \citep[e.g.][]{Maseda_WIDE_2024}. In this approach, only two of the nods—specifically, the ones where the galaxy is in the outermost shutter—are used. This minimizes the risk of self-subtraction, where the galaxy’s emission would be subtracted along with the background, while maintaining the advantage of pixel-by-pixel subtraction which mitigates detector artifacts. However, this strategy comes at the cost of sacrificing one-third of the total exposure time, as the exposure in which the galaxy is located in the central shutter is excluded (Fig. \ref{fig:observational_strategies}, middle). 

An alternative approach to background subtraction consist in employing the combined, non-background subtracted 2D spectra from all three nodded exposures to extract and construct a `local' background as the average of the two outermost shutters. 
Specifically, we 
extracted the background spectrum from the two outermost shutters in the nodding configuration, summing the flux across all pixels in each shutter and then averaging the results to obtain a representative background spectrum (Fig. \ref{fig:observational_strategies}, right). The galaxy spectrum was extracted from 
a boxcar aperture centred on the peak of the signal-to-noise in the spectral trace of the target and of width of 5 pixels (roughly equivalent to the size of one shutter), although in some cases extended line emission as visible in the 2D spectra suggested to increase the size of the extraction aperture to avoid siginificant flux losses and possible biases in the inferred line ratios. 
Such `local subtraction' strategy benefits on the one hand the full on-source integration time by co-adding all three nodded exposures while, on the other, suffers from the increased noise affecting the background spectrum, which is generated from exposures with shorter effective integration times than the central trace containing the target source. Moreover this strategy does not
leverage the pixel-by-pixel background subtraction of the standard `nodding' therefore sometimes resulting in spectra with worse noise properties due to detector artefacts.
For compact galaxies or those which do not show significant emission in adjacent shutters, the standard nodding background subtraction was applied and we checked that the fluxes obtained from the local background subtraction are comparable. Specifically, we applied the local background subtraction for galaxy MARTA\_3942, the nod2 method for MARTA\_4195, and the standard nodding background subtraction for the remaining sources. To compare the results of these different strategies and ensure that absence of self-subtraction, we assessed their impact on the extracted and measured fluxes of $\OIIIopt~\lambda5007$ in the G140M and on the $\rm H\alpha$ in the G235M gratings, respectively. 
We verified that, had the standard nodding procedure been used in the case of extended galaxies, these lines would have been significantly underestimated. In particular, we tested the effect on MARTA\_3942 and MARTA\_4195, and found that the \OIIIopt flux is affected by 13-22\% and the \Halpha flux by 7-20\%.

As some of our galaxies look morphologically extended, they might be affected by larger slit losses than modelled for simple point-like sources. On the other hand, in several cases the bulk of the emission captured by the slit comes from bright individual clumps, hence the assumption of constant surface brightness across the slit is also incorrect. The analysis presented in this work only relies on emission lines ratios, and is therefore not affected by uncertainties on the absolute flux calibration. However, a second-order correction to the relative line fluxes is expected due to the PSF-dependence of the path-loss correction, which is not modelled here. 
In order to test the reliability of the relative flux calibration in our spectra, we have measured the `intra-shutter' photometry from NIRCam imaging in the F115W, F150W, F200W, and F270W broadband filters (sampling the spectral range covered by G140M and G235M gratings), compared it to the synthetic photometry extracted by applying the transmission curve of each filter to the spectra, and verified that for the 16 galaxies analysed in the present study no significant relative flux calibration mismatch between the blue-end and red-end part of each grating was observed.

Finally, we verified the consistency of the relative flux calibration across gratings by analysing the overlapping region between G140M and G235M spectra, which roughly spans the 1.66 and 1.90 $\rm\mu m$ wavelength interval. Specifically, we calculated the ratio of continuum fluxes between the G140M and G235M gratings for seven galaxies in our sample (Sec. \ref{sec:golden}). For most galaxies, these ratios fell within 5\% of unity - i.e., between 0.95 and 1.05 $-$ demonstrating excellent agreement between the two gratings. For galaxies with ratios outside this range, the discrepancies were attributable to contaminants in the bluest or reddest regions of the spectra, as evident by inspecting their 2D spectra. We, therefore, do not perform a correction to the flux calibration among gratings.

\section{Spectral fitting}
\label{sec:fitting}

We measured spectroscopic redshifts for 73 out of the 126 targets by using strong lines, such as \OIIIopt $\lambda$5007 and H$\rm \alpha$. A detailed description of the spectral properties for the full sample will be presented in a future publication.

We performed spectral fitting on the 1D data using the \textsc{python} module of the \textsc{pPXF} (penalized pixel-fitting, \citealt{cappellari_parametric_2004}) routine, which allows for a simultaneous fit of both galaxy emission lines and the underlying stellar continuum. Since the emission lines are not spectrally resolved, we model them using a single Gaussian component per line. The emission lines included in our analysis are listed in Table \ref{tab:lines}.
For the continuum fitting, we used eMILES stellar population templates \citep{2016MNRAS.463.3409V, vazdekis_evolutionary_2010} with BASTI isochrones (\citealt{2004ApJ...612..168P}) and a \cite{chabrier_galactic_2003} initial mass function. We modelled the line spread function (LSF) with a Gaussian profile varying as a function of wavelength, using the spectral resolution profiles provided in the JWST documentation for the G140M and G235M filters\footnote{\url{https://jwst-docs.stsci.edu/jwst-near-infrared-spectrograph/nirspec-instrumentation/nirspec-dispersers-and-filters}}.

We assessed the impact of different stellar population libraries by systematically comparing the measured fluxes of key emission lines, focusing in particular on \OIIIopt$\lambda$4363. Although the strength of Balmer absorption features varies noticeably across templates, the effect on the \OIIIopt$\lambda$4363 flux remains minor—typically less than $\sim$10\%, with a median variation of $\sim$6\%. A more detailed description of these tests is provided in Appendix~\ref{sec:appB}.

A 24-degree multiplicative polynomial was applied to modify the continuum templates, accounting for potential dust reddening, residual flux calibration uncertainties, and template mismatch effects. The choice of polynomial degree was determined through a systematic analysis comparing different orders (5, 10, and 24) by evaluating the quality of the fit and the statistical properties of the residuals, with specific focus on the line-free region between rest-frame 4125 and 4325 \AA, chosen to assess the goodness of the fit in proximity to the auroral \OIIIopt$\lambda$4363 line. The 24-degree polynomial consistently yielded the best results, indicating an improved match between the model and the observed spectrum. Nonetheless, we verified that the variations in the emission lines driven by the choice of different polynomial degrees do not significantly impact the results presented in this work. 
Further discussion of the impact of the continuum subtraction on the measurement of weak auroral lines and an example spectral fit is presented in Sec. \ref{sec:golden}.
Within each grating, the velocity ($v$) and velocity dispersion ($\sigma$) parameters of the various emission lines were tied together, forcing a consistent kinematic structure for the emission regions. 
We performed the fitting separately for each of the two gratings, G140M and G235M. Where some emission lines were serendipitously observed in both gratings, we found that their fluxes were consistently in agreement within the uncertainties.

\begin{figure}[t]
    \centering
    \includegraphics[width=0.99\linewidth]{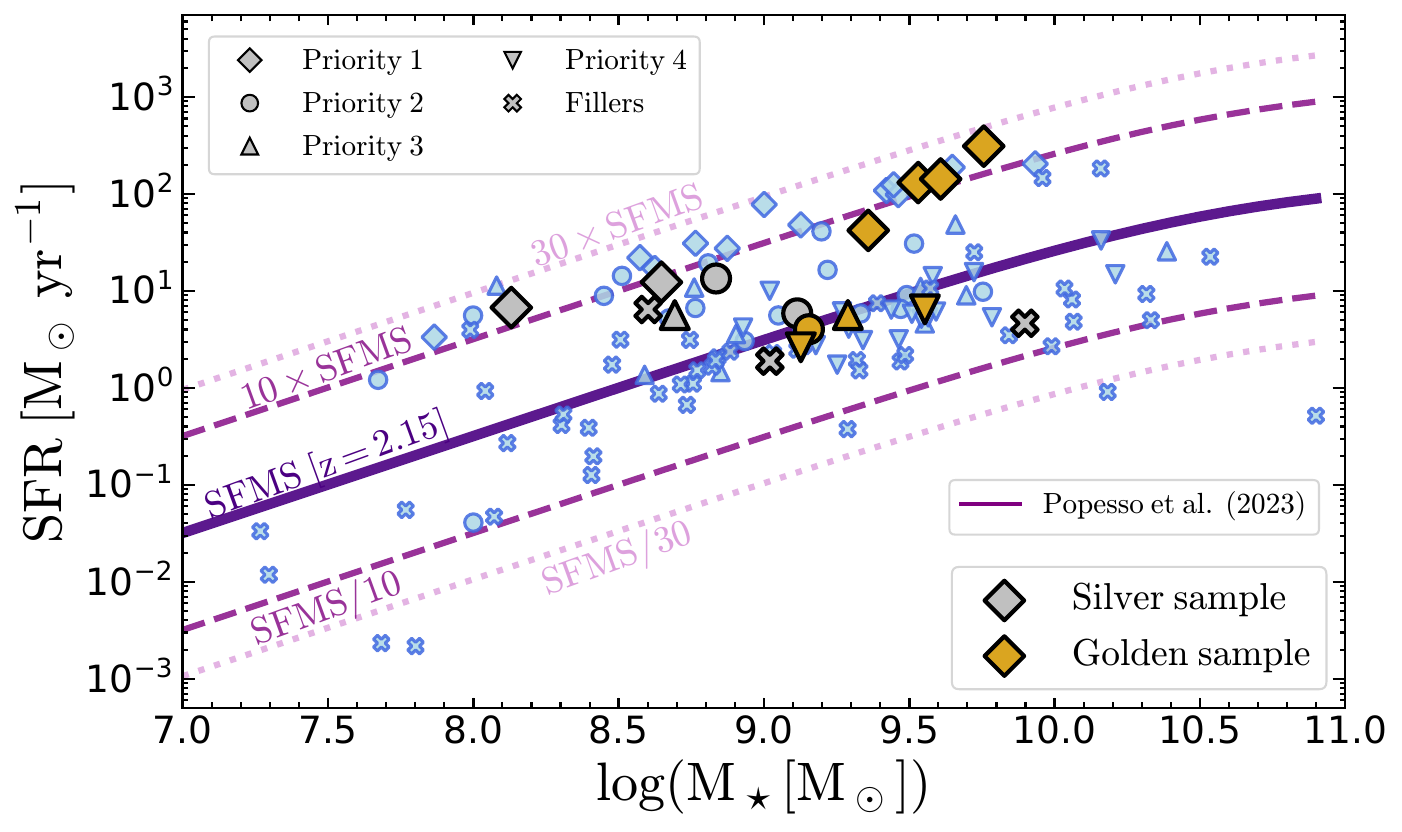}
    \caption{Star formation main sequence for the MARTA  sample. Marker shapes indicate different priority classes as described in Table ~\ref{tab:priority_classes}. The colour-coding highlights the golden sample (Sec. \ref{sec:golden}), the silver sample (Sec. \ref{sec:silver}), and the rest of the galaxies in blue. For comparison, we plot the parametrization of the star-forming main sequence (SFMS) at z$\sim$2.15 (median redshift of the full MARTA sample) by \citealt{popesso_SFMS_2023}.}
    \label{fig:MS}
\end{figure}

\begin{figure*}
    \centering
    \includegraphics[width=0.99\linewidth]{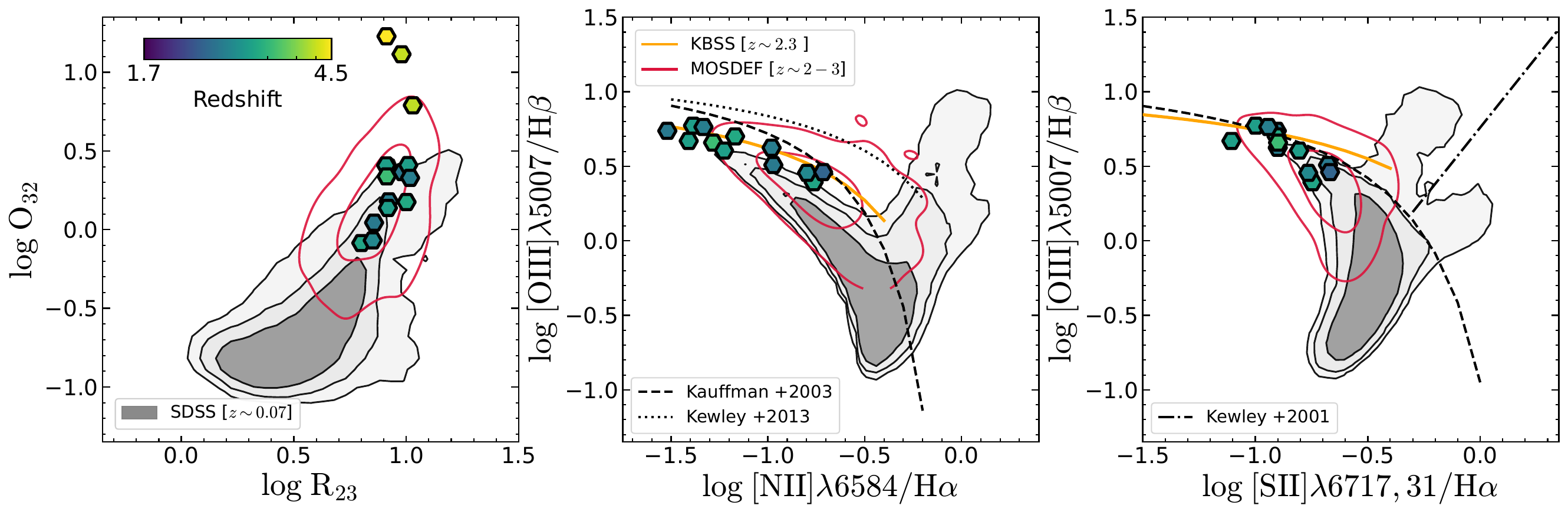}
    \caption{Diagnostic diagrams for MARTA galaxies analysed in this work. From left to right, we plot the distribution of the combined golden and silver samples on the O$_{32}$ vs R$_{23}$, \OIIIopt/\Hbeta vs \NII/\Halpha, and \OIIIopt/\Hbeta vs \SII/\Halpha diagrams, respectively, colour-coded by galaxy redshift.
    The gray contours indicate the region populated by local galaxies from SDSS, with darker shades denoting higher densities.
    Red contours mark the distribution of galaxies from the MOSDEF survey \citep{kriek_mosfire_2015}, whereas the orange curve is the fit to KBSS galaxies from \citep{strom_nebular_2017}, both surveys targeting primarily $z\sim2-3$ systems. 
    The dividing lines between star-forming galaxies, AGN, and LINER galaxies in both the \NII-BPT and \SII-BPT diagrams from \cite{kewley_theoretical_2001}, \cite{kauffmann_dependence_2003}, and \cite{kewley_cosmic_2013} are also drawn. Median error bars are not shown, as they are smaller than the symbol size and not visible in the plot; this is due to the fact that these are among the strongest lines, measured with very high signal-to-noise ratios.}
    \label{fig:BPT}
\end{figure*}

\begin{table*}
\caption{Physical properties of MARTA galaxies in the golden (top) and silver (bottom) samples.}
\centering
\begin{tabular}{cccccccccc}
\hline
\hline
Galaxy ID & $z$ & $\log{M_*}$ & $\log{\mathrm{SFR}}$ & $E(B-V)$ & $\sigma(E(B-V))$ & $T_2$$^{\dagger}$ & $T_3$ & $n_e$$^{\star}$ & 12+log(O/H) \\
 &   & $[M_\odot]$ & $[ \mathrm{M_\odot~yr^{-1}}]$ & [mag] & [mag] & [10$^4$ K] & [10$^4$ K]  & [$\rm cm^{-3}$] & \\
\hline
2387 & 1.836 & 9.76 & 2.49 & 0.00$\pm$0.00 & 0.10 & 0.97$\pm$0.13 & 1.23$\pm$0.14 & 718$\pm$193 & 8.28$\pm$0.10 (0.11)\\
5014 & 1.837 & 9.15 & 0.60 & 0.02$\pm$0.01 & 0.03 & 1.17$\pm$0.05 & 1.00$\pm$0.07 & 142$\pm$37 & 8.19$\pm$0.01 (0.02)\\
4502 & 1.854 & 9.29 & 0.75 & 0.08$\pm$0.01 & 0.11 & 1.33$\pm$0.06 & 0.94$\pm$0.05 & 75$\pm$18 & 8.31$\pm$0.03 (0.04) \\
552  & 1.924 & 9.53 & 2.12 & 0.02$\pm$0.01 & 0.03 & 0.94$\pm$0.05 & 1.16$\pm$0.03 & 263$\pm$56 & 8.35$\pm$0.03 (0.03)\\
3942 & 2.098 & 9.13 & 0.42 & 0.24$\pm$0.01 & 0.05 & 1.07$\pm$0.05 & 1.01$\pm$0.05 & 104$\pm$37 & 8.27$\pm$0.02 (0.03) \\
4195 & 2.178 & 9.55 & 0.81 & 0.18$\pm$0.01 & 0.13 & 0.91$\pm$0.07 & 0.98$\pm$0.16 & 241$\pm$61 & 8.40$\pm$0.05 (0.09) \\
4327 & 2.224 & 9.36 & 1.62 & 0.20$\pm$0.01 & 0.15 & 1.08$\pm$0.03 & 1.04$\pm$0.02 & 210$\pm$31 & 8.35$\pm$0.01 (0.02) \\
\hline
3408 & 1.861 & 9.02 & 0.28 & 0.10$\pm$0.02 & 0.12 & 1.00$\pm$0.03 & 0.97$\pm$0.05 & -- & 8.46$\pm$0.02 (0.03) \\
414  & 2.093 & 8.65 & 1.09 & 0.07$\pm$0.02 & 0.10 & 1.23$\pm$0.03 & 1.39$\pm$0.05 &  171$\pm$32 & 7.97$\pm$0.01 (0.02) \\
4471 & 2.096 & 9.11 & 0.78 & 0.12$\pm$0.02 & 0.06 & 1.40$\pm$0.06 & 1.70$\pm$0.11 & 1400$\pm$200 & 7.67$\pm$0.01 (0.03) \\
329  & 2.106 & 8.13 & 0.83 & 0.00$\pm$0.01 & 0.03 & 1.99$\pm$0.10 & 2.79$\pm$0.18 & -- & 7.21$\pm$0.02 (0.02) \\
1084 & 1.764 & 9.61 & 2.15 & 0.06$\pm$0.01 & 0.08 & 0.94$\pm$0.03 & 0.86$\pm$0.05 & -- & 8.55$\pm$0.05 (0.06) \\
3115 & 2.492 & 8.84 & 1.13 & 0.01$\pm$0.07 & --  & 1.21$\pm$0.10 & 1.35$\pm$0.18 & -- & 7.87$\pm$0.06 (0.07) \\
1374 & 3.713 & 9.90 & 0.67 & 0.00$\pm$0.34 & --  & 1.05$\pm$0.04 & 1.07$\pm$0.08 & -- & 8.32$\pm$0.26 \\
3887 & 3.796 & 8.69 & 0.75 & 0.00$\pm$0.39 & --  & 1.29$\pm$0.02 & 1.51$\pm$0.03 & -- & 7.84$\pm$0.17 \\
3926 & 4.658 & 8.60 & 0.81 & 0.00$\pm$0.24 & --  & 1.30$\pm$0.09 & 1.52$\pm$0.17 & -- & 7.75$\pm$0.23 \\
\hline
\end{tabular}
\tablefoot{Masses and SFR are from the COSMOS2020 catalogue \citep{Weaver_COSMOS2020_2022}. The $E(B-V)$ is determined from a fit to Balmer decrements as described in Section~\ref{sec:dust_attenuation}, while $\sigma(E(B-V))$ reports the standard deviation of the values obtained from individual Balmer line ratios. The uncertainties in parentheses in the metallicity column correspond to the values obtained by propagating the larger $\sigma(E(B-V))$ uncertainty into the metallicity derivation.\\
\tablefoottext{$\dagger$}{Obtained from the $T_2$–$T_3$ relation of Eq.~\ref{eq1} for the silver sample.}\\
\tablefoottext{$\star$}{Where the \SII\ doublet is not available or undetected (S/N < 3), a fixed density of 300,cm$^{-3}$ is assumed.}
}
\label{tab:galaxy_properties_updated}
\end{table*}

\section{Auroral emission lines in the MARTA sample}
\label{sec:MARTA_sample}
\subsection{Properties of galaxies with auroral line detections}

In this work, we focus on a subset of 16 galaxies with one or more auroral line detections. 

We note that this subsample has been explicitly selected and prioritised on the basis of the likelihood of \OIIIopt$\lambda$4363 detection. As a consequence, it could be biased toward galaxies with elevated auroral \OIIIopt fluxes, which in turn correspond to higher electron temperatures, lower metallicities, and higher SFRs. This selection effect must be kept in mind when interpreting the physical properties of these galaxies in a broader context. However, as we will show later in this section, strong-line diagnostic analyses do not reveal significant systematic biases. Nonetheless, potential biases could emerge in scaling relations, which are beyond the scope of this paper but should be carefully considered in future studies.

We classify these 16 galaxies into two groups: we consider a sample of seven galaxies with both \OIIIopt$\lambda$4363 and \OII$\lambda$7320,7330 detections as our `golden sample', and an additional nine galaxies with detection of \OIIIopt$\lambda$4363 only, as the `silver sample'. 
For five galaxies in the `golden sample' additional auroral lines from both high-ionisation and low-ionisation ions are detected.

Figure \ref{fig:MS} shows the distribution of all observed MARTA targets (marked according to their different priority classes) in the $M_\star$–SFR plane. 
The parametrisation of the star-forming main sequence \citep[SFMS,][]{noeske_star_2007} at $z=2.15$ (median redshift of the sample) from \citet{popesso_SFMS_2023} is also shown for comparison.
Galaxies with with auroral lines detections in MARTA sit preferentially above the SFMS at $z=2.15$, having SFRs between 5 and 30 times higher the average at fixed $M_\star$.

In Figure \ref{fig:BPT} we plot the position of the MARTA galaxies analysed in the present study (the `golden' and `silver' sample) onto different emission line-based diagnostic diagrams, namely R$_\text{23}$ vs O$_\text{32}$, the \NII-BPT diagram, and the \SII-BPT (\SII/\Halpha vs \OIIIopt/\Hbeta). 
In the R$_\text{23}$ vs O$_\text{32}$ diagram MARTA galaxies at $z>3$ are offset from both the SDSS and MOSDEF ($z\sim2-3$, \citealt{kriek_mosfire_2015}) contours, occupying a region generally associated with high-ionisation, low-metallicity systems, as recently probed by large (though shallower) JWST surveys such as JADES \citep{cameron_jades_bpt_2023} and CEERS \citep{sanders_ceers_2023}. In the BPT diagrams, MARTA galaxies at $z\sim2$ occupy the upper envelope of the local SDSS distribution, showcasing also the known offset in the \NII-BPT, \citep{steidel_strong_2014, shapley_mosdef_2015}. They overlap with the regions populated by  surveys targeting $z\sim2-3$  galaxies, such as MOSDEF and KBSS \citep{strom_nebular_2017}.

Based on classical classification schemes, such as those provided by \cite{kewley_theoretical_2001, kewley_cosmic_2013, kauffmann_dependence_2003}, we find no evidence for potential active galactic nuclei (AGN) contamination. This conclusion is corroborated also by the absence of specific AGN spectral features in our sample, such as broad components in the Balmer emission lines, very high-ionization species, or \HeII\ emission.\footnote{With the exception of one galaxy, MARTA\_4327, where such emission is interpreted as stellar in nature and associated to the presence of Wolf-Rayet stars, see Curti et al. (in prep.).} 
Furthermore, as described in Sec. \ref{sec:tt_relation}, we do not find unusually high ($\gtrsim$25000 K) electron temperatures which could be associated with AGN ionisation \citep[e.g.,][]{Mazzolari2024New}.

\subsection{The MARTA golden sample: multiple auroral line detections}
\label{sec:golden}

\begin{figure*}
    \centering
    \includegraphics[width=0.8\linewidth]{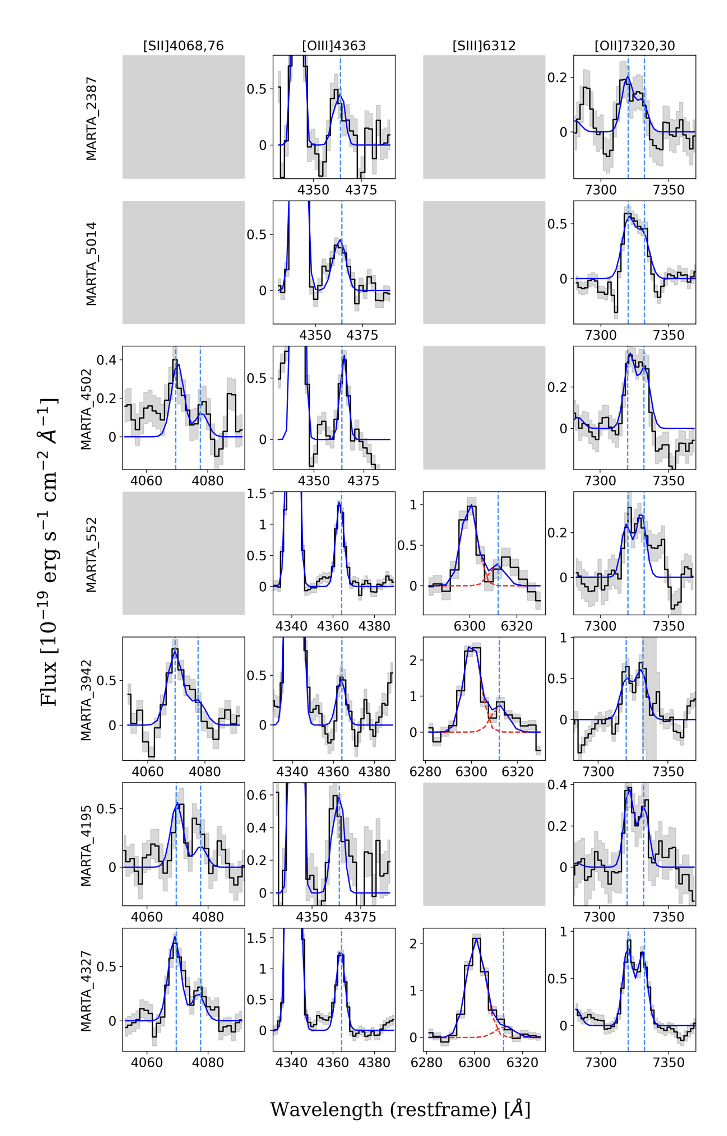}
    \caption{Auroral emission lines detected in the MARTA golden sample. The panels show emission line fits of the \SII$\lambda$4068,4076, \OIIIopt$\lambda$4363, \SIII$\lambda$6312, and \OII$\lambda$7320,7330 transitions, with their rest-frame wavelengths marked as blue dashed lines. We show the continuum-subtracted spectra in black, and the Gaussian emission-line fits in blue. In the case of \SIII$\lambda$6312, two single Gaussians are shown (red dashed lines), since the \SIII line is blended with the \OIL.}
    \label{fig:all_aurorals}
\end{figure*}

\begin{figure*}[ht]
    \centering
    \includegraphics[width=\linewidth]{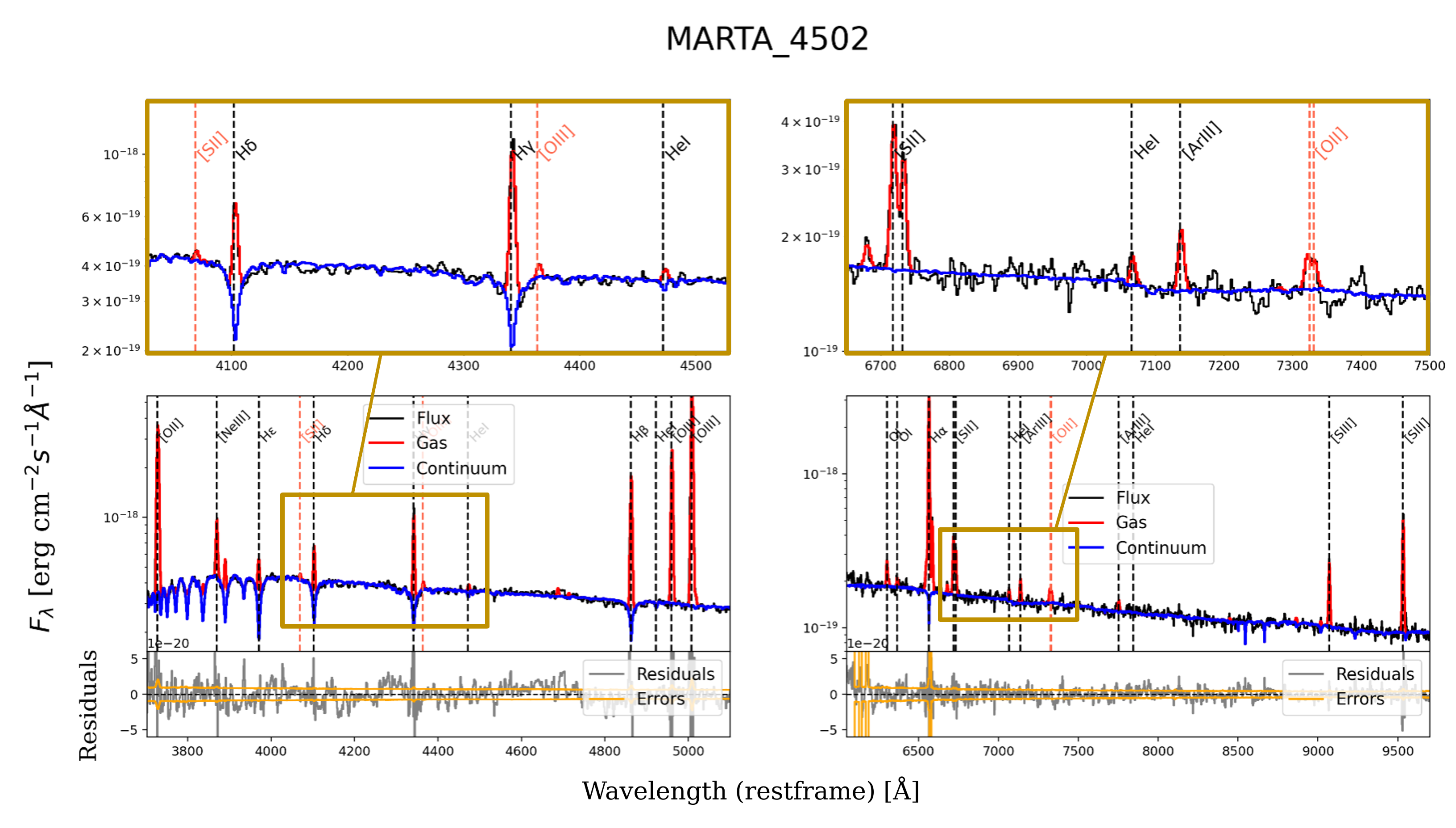}
    \caption{Spectral fit for MARTA\_4502, observed using grating G140M (left) and G235M (right). The black line represents the observed flux, the blue line represents the stellar continuum fit, and the red lines show the emission-line Gaussian fits. In the bottom panels we show in gray the residuals from the fit, while the orange line represents the pipeline error estimates. The zoom-in panels at the top provide a more detailed view of specific spectral features: on the left the auroral lines \SII$\lambda$4068 and \OIIIopt$\lambda$4363; on the right the spectral range around the \OII$\lambda\lambda$7320,7330 auroral lines.}
    \label{fig:spectral_fit}
\end{figure*}

\begin{figure*}
    \centering
    \includegraphics[width=0.75\linewidth]{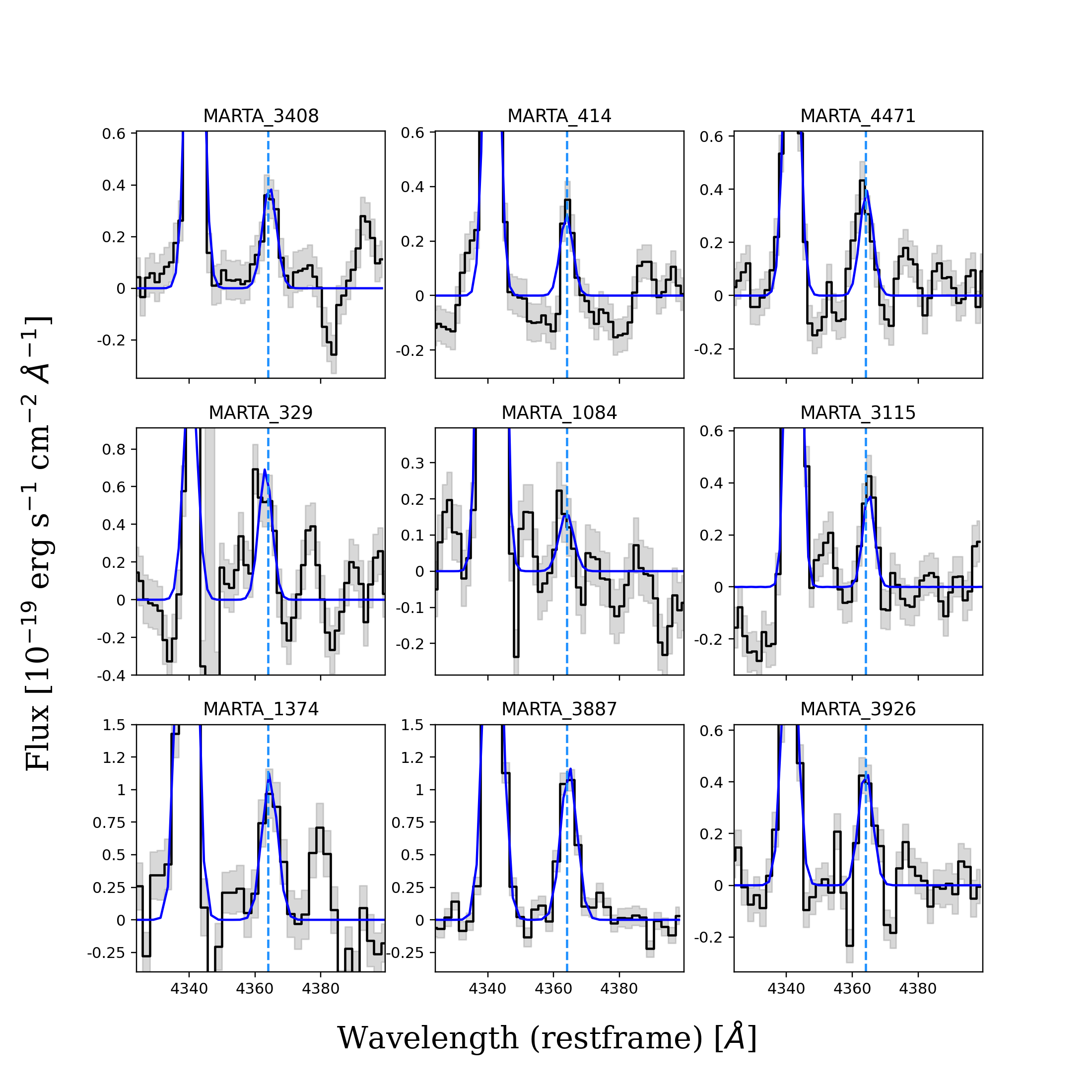}
    \caption{\OIIIopt$\lambda$4363 auroral emission line detected in the MARTA silver sample. The continuum-subtracted spectra are shown in black, with Gaussian emission-line fits overlaid in blue. The dashed light-blue lines mark the rest-frame wavelength of the \OIIIopt.}
    \label{fig:aurorals_silver}
\end{figure*}

The MARTA golden sample consists of seven galaxies with S/N $\gtrsim$ 5 detections of the \OIIIopt$\lambda$4363 auroral line and S/N $\gtrsim$ 3 for the \OII$\lambda\lambda$7320,7330 auroral doublet (where each line in the doublet is detected at $\gtrsim 3\sigma$).
The S/N on emission lines is estimated both from the output of the \textsc{ppxf} fitting, as well as from the RMS of the spectra within a featureless region around the emission line of interest; a comparison between these two methods is presented in Appendix \ref{sec:AppA}. 
As reported in Table \ref{tab:galaxy_properties_updated}, the golden sample spans $z=[1.8-2.2]$,  $\log( M_*/M_{\sun}) = [9.2-9.8]$ and SFRs ranging from 2.6 to 310 $M_{\odot}$/yr (with a mean value of 80 $M_{\odot}$/yr). 
These galaxies lie on average significantly above the median SFMS at $z\sim2$ (Fig.~\ref{fig:MS}). We quantified the \(\Delta\)MS as the distance in log(SFR) from the parametrisation of \cite{popesso_SFMS_2023}, and found \(\Delta\)MS values spanning from $-0.2$ to $+1.3$~dex, with a median value of $+0.47$~dex.

Galaxy MARTA\_1084 was excluded from this sample, despite its detection of both oxygen auroral lines, because its \OII$\lambda$3726,3729 line falls within the detector gap, precluding its use for O$^+$ electron temperature calculation and thus for direct metallicity determination. We therefore include this galaxy in the silver sample below.

For some of these galaxies we also report the detection of additional auroral lines. In particular, the \SII$\lambda$4068 auroral line is detected with S/N $>$ 3 in four cases. Similarly, the \SIII$\lambda$6312 auroral line is detected at $\gtrsim 3\sigma$ in three galaxies, where a proper de-blending from the nearby nebular \OIL[6300] line can be performed.
Figure \ref{fig:all_aurorals} shows a compilation of the auroral line detections across this sub-sample. Although MARTA\_1084 was excluded from the golden sample, it is included in the plot as it displays two auroral lines. In this figure, each panel corresponds to one of the auroral emission lines, with the continuum-subtracted spectrum shown along with the fitted Gaussian model. In some cases bespoke line fits were performed for individual objects, as detailed in Appendix \ref{sec:AppB}. 

Figure \ref{fig:spectral_fit} reports instead the complete spectral fitting, for both G140M and G235M gratings, for one of the galaxies in the golden sample - MARTA\_4502. 
This spectrum is representative of the quality and S/N of other objects in the golden sample, which are shown in Appendix \ref{sec:AppB}. 
We observe strong continuum with clear stellar absorption features under the Balmer lines. 
We observe multiple high-S/N auroral lines, including \OII$\lambda\lambda$7320,7330, \OIIIopt$\lambda$4363, and \SII$\lambda$4068. The Balmer series is observed up to H10, and generally detected up to H7 in all golden sample galaxies. In several objects, Paschen 10 and Paschen 11 are also visible. We note the presence of other semi-strong emission lines, such as \NeIIIL, \ArIIIL, and \SIII$\lambda\lambda$9071,9533, which are particularly valuable as they provide insights into the chemical abundances and ionization level of the elements in the gas, enabling more detailed studies of chemical abundance patterns \citep[e.g.][]{Stanton_excels_2025}.

\subsection{The MARTA silver sample: single auroral line detections}
\label{sec:silver}

The MARTA silver sample consists of nine galaxies, including eight additional detections of the auroral \OIIIopt$\lambda$4363 line with S/N $>$ 5 (Table \ref{tab:galaxy_properties_updated}). Additionally, we include the previously discussed MARTA\_1084 in this sample.
In Fig. \ref{fig:aurorals_silver} we show a zoom-in of the spectral region around the \OIIIoptL[4363] auroral line in these systems, with the best-fit model overlaid.

These galaxies span a broader redshift range than the golden sample objects, with two objects at $z>3$ and one at $z>4$. The silver sample also has generally lower stellar masses, $\rm{log(M_{\star}/M_{\odot})} = [8.13- 9.9]$, and a $\Delta$MS spanning from $-0.66$ to $+1.2$ dex, with a median value of $+0.43$ dex, so slightly less star forming, on average, than the golden sample.
Although these galaxies cannot be used to evaluate the relationships between the temperatures of different ionic species, they are employed in Section \ref{sec:strong_lines} to recalibrate the relationships with strong-line diagnostics.

\section{Balmer decrements and dust attenuation}
\label{sec:dust_attenuation}

To correct the measured emission line fluxes for dust attenuation, we primarily relied on the ratios among the strongest Balmer lines observed in MARTA spectra, 
namely $\rm H\alpha/H\beta$, $\rm H\gamma/H\beta$, and $\rm H\delta/H\beta$, noting that $\rm H\alpha/H\beta$ typically involves computing a ratio between lines measured across the G235M and G140M gratings. We employed the \cite{cardelli_relationship_1989} extinction curve and computed the $E(B-V)$ utilizing the \textsc{PyNeb} routine \citep{luridiana_pyneb_2012} assuming intrinsic Balmer line ratios of $\rm H\alpha/H\beta$ = 2.86, $\rm H\gamma/H\beta$ = 0.468, and  $\rm H\delta/H\beta$ = 0.259, as predicted by Case B recombination assuming an electron temperature $T_e=10^4$ K and electron density $n_e=100 ~\mathrm{cm^{-3}}$ \citep{osterbrock_astrophysics_2006}. To verify the impact of these assumptions, we implemented an iterative procedure in which the extinction correction was recomputed at each step using updated values of $T_e$ and $n_e$ derived from the previous iteration. This process was repeated until convergence. We found that the resulting changes in $E(B-V)$ and in the derived physical properties are minimal: both $T_e$ and $n_e$ vary by less than 2\% across the entire sample. These small differences confirm that the use of fixed values for the extinction correction does not introduce any significant bias in our analysis.

The E(B-V) was determined by minimizing a $\chi^2$ function defined by co-adding in quadrature the differences between the three observed Balmer line ratios and their theoretical values, given the adopted reddening curve. 
The resulting best-fit value was adopted as our fiducial E(B-V) and used to correct all line fluxes. The use of an alternative attenuation curve, such as the \cite{calzetti_dust_1994} law, the Small Magellanic Cloud (SMC) law \citep{gordon_LMC_attenuation_2003}, or the parametrization of \cite{2020ApJ...902..123R} calibrated on z$\sim$1-3 galaxies does not significantly affect our results. \cite{2023ApJ...948...83R} recently analysed galaxies at similar redshifts to our MARTA sample using Paschen line ratios and found a dust attenuation behaviour consistent with that of the SMC.

However, the $E(B-V)$ values derived individually from different Balmer line ratios within the same galaxy were not always consistent among themselves, with differences reaching up to 0.3 dex in some cases. 

We therefore also compute the dispersion among $E(B-V)$ values estimated from $\rm H\alpha/H\beta$, $\rm H\gamma/H\beta$, and $\rm H\delta/H\beta$ individually, which we report in Table \ref{tab:galaxy_properties_updated}; however, this systematic uncertainty is not propagated throughout the rest of the analysis.

Discrepancies among $E(B-V)$ as derived from different ratios of Balmer lines have been observed before at high redshift. 
For instance, \cite{2024arXiv240515859M} modelled Balmer decrements in JADES observations at $z > 2$ by studying the differential impact of radiation- and density-bounded nebulae, showing how, in the latter case, Balmer line ratios deviate from Case B values in a way that could partially explain the observed discrepancies in $E(B-V)$ when different ratios are used. In that scenario, for moderate-to-high metallicity conditions (such as those probed by MARTA galaxies), the $\rm H\gamma/H\beta$ ratio is predicted to take higher values than in the case B scenario. Hence, any dust reddening would lower that ratio closer to the case B theoretical value (mimicking no dust), while increasing $\rm H\alpha/H\beta$.

Another possibility to explain such discrepancies involves a different shape of the extinction curve in high-$z$ galaxies. Recent work by \citet{2024arXiv240805273S} showed that for a galaxy at $z=$4.41, the inferred extinction curve deviates significantly from both the \cite{calzetti_dust_1994} and \cite{cardelli_relationship_1989} laws, exhibiting a steeper slope at long wavelengths and a shallower slope in the ultraviolet. This finding suggests that commonly assumed dust laws may not be universally applicable to high-redshift galaxies. If a similar effect is present in our sample, the adoption of a different extinction curve could partially account for the discrepancies observed in $E(B-V)$ estimates.
A more detailed investigation into the behaviour of $E(B-V)$ and the disagreement between different individual Balmer line ratios is beyond the scope of this work, and will be addressed in future work.

\section{The temperature-temperature relations}
\label{sec:tt_relation}

To determine the chemical composition of a photoionised nebula, one needs to measure (or assume) its temperature and density structure \citep{stasinska_abundance_2002}.
A commonly adopted approach models the nebula as constituted by two main ionization zones: a high-ionization and a low-ionization zone. In this framework, the electron temperature of the O$^{2+}$ ion, referred to as $T_3$, characterizes highly ionized species.  
In contrast, the electron temperature of the O$^+$ ion, denoted $T_2$, represents low-ionization species such as S$^+$, or N$^+$. Some species, like S$^{2+}$ have an intermediate ionization potential between the two zones, and may therefore overlap both \citep[e.g.][]{berg_chaos_2020}.

In this section, we leverage some of the first available measurements of electron temperatures from different ionic species as delivered by JWST (including $T_2$, $T_3$, and, for a subset of targets, the temperatures associated with sulphur ions S$^{+}$ and S$^{2+}$) to study temperature-temperature relationships at high-redshift, assess possible evolution from their low-$z$ counterparts, and analyse trends with different physical properties.
Throughout this section and the rest of the paper, we have adopted the following atomic parameters and collisional strengths: \cite{palay_improved_2012} for \OIIIopt, \cite{Kisielius_2009} and \cite{10.1093/mnras/198.1.111} for \OII, \cite{2019A&A...623A.155R} and \cite{2010ApJS..188...32T} for \SII, \cite{2009JPCRD..38..171P} and \cite{1998APS..GECOWP705G} for \SIII, for collisional strength and atomic parameters, respectively.

To estimate the uncertainties on the computed properties we randomly perturbed the best-fit line fluxes by adding noise sampled from a Gaussian distribution with sigma equal to the flux uncertainties, and derived the corresponding electron temperature and density 100 times.
The fiducial values and associated uncertainties of these quantities were then quantified as the mean and standard deviation of the distributions obtained, respectively, as reported in Table \ref{tab:galaxy_properties_updated}.
Given the high signal-to-noise ratios of the Balmer lines (typically >30–300 for \Halpha and \Hbeta, and >10–100 for \Hdelta and \Hgamma across the sample), the impact of flux perturbations on the derived $E(B-V)$ is negligible. Therefore, for computational efficiency, $E(B-V)$ was held fixed during the Monte Carlo realizations, focusing the flux perturbations on the auroral and nebular oxygen lines and the \SII doublet, which have a more direct influence on the electron temperature and density determinations.

\subsection{Measuring electron temperatures and densities}

Both $T_3$ and $T_2$ were inferred from the respective nebular-to-auroral line ratio. Specifically, we simultaneously determined $T_2$ and $n_{\rm e}$ from the \OII$\lambda\lambda$7320,7330/\OII$\lambda\lambda$3726,3729 ratio and the \SII$\lambda\lambda$6716,6730 density-sensitive doublet adopting the iterative procedure implemented in the \texttt{getCrossTemDen} function of \texttt{PyNeb}. This allows us to take into account the non-negligible density dependence of the \OII\ nebular-to-auroral ratio in the $T_2$ derivation. We could not infer $n_{\rm e}$ from the \OII$\lambda\lambda$3726,3729 doublet, as these lines are not accurately deblended in our data. However, on average, \SII\ and \OII\ yield consistent estimates of electron density, as shown by, e.g. \citet[]{sanders_mosdef_2016}.

We then used the inferred electron density to calculate the electron temperature of the O$^{2+}$ ionized region, using the \OIIIopt$\lambda$4363/\OIIIopt$\lambda$5007 ratio.
For some of our galaxies the \SII doublet, necessary for determining electron density, was unavailable. In 3/9 galaxies of the silver sample (at $z>3$) the doublet fell outside the observed spectral range. In other cases, the lines fell into the detector gap (2/9 galaxies), or the spectrum was too noisy in its red part to detect the doublet (MARTA\_329). In these cases we assumed an electron density of 300 cm$^{-3}$, consistent with the typical densities measured in high-$z$ galaxies \citep[e.g.][]{sanders_mosdef_2016, Stanton_excels_2025, Topping_AURORA_density_2025}
In one case (MARTA\_3408), the ratio between the two sulphur lines placed the galaxy in the low-density limit, and we therefore assumed $n_e=100$~cm$^{-3}$.
Nonetheless, the derivation of T$_3$ is only marginally affected by density variations, and we verified that repeating the calculations with density values in the range between 10 and 1000 cm$^{-3}$ does not change the results. 

The temperature of the singly ionized sulphur (T\SII) and of the doubly ionized sulphur (T\SIII) were determined following a similar approach. T\SII was computed using the auroral-to-nebular ratio \(\text{\SII}\,\lambda\lambda4068,4076 / \text{\SII}\,\lambda\lambda6716,6730\), while T\SIII was obtained from the ratio \(\text{\SIII}\,\lambda6312 / \text{\SIII}\,\lambda9071\). Both were inferred using the \texttt{getTemDen} function in \texttt{PyNeb}, adopting the electron density previously determined in the T$_2$ calculation. 
We have also tested that measuring T\SII and $n_{\rm e}$ simultaneously with \texttt{getCrossTemDen} (as done for T\OII), instead of fixing the density to the previously inferred value, provides fully consistent outcomes.
The fluxes of all measured auroral lines in MARTA galaxies, including sulphur ones, are listed in Table \ref{tab:aurorals} in the Appendix.

\subsection{The $T_2$-$T_3$ relation}

\begin{figure}

    \centering
    \includegraphics[width=\linewidth]{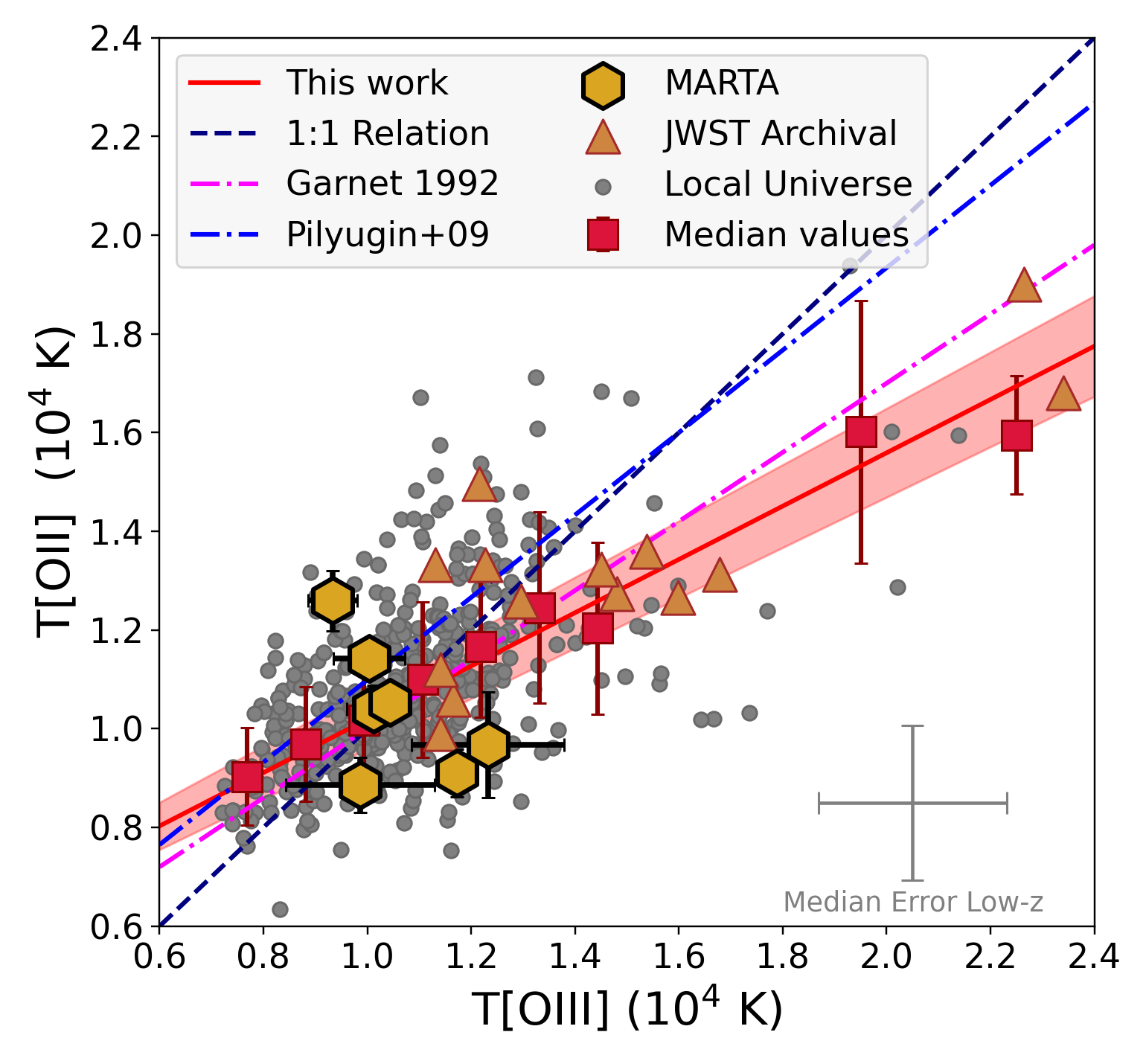}
    \caption{Relation between T\OIIIopt and T\OII for the MARTA sample (gold hexagons), compared with various samples from the literature at low (grey dots) and high redshift (from JWST archival data, orange triangles). 
    The plot also shows the 1:1 relation (blue dashed line) and two empirical calibrations of the $T_2$-$T_3$ relation, from \cite{pilyugin_electron_2009} (purple dashed line) and \cite{garnett_electron_1992} (magenta dashed line). The red squares represent the median values of the entire sample, binned in $T_3$, with uncertainties given by the standard deviation within each bin. The red line shows the best-fit relation derived in this work.
    }
    \label{fig:tt}
\end{figure}

In Fig. \ref{fig:tt} we present the measurements of the MARTA golden sample on the $T_2$-$T_3$ relation (golden hexagons), and compare them with a compilation of literature data at both low and high redshift (orange triangle markers). Currently, only a small number of high-redshift (i.e. $z>1$) galaxies have detections of both the auroral lines required for the calculation of $T_2$ and $T_3$. In particular, the high-redshift comparison sample includes SGAS 1732 from \cite{Welch2024TEMPLATES:} and the Sunburst Arc from \cite{welch2024sunburst}, as well as CEERS 11088 and CEERS 3788 from \cite{sanders_calibrations_2023}, with redshifts ranging from $\sim$ 1 to 3.5, and temperatures from the compilation of \cite{2025ApJ...985...24C} at redshift $\sim$ 3-10 taken from JADES \citep{d_eugenio_jades_DR3_2024} and \cite{2024ApJ...971...43M, Morishita2024Diverse, Welch2024TEMPLATES:, welch2024sunburst, 2024ApJ...964L..12R, 2024ApJ...962...24S}, leading to a total sample of 21 objects. 

For high-$z$ studies, we compiled the emission line fluxes where available and re-derived the electron temperatures employing the same methodology as for MARTA galaxies, which includes adopting a self-consistent set of atomic parameters.
In fact, using different sets of collision strengths results in temperature estimates that can differ by $\gtrsim 500$~K, especially for the O$^{++}$ ion \citep{nicholls_measuring_2013}. 
Therefore, when only temperatures were tabulated but we had no access to the measured line fluxes, we corrected the compiled temperatures to account for such a discrepancy, if needed.

In particular, we used \textsc{pyneb} to model the difference in the inferred $T_3$ from the \cite{palay_improved_2012} dataset (employed in this work) and
the datasets from \cite{Storey_atomic_2014} and \cite{Aggarwal_atomic_1999} (which provide comparable results to each other), and applied such an offset to the $T_3$ values from the literature that were derived adopting either of the two latter atomic parameters \citep[as in e.g.][]{2025ApJ...985...24C}. This approach brings all $T_3$ values analysed in this work to a consistent scale. Such correction was not required for $T_2$, as all literature studies considered adopted the same atomic parameters.

We also compare these data with a large compilation of measurements in low-redshift galaxies (small grey points in Figure~\ref{fig:tt}), for which we compiled emission-line fluxes and re-measured temperatures self-consistently with the high-$z$ sample.
In particular, we include data from low-metallicity local emission-line galaxies from \cite{2009A&A...503...61I, 2009A&A...505...63G, guseva_vlt_2011}; the \cite{curti_new_2017} sample, consisting of composite spectra of SDSS Data Release 7 star-forming galaxies
stacked based on their combined \OII/H$\beta$ and \OIIIopt/H$\beta$ flux ratios; a sample of SDSS galaxies where both oxygen auroral lines were detected in individual objects \citep[see e.g.][]{pilyugin_abundance_2012};  the CHAOS data targeting \hii\ region in massive local spiral galaxies with the MODS spectrograph on the LBT \citep{berg_chaos_2015, berg_chaos_2020, rogers_chaos_2021, rogers2022chaos};
the stacks of \cite{Khoram_2025}, which consist of spatially resolved spectra from SDSS-IV MaNGA stacked in bins across the $M_\star$-SFR plane;
and the \cite{2021MNRAS.500.2359Z} compilation of 2831 published \hii region emission-line fluxes in 51 nearby spiral galaxies. In all these cases we recovered the measured extinction-corrected fluxes and re-calculated temperatures following the same approach used for the MARTA data, applying a uniform S/N $>$ 3 cut on the auroral lines.

Furthermore, in Fig. \ref{fig:tt} we also plotted some empirical calibrations of the $T_2$-$T_3$ relation. In particular, \cite{pilyugin_electron_2009} calibrated the relation on a sample of \hii\ regions from the local Universe, while \cite{garnett_electron_1992} provided a fit based on photoionization models.

Some galaxies in the local sample exhibited evident contamination from \FeII \citep[e.g.][see also Moreschini et al., in prep]{curti_new_2017, Rogers_CHAOS_NGC2403_2021, Hamel-Bravo_DUVET_2024}.
This contamination led to artificially elevated $T_3$ values due to the consequent overestimation of the \OIIIopt$\lambda$4363 flux. To exclude contaminated galaxies in our analysis, we applied an empirical cut (discussed in Moreschini et al., in prep) defined as $\log(R/R3) > -2.1 + \log({\rm O}32)$, where $R \equiv \OIIIopt \lambda 4363/ \Hbeta $, as contaminated. This cut ensured the exclusion of the vast majority of the SDSS stacks flagged as contaminated by \cite{curti_new_2017} through visual inspection. 

The MARTA data shown in Fig. \ref{fig:tt} (orange hexagons) exhibits significant scatter, which is also evident in the data from local galaxies in the literature (grey points, typical errors shown in light grey). 
The red squares in Fig. \ref{fig:tt} indicate the median values for the whole sample, binned in $T_3$, with error bars reporting the standard deviation within each bin. The binning was chosen to ensure a minimum of five galaxies per bin, adapting the bin width in the high-$T_3$ regime, where the data is more sparsely sampled.
The red line represents a linear fit derived in this work, based on the entire sample. This fit was performed using an orthogonal linear regression algorithm, applied to the median values rather than individual points, to reduce the influence of sampling effects in the diagram. The best-fit linear relation is given by:
\begin{equation}
    T_2 = (0.54 \pm 0.03)~ T_3 + (4790 \pm 300) ~\mathrm{K}.
\label{eq1}
\end{equation}

The slope obtained from our fit is significantly shallower than those reported by e.g. \cite{garnett_electron_1992} and \cite{pilyugin_electron_2009}. This effect arises from the inclusion of high-temperature points, which tend to be systematically lower in $T_2$ than in $T_3$ when $T_3 > 14\,000 $K. This behaviour is seen consistently in both low- and high-redshift data. This high-$T_3$ range is however populated by only $\sim$ 30 data points in our large compilation and has therefore been previously hard to confirm observationally in the literature. Hints of this behaviour were however already observed by \cite{mendez_delgado_density_biases_2023} and \cite{2020MNRAS.497..672A} when studying the T\NII-T\OIIIopt relation.

We performed an alternative fit considering only the range $T_3 = [7000-14\,000]$K, and we obtained a steeper slope of $0.64 \pm 0.04$, closer to previous calibrations. 
Furthermore, we compared our orthogonal fit with the median values of the bin with a Bayesian fit performed on individual points using the \textsc{linmix} method \citep{Kelly_2007}. The results are consistent within two sigmas, with the Bayesian fit yielding a slope of $0.50 \pm 0.03$ and an intercept of $5510 \pm 330$ K.
 
We find that the $T_2$-$T_3$ relation  is affected by significant scatter across redshift. This scatter could reflect both observational uncertainties, the adoption of different atomic datasets, as well as be driven by the combination of different physical processes (see Sec.~\ref{sec:te_te_trends}).  
For instance, the use of \OII as a tracer of the low-ionization zone temperature is particularly controversial. While models predict that T\OII should be comparable to T\SII and T\NII \citep{campbell_stellar_1986, pilyugin_relation_2006}, in practice T\OII tends to overestimate the temperature compared to T\NII and T\SII \citep{zurita_chemical_2012, curti_new_2017, berg_chaos_2020, mendez_delgado_t_inhomogeneities_2023}. This discrepancy may be explained by density inhomogeneities within \hii regions, which affect the \OII and \SII ratios but have little impact on the \NII lines \citep{mendez_delgado_t_inhomogeneities_2023}, as well as from the effect of recombination on the emissivity of the \OII auroral doublet.
 
Nonetheless, our observations shows that high-$z$ systems are overall consistent with the distribution of local galaxies on the T$_2$-T$_3$ relationship, with no systematic offset observed. This finding is in good agreement with the independent analysis by \citet{2025arXiv250810099S}, who likewise report no evidence for an evolution of the $T_2$–$T_3$ relation with redshift.
This consistency suggests that, while the relationship is not particularly tight, the underlying physical processes governing $T_2$ and $T_3$ do not vary significantly across cosmic time.

\begin{figure*}[th]

    \centering
    \includegraphics[width=\linewidth]{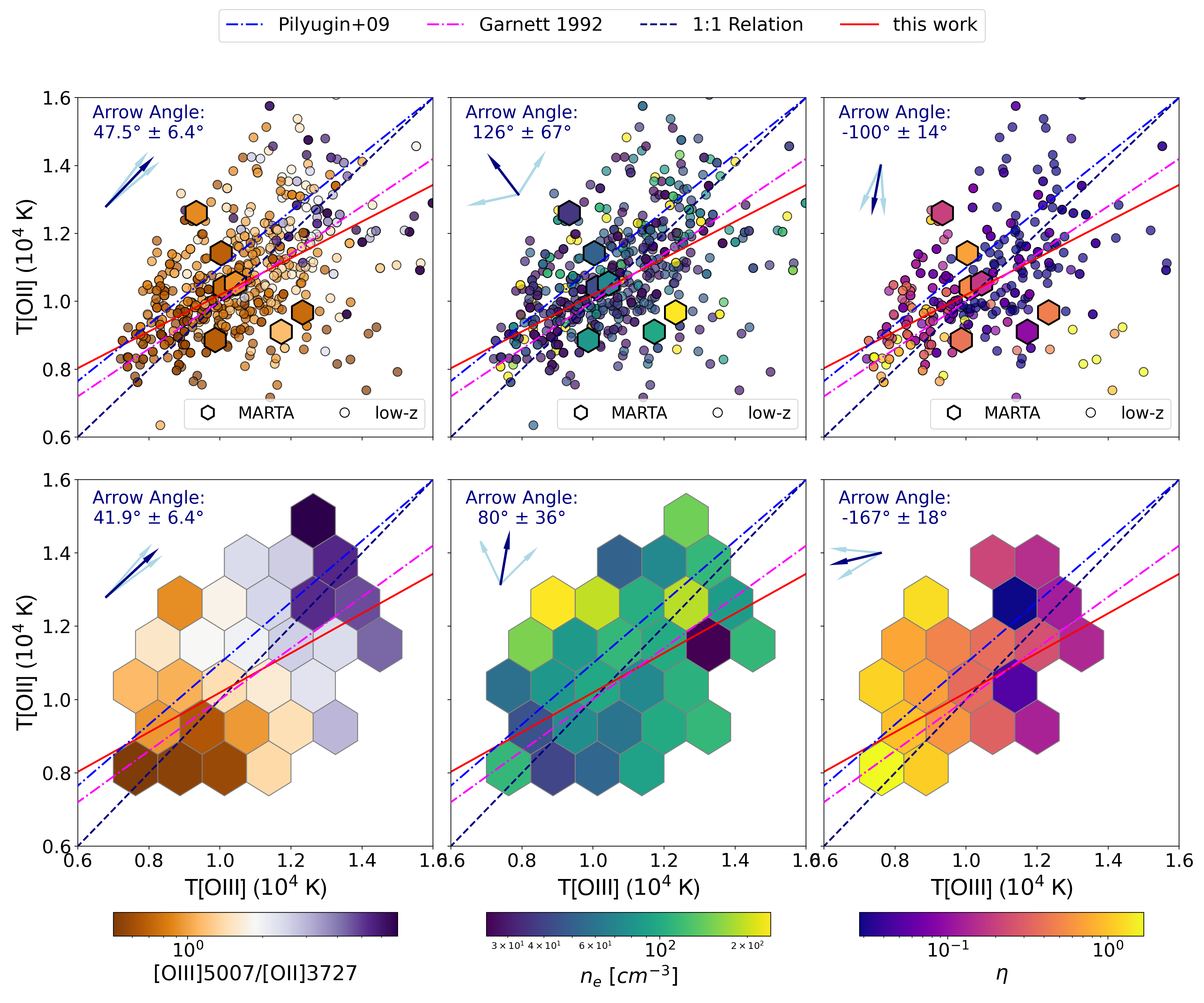}
    \caption{Dependence of the $T_2$–$T_3$ relationship on different physical properties. The figure consists of six panels: the top row presents individual data points, while the bottom row shows the corresponding hexagonal binning. The colour coding is consistent across both rows and represents the three selected physical properties, computed as median values within each bin in the bottom panels: the $\OIIIopt/\OII$ line ratio (left), the electron density $n_{\rm e}$ (centre), and the softness parameter $\eta$ (right). In the top panels, MARTA objects are marked with hexagons, while low-redshift galaxies are shown as circles. The plots also include the \citet{pilyugin_electron_2009} relation, the \citet{garnett_electron_1992} relation, the 1:1 relation, and our best-fit model. The direction of maximum variation in the colour-coded quantity, determined by the gradient angle, is indicated by navy arrows, with the associated uncertainty shown for reference as light-blue arrows. This provides a visual representation of how each parameter varies across the $T_2$–$T_3$ plane.}
    \label{fig:tt_hexbin}
\end{figure*}

\subsection{Dependence of the $T_2$-$T_3$ relation on additional physical parameters}
\label{sec:te_te_trends}

We explore the possible dependence of the $T_2$-$T_3$ relation on additional physical properties to probe the origin of the observed scatter. In particular, we consider three physical parameters: electron density derived from the \SII ratio ($n_{\rm e}$), the \OIIIopt/\OII line ratio as a proxy for the ionization parameter, and the softness parameter \citep{1988MNRAS.231..257V}, defined as:
\begin{equation}
\eta \equiv \frac{ (\OII3726 + \OII3729)\, / \, (\OIIIopt4959+\OIIIopt5007)}{( \,\SII 6730 + \SII 6716 \,) /( \, \SIII 9069 + \SIII 9532 \, )}.
\label{eq:eta}
\end{equation}
This parameter is used as a proxy for the hardness of ionizing radiation, with a secondary dependence on gas-phase metallicity and other nebular parameters. Since galaxies at high redshift are known to show a different distribution in these three parameters compared to their low-redshift counterparts \citep[e.g.,][]{sanders_mosdef_2016, kaasinen_ionization_2018}, a dependence of the scatter or the residuals of the $T_2$-$T_3$ relation on these parameters would provide evidence for a potential redshift evolution.


Figure \ref{fig:tt_hexbin} presents six panels displaying the $T_2$-$T_3$ relation colour-coded by \OIIIopt/\OII, $n_{\rm e}$, and $\eta$ for both individual points and hexagonal bins. The hexagonal bins were computed with a minimum of five objects per bin to ensure statistical robustness. The $\eta$ diagram contains fewer data points compared to the first two, as only some galaxies in our sample had detections of all four emission lines required to compute $\eta$. In particular, the \SIII$\lambda$9070, 9532 are not available for SDSS galaxies, as they fall outside the observed spectral coverage.

To assess potential correlations between the temperatures, gas density, \OIIIopt/\OII{} ratio, and $\eta$, we conducted a partial correlation analysis on the full sample, including both local and high-redshift galaxies, using both individual data points and hexagonal bins. The corresponding partial correlation coefficients ($r$), 95\% confidence intervals (CI95\%), and p-values are summarized in Table~\ref{tab:partial_corr}. The table reports all partial correlation coefficients computed for both binned and non-binned data, allowing for a direct comparison of the two approaches.

The results indicate that both $T_3$ and $T_2$ exhibit a strong positive correlation with the \OIIIopt/\OII ratio.
This correlation is significantly stronger when computed on the hexagonal bins compared to individual data points, suggesting that binning reduces scatter and enhances the underlying trend. However, the correlation is statistically significant in both cases, confirming its robustness.

In contrast, $T_2$ and $T_3$ exhibit weak correlations with gas density, both from single points and binned data. However, a marginally significant ($\sim 2\text{--}4\sigma$) positive correlation is observed between $T_2$ and density. The stronger correlation between $T_2$ and density (compared to $T_3$ and density) may arise because the \SII and \OII diagnostics, being low-ionization tracers, probe similar gas regions more effectively than \OIIIopt and \SII. Moreover, the \OII auroral-to-nebular line ratio used for temperature diagnostics is highly sensitive to density variations, potentially inducing a spurious trend. A more comprehensive assessment of this trend would require density diagnostics for higher-ionization lines, which could provide a more detailed representation of the emitting gas distribution and allow us to test a potential dependence of $T_3$ on density.

Regarding $\eta$, the results exhibit a strong dependence on binning, with substantial differences between binned and non-binned analyses. In the non-binned data, strong and seemingly significant negative correlations are observed with both temperatures. However, the correlation with $T_2$ is likely driven by a sparsely populated region at high $T_3$ and high $\eta$, which skews the overall trend downward. These high-$T_3$, high-$\eta$ data points are absent in the binned analysis, leading to a much weaker correlation in the hexbin results. In contrast, the correlation with $T_3$ remains more robust, as it persists in both the binned and non-binned cases, suggesting a more intrinsic relationship between $T_3$ and $\eta$.

To further explore these trends, we examined the gradient angle of the colour-coded parameters within the $T_2$-$T_3$ plane, which is displayed as a navy arrow in each panel of Fig. \ref{fig:tt_hexbin}. The gradient angle represents the direction of maximum variation in a given physical parameter (e.g., \OIIIopt/\OII) and is computed as:
\begin{equation}
\theta_{\text{grad}} = \arctan\left(\frac{r_{T_2, Z}}{r_{T_3, Z}}\right),
\end{equation}
where $r_{T_2, Z}$ and $r_{T_3, Z}$ are the partial correlation coefficients of $T_2$ and $T_3$ with a given property $Z$ \citep[see e.g.,][]{bluck_are_2019, Curti_BPT_2022, baker_resolved_fmr_2023}. We quantified the uncertainty in the gradient angle via a bootstrapping procedure, randomly resampling the dataset (with replacement) 300 times and recalculating the angle for each iteration. The standard deviation of the resulting angle distribution provides the error, represented by the light blue arrows in Figure \ref{fig:tt_hexbin}. We tested using a larger number of samples for bootstrapping and the inferred uncertainty remains consistent.

Notably, the large errors observed in the central panels of Figure \ref{fig:tt_hexbin} reflect the weak statistical significance of the partial correlation analysis for $n_{\rm e}$. The visual near-alignment of the \OIIIopt/\OII gradient angle with the best-fit $T_2$-$T_3$ relation seems to point that the orthogonal scatter within this relation is not strongly correlated with this parameter. Such correlation could arise from a metallicity-driven effect. In fact, higher metallicity leads to softer ionizing spectra \citep{mcgaugh_h_1991, Bresolin1998The}, lowering the ionization parameter and reducing the \OIIIopt/\OII\ ratio. Conversely, lower metallicity results in higher ionization parameters and \OIIIopt/\OII\ ratios \citep{Ji2021Correlation, Grasha2022Metallicity}. Since lower metallicity environments also exhibit higher temperatures, this trend naturally explains the observed correlation. Nevertheless, a more quantitative assessment is needed to robustly evaluate the contribution of the ionization parameter to the scatter around the $T_2$–$T_3$ relation.

In addition, we performed an alternative correlation analysis based on the residuals of the $T_2$-$T_3$ relation relative to our best-fit linear fit in Eq. \ref{eq1}. 
In particular, we defined the quantity $\Delta T$ for each point in the sample as its orthogonal deviation from Eq. \ref{eq1}, measured as ($\Delta T$)$^{2}$ = ($\Delta T_3$)$^{2}$ + ($\Delta T_2$)$^{2}$.
We then calculated the Spearman correlation coefficient between $\Delta T$ and each of the physical parameters involved in the analysis.
The results reveal a mild yet statistically significant positive correlation (coefficient of 0.15, significant at $\sim 3\sigma$) between the amplitude of the offset from the temperature relation $\Delta T$ and the \OIIIopt/\OII ratio. This suggests that, although the effect is not dominant, variations in the ionisation parameter -as traced by \OIIIopt/\OII- do contribute to the scatter around the $T_2$–$T_3$ relation.

In contrast, both $n_e$ and $\eta$ show minimal effect and weak statistical significance. These results suggest that the orthogonal scatter relative to the best-fit relation does not exhibit a clear dependence—at least not one detectable from the residuals—on these two quantities.

\begin{table*}
\caption{Results of the partial correlation analysis on the $T_2$-$T_3$ relation.}
\centering
\begin{tabular}{lccc}
\hline
\hline
Correlation & $r$ & CI95\% & p-value ($\sigma$) \\ 
\hline
\multicolumn{4}{c}{Individual Data Points} \\
\hline
$T_2$ vs $n_{\rm e}$ | $T_3$ & 0.15 & [0.08, 0.23] & $1 \times 10^{-4}$ ($\sim3.9\sigma$) \\ 
$T_3$ vs $n_{\rm e}$ | $T_2$ & 0.03 & [-0.04, 0.11] & 0.4 ($<1\sigma$) \\
$T_2$ vs \OIIIopt/\OII | $T_3$ & 0.57 & [0.51, 0.62] & $< 10^{-6}$ ($>5\sigma$) \\ 
$T_3$ vs \OIIIopt/\OII | $T_2$ & 0.34 & [0.28, 0.41] & $< 10^{-6}$ ($>5\sigma$) \\ 
$T_2$ vs $\eta$ | $T_3$ & -0.38 & [-0.46, -0.29] & $< 10^{-6}$ ($>5\sigma$) \\ 
$T_3$ vs $\eta$ | $T_2$ & -0.50 & [-0.57, -0.42] & $< 10^{-6}$ ($>5\sigma$) \\  
\hline
\multicolumn{4}{c}{Hexagonal Bins} \\
\hline
$T_2$ vs $n_{\rm e}$ | $T_3$ & 0.35 & [-0.03, 0.64] & 0.07 ($\sim1.8\sigma$) \\ 
$T_3$ vs $n_{\rm e}$ | $T_2$ & 0.02 & [-0.35, 0.39] & 0.9 ($<1\sigma$) \\ 
$T_2$ vs \OIIIopt/\OII | $T_3$ & 0.80 & [0.61, 0.90] & $< 10^{-6}$ ($>5\sigma$)  \\ 
$T_3$ vs \OIIIopt/\OII | $T_2$ & 0.86 & [0.71, 0.93] & $< 10^{-6}$ ($>5\sigma$)  \\ 
$T_2$ vs $\eta$ | $T_3$ & -0.05 & [-0.49, 0.41] & 0.8 ($<1\sigma$) \\ 
$T_3$ vs $\eta$ | $T_2$ & -0.88 & [-0.95, -0.71] & $< 10^{-6}$ ($>5\sigma$)  \\  
\hline
\multicolumn{4}{c}{Orthogonal offsets from the $T_2$-$T_3$ relationship best fit} \\
\hline
$\Delta$T vs \OIIIopt/\OII & 0.151 & [0.05, 0.24] & $2.2 \times 10^{-3}$ ($\sim 3\sigma$) \\ 
$\Delta$T vs $n_{\rm e}$ & 0.112 & [0.02, 0.21] & $0.02$ ($\sim 2.3\sigma$) \\  
$\Delta$T vs $\eta$ & 0.088 & [-0.01, 0.18] & $0.07$ ($\sim 1.8\sigma$) \\  
\hline
\end{tabular}

\tablefoot{The table reports the Spearman correlation coefficients ($r$), their 95\% confidence intervals (CI95\%), and p-values (converted into significance in $\sigma$ where appropriate). The first set of values corresponds to the analysis performed on the full sample, while the second set refers to the analysis conducted on median values within hexagonal bins.
The bottom part of the table reports instead the
analysis of the correlation between the same physical parameters and the orthogonal offset $\Delta T$ from the best-fit relation of Eq.~\ref{eq1}, defined as in Section~\ref{sec:te_te_trends}. 
}
\label{tab:partial_corr}
\end{table*}

\begin{figure*}
    \centering
    \includegraphics[width=\linewidth]{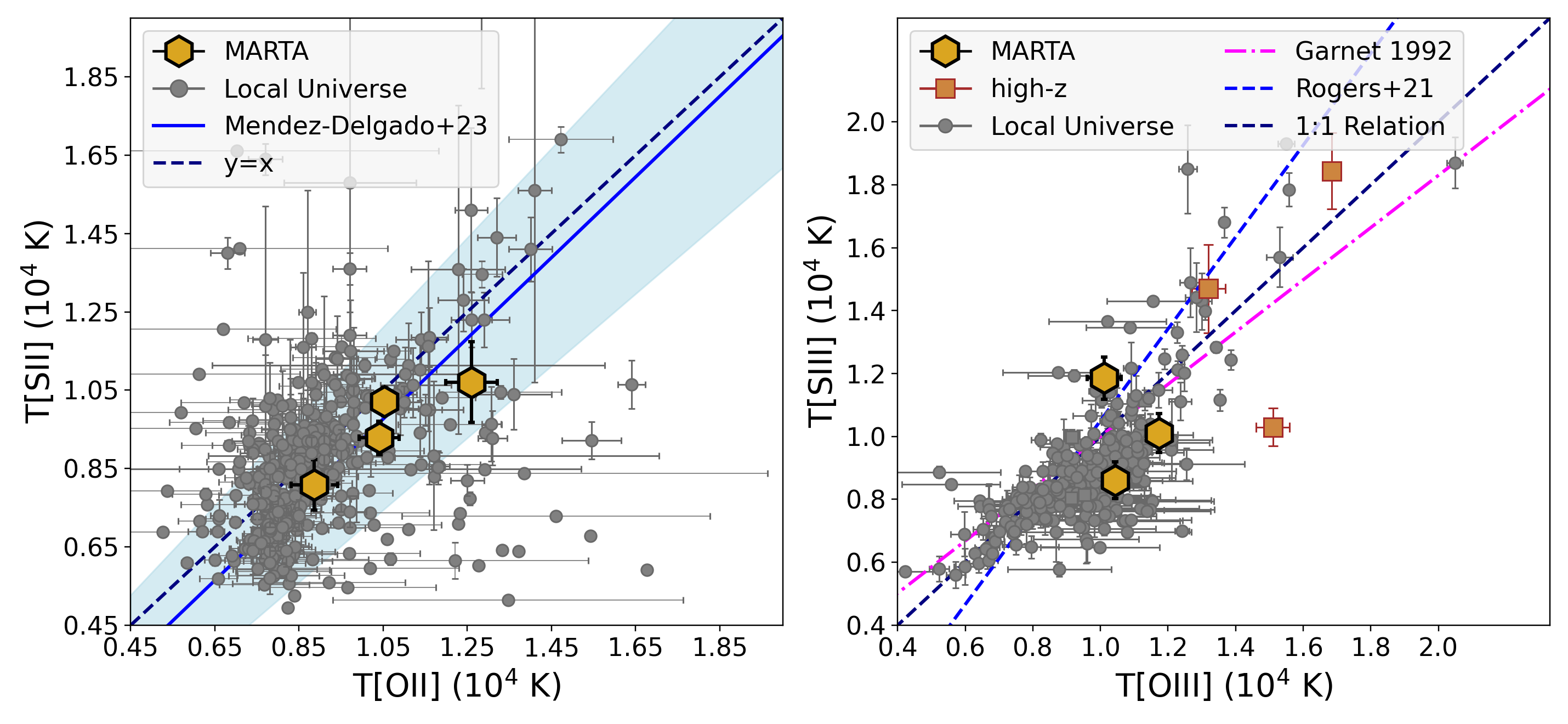}
    
    \caption{Left Panel: T\SII - T\OII relation for MARTA objects, compared with extragalactic \hii regions 
    together with the empirical calibration by \cite{mendez_delgado_t_inhomogeneities_2023}, based on local \hii regions. 
    Right Panel: T\SIII - T\OIIIopt relation for MARTA objects, along with local \hii region data 
    and two high-redshift galaxies with archival JWST data \citep{welch2024sunburst,2024ApJ...964L..12R}. The plot also shows the calibrations from \cite{garnett_electron_1992} and \cite{rogers_chaos_2021}. }
    \label{fig:TT_sulfur}
\end{figure*}

\subsection{Sulphur temperatures}

Sulphur is a valuable tracer of ionized gas conditions in galaxies, as its ionization states probe different regions of the ISM. The \SIII\ and \SII\ ions have ionization potentials slightly different from those of \OIIIopt\ and \OII, allowing sulphur lines to trace intermediate and low-ionization regions, respectively \citep{berg_4zones_2021}. Comparing sulphur and oxygen temperatures is therefore a useful diagnostic for understanding the ionization structure of \hii\ regions and testing photoionization models.

The availability of reliable detections of sulphur auroral lines in some of our selected high-redshift galaxies presents a valuable opportunity to compare such temperatures. 
Our work marks one of the first instances where this relationship is examined for galaxies at high redshift \citep{2024ApJ...964L..12R, welch2024sunburst, Welch2024TEMPLATES:, 2025ApJ...979...87M}.
Photoionization models predict that $T\SII \approx T\OII$ \citep{campbell_stellar_1986, pilyugin_relation_2006, M_ndez_Delgado_2023}. $T\OIIIopt$, on the other hand, is expected to be higher than $T\SIII$ for \OII temperatures higher than $8000$K \citep{garnett_electron_1992}. Substantial deviations from both these trends have, however, been observed in extragalactic \hii\ regions \citep{kennicutt_composition_2003, esteban_keck_2009, berg_chaos_2015, croxall_chaos_2016, rogers_chaos_2021}.

As shown in Fig. \ref{fig:TT_sulfur}, the observed trends between the temperatures of sulphur and oxygen ions are consistent with the previous literature and empirical calibrations. In particular, we compared the MARTA objects with extragalactic \hii\ regions \citep{esteban_keck_2009, esteban_carbon_2014, esteban2020carbon,dominguez2022homogeneity, rogers2022chaos}. The left panel of Fig. \ref{fig:TT_sulfur} shows the T\SII - T\OII relation and the empirical calibration by \cite{mendez_delgado_density_biases_2023}, based on local \hii regions. Our high-redshift sample lies systematically below the 1:1 relation, but such a trend aligns with the findings of \cite{mendez_delgado_density_biases_2023}, who also report an offset in the relation, with \OII temperatures generally higher than \SII ones. They attributed this effect to the presence of density fluctuations.

In the right panel of Fig. \ref{fig:TT_sulfur} we show the the $T\SIII - T\OIIIopt$ relation. This plot also includes the sample of \hii\ regions from CHAOS \citep{berg_chaos_2015, 2015ApJ...808...42C, croxall_chaos_2016, berg_chaos_2020}, and three high-redshift galaxies: the Sunburst Arc (\citealt{welch2024sunburst}), Q2343-D40 from \cite{2024ApJ...964L..12R}, and ID60001 from \cite{2025arXiv250204817Z} which, to the best of our knowledge, are the only other galaxies at high redshift with simultaneous measurements of the \SIII and \OIIIopt auroral lines. 
The figure also shows two calibrations from the literature. The relation in \cite{garnett_electron_1992} was based on photoionization models and the one in \cite{rogers_chaos_2021} was calibrated on local \hii\ regions from the CHAOS survey.

The relation appears somewhat scattered but is still broadly consistent with results from the local literature and comparable to the other high-redshift galaxies. This result is also consistent with the recent findings of \citet{2025arXiv250810099S}, who reported no evidence for systematic redshift evolution in $T$–$T$ relations involving sulphur ions.
In models, these two temperatures are not always equal, since \SIII and \OIIIopt likely trace different ionization zones due to their significantly different ionization potentials, as discussed in \cite{berg_4zones_2021}. 

Quantitatively, our $T\SII$ measurements show an average negative offset of $\sim$ 930 K from the one-to-one relation and $\sim$ 470 K from the \cite{mendez_delgado_density_biases_2023} relation. These correspond to average deviations of 1.14$\sigma$ and 0.52$\sigma$ (considering our estimates of the temperatures error bars), respectively, indicating that our measurements remain well within expectations based on previous studies. Notably, three out of the four galaxies lie directly on the best-fit relation of \cite{mendez_delgado_t_inhomogeneities_2023}, while the fourth remains consistent within the best-fit uncertainty range. 

Similarly, for $T\SIII$, we find the average of the absolute value offset to be 1.9$\sigma$ from the 1:1 relation, while the offsets relative to \cite{garnett_electron_1992} and \cite{rogers_chaos_2021} are 1.9$\sigma$, and 2$\sigma$, respectively. This time the offset is not systematically negative but the data scatter around the relation. 

\section{Strong-line metallicity diagnostics and calibrations}
\label{sec:strong_lines}
Strong-line methods based on combination of nebular lines ratios including $\OII\lambda\lambda 3726,3729$, $\OIIIopt \lambda 4959$, $\lambda 5007$, and $\NII \lambda 6584$ are a valuable way to measure the metallicity of galaxies, in virtue to their brightness and ease of detection across a wide range of redshifts \citep{maiolino_re_2019}. 
Nonetheless, the redshift evolution in the physical conditions that are at the basis of the (either direct or indirect) dependence between the involved line ratios and metallicity possibly translates into biases when using such methods to derive the metallicity from high-$z$ spectra adopting diagnostics calibrated on local galaxy samples \citep{bian_ldquodirectrdquo_2018, sanders_mosdef_mzr_2021}.

The advent of JWST promised to directly tackle this issue by delivering robust auroral lines detections in high-$z$ galaxies,
and this was indeed one of the key objectives of several programmes within the first few cycles. 
In this Section, we combine T$\rm _e$-based metallicity measurements for MARTA galaxies with a compilation of measurements from the literature at different redshifts, and test the validity and evolution of various classical strong-line calibrations in the high-$z$ Universe.

\subsection{Oxygen abundance determination}
We derived the ionic abundances of the O$^+$ and O$^{++}$ ions in the MARTA golden sample galaxies leveraging the direct measurement of the respective temperatures and employing the \texttt{GetIonicAbundances} routine in \texttt{pyNeb}. 
To measure the O$^+$ abundance in the MARTA silver sample, we inferred the  $T_2$ from the $T_3$ as measured from \OIIIoptL[4363]/\OIIIoptL[5007] ratio adopting our newly derived fit of Equation~\ref{eq1}.
The total oxygen abundance is then computed as the sum of these two ionic abundances, with the contribution from higher ionization states like O$^{3+}$ assumed to be negligible. The results of these abundance calculations are presented in Table \ref{tab:galaxy_properties_updated}.

\begin{figure*}
    \centering
    \includegraphics[width=1.1\textwidth]{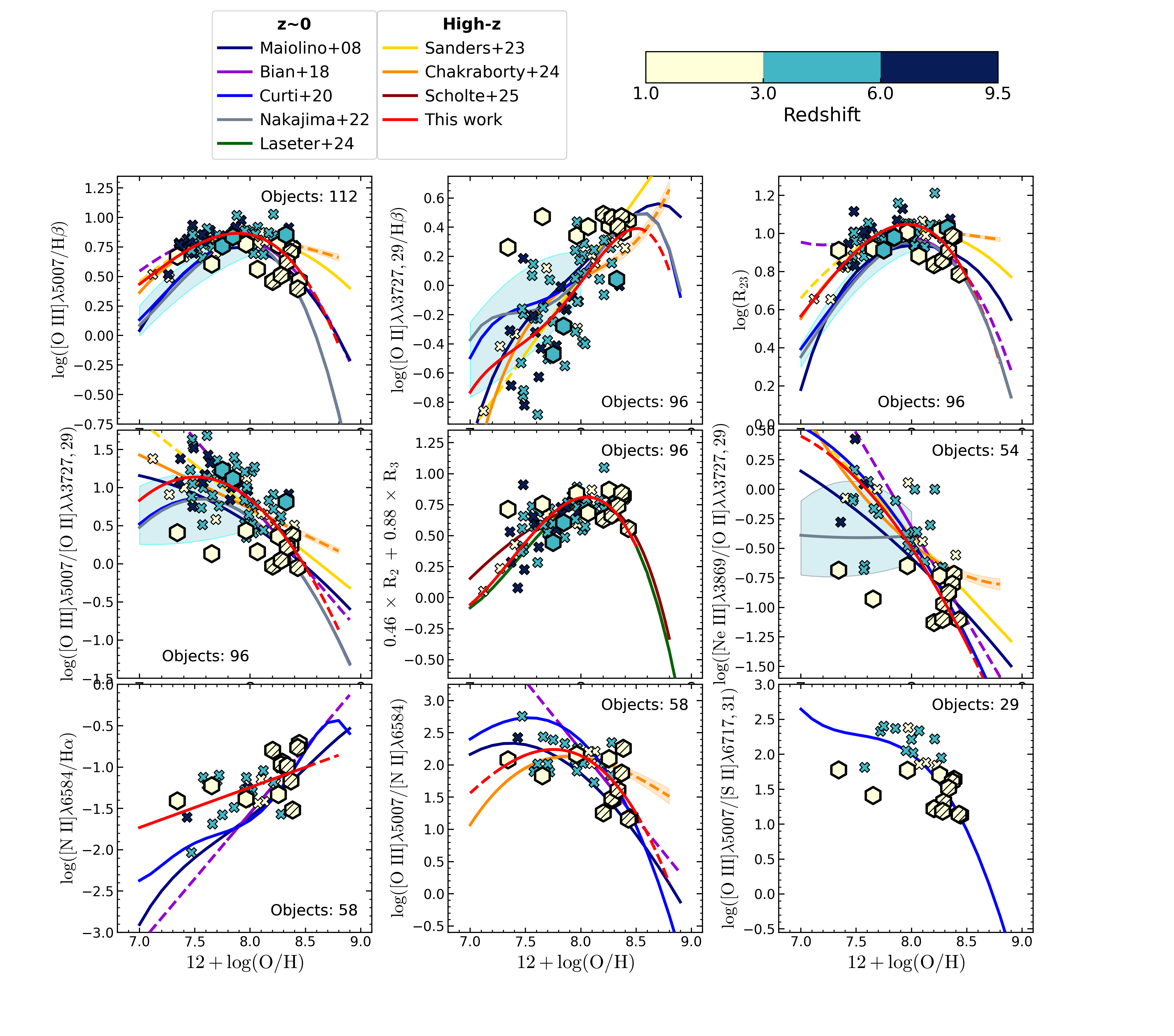} 
    \caption{Nine different strong-line diagnostics comparing our MARTA data (striped hexagons for the golden sample, plain hexagons for the silver) with high-z literature samples (crosses) and a set of empirical calibrations. The data are colour-coded in redshift bins. The high-redshift data include strong-line calibration measurements from JWST surveys \citep{2025ApJ...980..225T, 2025ApJ...981..136S, sanders_calibrations_2023, 2024arXiv241018763N, 2025A&A...697A..89C, 2025MNRAS.540.2991A, d_eugenio_gsz12_2023, 2024arXiv240714201N, 2024A&A...687L..11S, 2025ApJ...979...87M, 2025arXiv250210499S}. The empirical calibration curves from the local Universe are from \cite{maiolino_amaze_2008, curti_klever_2020, nakajima_empress_2022, bian_ldquodirectrdquo_2018}, color-coded in the plot using a dark color palette (blue, purple, green) while the high-redshift-based calibrations from \cite{sanders_calibrations_2023, 2025ApJ...985...24C} and they are represented with a warm color palette (yellow, orange, red). Dashed lines denote extrapolations of the calibrations outside the regime where they have been calibrated.
    Our best-fit relations to the high-redshift dataset are shown in red. In most cases, a third-degree polynomial was used; for the N2 and R2 diagnostics, we adopted a linear fit,  as there were no particular features requiring a higher-order polynomial. Furthermore, in the $\tilde{R}$ panel (central one), the \citep{laseter_auroral_jades_2023} and \cite{2025arXiv250210499S} empirical re-calibration are displayed as well.}
\label{fig:diagnostics}

\end{figure*}

\begin{table}
    \centering
    \caption{Polynomial coefficients of the best fit for the strong-line metallicity diagnostics.}
    \begin{tabular}{lccccc}
        \hline
        \hline
        Diagnostic & $A$ & $B$ & $C$ & $D$ & $E$\\
        \hline
        R3       & 0.158  & -1.977  & -1.629  & -0.329  & --\\
        R2       & 0.290  & -1.202  & -4.104  & -3.167  & -0.812 \\
        R23      & 0.515  & -1.614  & -1.421  & -0.286  & --\\
        $\Tilde{R}$ & -0.055 & -3.049 & -3.168 & -0.807 & --\\
        O32      & -0.537  & -2.832  & -1.200  & -0.002 & -- \\
        Ne3O2    & -1.679  & -2.034  & -0.458  & --     & -- \\
        N2       & -0.910  &  0.486  & --      & --     & -- \\
        O3N2     & 0.619   & -3.794  & -2.561  & -0.382 & -- \\
        \hline
    \end{tabular}
    \tablefoot{The fitted functional form is $f(x) = A + Bx + Cx^2 + Dx^3 + Ex^4$, where $f$ is the line ratio diagnostic defined as in Sect.~\ref{sec:intro} and $x$ represents the gas-phase metallicity, normalised to the solar value ($12 + \log(\mathrm{O/H})$ - 8.69).
    The polynomials are fitted over the range $12+\log(\mathrm{O/H}) \in [7.0, 8.4]$ and include a regularization at high and low metallicity based on the \citet{curti_new_2017} calibrations to stabilize the fit where direct $T_e$ measurements are not available. Caution is advised when extrapolating these relations beyond the fitted range}.
    
    \label{tab:strong_lines}
\end{table}

In addition, we complemented our sample with literature NIRSpec data of galaxies that were detected in the \OIIIopt$\lambda$4363 line. 
We compiled the fluxes (where publicly available) and self-consistently performed extinction correction, temperature, and abundance calculations following the same methodology applied for the MARTA silver sample. The literature sample includes data from \cite{d_eugenio_gsz12_2023, sanders_calibrations_2023, 2025ApJ...980..225T, 2025ApJ...981..136S,  2024arXiv241018763N, 2025A&A...697A..89C, 2025MNRAS.540.2991A,  2024arXiv240714201N, 2024A&A...687L..11S, 2025ApJ...979...87M,2025arXiv250210499S}. Our total sample, combining MARTA and literature sources, consists of 128 galaxies. 

As not all strong lines are always detected in literature objects (either because of low sensitivity, or due to limited spectral coverage), each strong-line ratio can be effectively measured in a different number of objects, as reported in each panel in Fig.  \ref{fig:diagnostics}. As an example, for R23 we present a calibration based on 96 objects, nonetheless nearly doubling previous largest compilation of \cite{2025ApJ...985...24C}.

\subsection{New JWST-based metallicity calibrations}

In Fig. \ref{fig:diagnostics}, we show the position of our combined JWST sample on different strong-line ratios diagnostics versus metallicity diagrams. 
MARTA galaxies primarily occupy the high-metallicity end of the distribution, a region largely unexplored at high redshift.
We compare our combined sample with empirical calibration derived from both local and high-redshift galaxy studies. 

The local calibrations we consider include those from \cite{maiolino_amaze_2008}, which are based on low-metallicity galaxies from \cite{nagao_gas_2006} and photoionisation model-based oxygen abundances for SDSS galaxies in the high-metallicity regime. We also include the calibrations by \cite{curti_new_2017}, later revisited in \cite{curti_massmetallicity_2020}, which are based on stacked spectra of SDSS galaxies; and those by \cite{nakajima_empress_2022}, who extended the \cite{curti_new_2017} SDSS stacks by including extremely metal-poor galaxies (EMPGs) from the EMPRESS survey. Additionally, we use the calibrations by \cite{bian_ldquodirectrdquo_2018}, derived from stacked spectra of local high-redshift analogues selected based on their location on the \NII-BPT diagram.

For comparison, we also include recent works that recalibrate the same strong-line diagnostics using JWST data. In particular, we consider the calibrations from \cite{sanders_calibrations_2023}, based on galaxies at $z = 2$–9 from the CEERS survey \cite{Finkelstein_ceers_paper1_2023}, and those from \cite{2025ApJ...985...24C}, based on galaxies from the CEERS, JADES \citep{d_eugenio_jades_DR3_2024}, and UNCOVER \citep{Bezanson_UNCOVER_paper_2024} surveys, with fluxes compiled by the PRIMAL project \citep{Heintz_primal_2024}.

All curves corresponding to the different calibrations shown in Figure~\ref{fig:diagnostics} are plotted as solid lines within the directly probed metallicity range, while polynomial extrapolations beyond the calibration range are shown as dashed lines.
We also provide new fits to the strong-line ratios as a function of metallicity based on the full JWST sample compiled in this work. The fitting procedure was implemented as a direct polynomial regression, where the coefficients were optimized using the \texttt{least\_squares} routine from the \textsc{SciPy} library. To limit the influence of individual outliers, we adopted the Huber loss function\footnote{The Huber loss behaves as a standard $\chi^2$ (L2 norm) for small residuals while reducing the weight of points with large deviations by transitioning to a linear (L1) form.}. This approach retains the efficiency of least-squares fitting in the well-sampled regime, while improving robustness in the presence of non-Gaussian scatter.
In most cases, we adopted a third- or a forth-order polynomial, while for the N2 diagnostic, which does not exhibit significant curvature, we opted for a linear model. 

These polynomial fits are constrained by observations in the range 12+log(O/H)~$\in$~[7.0--8.4]. Outside this interval, the lack of direct constraints makes extrapolation highly uncertain. To stabilize the behavior at low and high metallicity, the fits were regularized by anchoring them to the analytic calibrations of \citet{curti_new_2017}. In practice, we introduced additional constraint points at metallicities beyond the observed range, with target values set by these empirical relations. These points were assigned enhanced weights in the loss function, ensuring that the fits remain driven by the data where observations are available, while gradually converging to the expected trends at the boundaries. The regression was performed on metallicities normalized to the solar abundance (12+log(O/H)~=~8.69), so that the best-fit coefficients reported in Table~\ref{tab:strong_lines} are expressed as functions of the normalized abundance $Z_{\mathrm{norm}} = (12+\log(\mathrm{O/H})) - 8.69$. The resulting calibrations are shown in red in Fig.~\ref{fig:diagnostics}. We did not report fits for the O3N2 diagnostic, for reasons discussed below. The calibrations presented in this work are based on a dataset that is potentially biased, since we only calibrate on a subset of galaxies for which auroral lines have been detected, favoring systems with elevated star-formation rates. Such biases are intrinsic to all calibrations based on auroral lines, including those derived from local objects. We address this issue in more detail in Appendix~\ref{sec:appE}.

\subsubsection{The R23, R3, and R2 diagnostics}

Among the newly calibrated relationships, R23 is characterised by the lowest scatter, in agreement with previous analyses \citep{nakajima_empress_2022, sanders_calibrations_2023, 2025ApJ...985...24C}. However, the R23 calibration exhibits a well-known double-branched structure, with most galaxies populating the plateau region where the relationship flattens, severely limiting its accuracy at intermediate metallicities. A similar behaviour is also observed in R3, where the turnover occurs around 12 + log(O/H) $\sim$ 8.0, leading to two distinct metallicity solutions.

R2 is generally not employed alone for metallicity measurements due to its sensitivity to the ionization parameter. However, it can assist in resolving ambiguities in the R23 and R3 calibrations. In our sample, R2 shows an increasing trend with metallicity, although with a much larger scatter in the line ratios at fixed O/H than observed for R3 and R23.

\subsubsection{O32, Ne3O2, O3N2, and O3S2 diagnostics}

Alternative diagnostics such as O32, Ne3O2, O3N2, and O3S2 also display large scatter, highlighting the challenges in metallicity estimation when these indicators are used individually. 
While O32 primarily traces the ionization parameter, which is correlated with metallicity \citep{Ji_U_vs_Z_2022}, other ratios are also sensitive to possible variations in the neon, nitrogen, and sulphur abundance patterns.
The scatter is particularly pronounced in the low-metallicity regime.
Nonetheless, O32 and Ne3O2 show a monotonic trend with metallicity (with Ne3O2 being also almost insensitivity to reddening), hence remain a valuable tool for differentiating between metallicity solutions, especially when used alongside diagnostics such as R23 \citep[e.g.][]{kewley_using_2002}. O3S2 was excluded from the fitting procedure due to the limited number of available data points and the high level of scatter observed.

Interestingly, we identify a subset of $z\sim2$ low-metallicity galaxies in the MARTA sample showcasing intense emission in the low-ionisation species which makes them deviate significantly from the average relation, displaying enhanced \OII/H$\beta$ (and \SII/H$\beta$) ratios given their O/H. These galaxies also exhibit elevated \NII/H$\alpha$ ratios (as also mentioned in Section \ref{sec:N2}), and correspondingly low \OIIIopt/\NII and \OIIIopt/\SII values, suggesting a systematic offset in multiple diagnostic line ratios. While dust attenuation uncertainties—such as deviations from the assumed attenuation law—could affect line ratios involving widely separated wavelengths (e.g. \OII/H$\beta$), they are unlikely to fully explain the broad pattern of anomalies observed across both nearby and closely spaced lines.
Conversely, MARTA galaxies at $z>3$ are characterised by some of the lowest \OII/H$\beta$ values measured in the JWST sample.
Such a spread, observed also in EMPGs in the local Universe \citep{nakajima_empress_2022}, highlights the role of intense star formation and varying ionization conditions in shaping the relationship between strong-line ratios and metallicity.

At low metallicities, galaxies typically exhibit higher ionization parameters; however, the ionization conditions are not uniform across different systems \citep[e.g.,][]{2022ApJ...929..118G}, but they are strongly modulated by variations in SFR \citep[e.g.,][]{Ji2021Correlation}, the geometric distribution of the gas, and the presence of young, hot stellar populations. Consequently, low-ionization tracers such as \NII\ and \SII\ display significant scatter, reflecting the intrinsic diversity in ionization conditions within low-metallicity environments. 
Given that our sample is relatively small, the presence of multiple such cases suggests that these conditions may be more common at high redshift than previously thought.

\subsubsection{The N2 diagnostic}
\label{sec:N2}

Thanks to JWST we can start populating the low-metallicity region of the N2 diagnostic. 
In line with well-established findings over the past decade from ground-based spectroscopic surveys such as MOSDEF \citep{shapley_mosdef_2015}, KBSS \citep{ steidel_strong_2014, strom_measuring_2017}, KLEVER \citep{hayden-pawson_NO_2022}, and FMOS \citep{kashino_fmos-cosmos_2017}, high-$z$ galaxies exhibit an offset in the \NII-BPT diagram (Figure~\ref{fig:BPT}), with enhanced \NII/\Halpha compared to their local counterparts. 
One of the most widely accepted interpretations of this behaviour is the hardening of the ionizing radiation field at high redshift, which—at fixed metallicity—shifts the \NII/\Halpha ratio to higher values without necessarily implying variations in chemical abundance patterns \citep{topping_mosdef-lris_2020_i}.

We now observe the same deviation in the \NII/\Halpha vs O/H relation compared to local calibrations, at least at low metallicity (12+log(O/H)$<$8).
The possible role of enhanced N/O ratios in driving this offset has been recently investigated using the EXCELS dataset \citep{2025arXiv250210499S}, where a correlation between N/O and the degree of offset from the median calibration was reported, both at high and low redshift. An analysis of the N/O abundance in our combined sample will be the subject of a future study.

At $z>6$ the optical \NII is rarely detected because it falls outside the wavelength range probed by JWST/NIRSpec, but also because galaxies at these redshifts tend to have low metallicity and high ionization parameters, possibly causing most of the nitrogen to reside in higher ionization states.
Thus, the absence of \NII in high-$z$ sources may also result from nitrogen being predominantly in higher ionization states (probed by \NIII, \NIV), as suggested by recent JWST findings \citep{curti_gs_z9_2024, bunker_gnz11_2023, cameron_gnz11_2023, charbonnel_gnz11_2023, marques-chaves_Nitrogen_2024, Schaerer_nitrogen_z94_2024, topping_z6_lens_2024, ji_nitrogen_AGN_z5_2024, isobe_CNO_2023}.

\subsubsection{The $\tilde{R}$ diagnostic}

Finally, we test the $\tilde{R}$ diagnostic proposed by \cite{laseter_auroral_jades_2023} and recently re-calibrated by \cite{2025arXiv250210499S} based on a combined sample of data from the JWST/EXCELS survey \citep{2024MNRAS.534..325C} and new local galaxy measurements from the DESI Early Data Release \citep{2022AJ....164..207D}. 
Although at high metallicity (Z$\approx$Z$_{\odot}$) the high-$z$ dataset lacks sufficient coverage to constrain the turnover of the functional form, performing a fit with the same functional form as in \cite{laseter_auroral_jades_2023} to the combined JWST sample provides very similar behaviour to the latter in the low-metallicity branch (which is the one primarily populated by high-$z$ galaxies). 

While very similar in shape, the \cite{2025arXiv250210499S} calibration appears slightly offset towards higher $\tilde{R}$ ratios at low metallicities compared to both the original \cite{laseter_auroral_jades_2023} parametrisation and to our current fit.
Such discrepancy may be caused by intrinsic differences in the calibration samples, but differences in the set of adopted atomic parameters for O$^{++}$ can also possibly play a role. Variations of up to $\sim 500$~K in the $T_3$ derivation (see discussion in Section~\ref{sec:tt_relation}) translate into a difference in the final inferred O/H of $\sim 0.02-0.04$~dex at fixed line ratios and at low-metallicity (12+log(O/H)$\sim 7.5$), where the total abundance budget is dominated by doubly ionised oxygen.
Ultimately, we do not provide the coefficients for the new fit to the $\tilde{R}$ diagnostic, as our data do not offer significant additional constraints compared to \cite{laseter_auroral_jades_2023}.

\begin{table*}
\caption{Median offset and dispersion (scatter) in line ratio at fixed O/H for each diagnostic and calibration considered in this work.}
\label{tab:median_offsets}
\centering
\begin{tabular}{l c c c c c c c c c}
\hline\hline
Calibration & \multicolumn{8}{c}{Median offset (dex)} \\
            & R3 & R2 & R23 & O32 & O3N2 & $\tilde{R}$ & N2 & Ne3O2 & O3S2 \\
\hline
Maiolino+08 & -0.13 & 0.055 & -0.084 & -0.24 & -0.21 & -- & -0.29 & -0.068 & -- \\
Bian+18 & -0.016 & -- & 0.034 & 0.12 & 0.41 & -- & -0.37 & 0.21 & -- \\
Curti+20 & -0.11 & 0.074 & -0.080 & -0.12 & 0.34 & -- & -0.34 & 0.041 & -0.039 \\
Nakajima+22 & -0.11 & 0.065 & -0.077 & -0.22 & 0.31 & -- & -0.35 & 0.11 & -- \\
Laseter+23 & -- & -- & -- & -- & -- & -0.039 & -- & -- & -- \\
Sanders+23 & -0.015 & -0.010 & 0.011 & 0.034 & -- & -- & -- & 0.12 & -- \\
Chakraborty+24 & 0.044 & -0.041 & 0.018 & -0.21 & -0.16 & -- & -- & 0.092 & -- \\
Scholte+25 & -- & -- & -- & -- & -- & 0.067 & -- & -- & -- \\

This work & 0.0071 & -0.011 & 0.012 & 0.0089 & 0.010 & -- & 0.038 & -0.049 & -- \\
\hline\
 & \multicolumn{8}{c}{Scatter (dex)} \\
\hline
Maiolino+08 & 0.13 & 0.85 & 0.12 & 0.36 & 0.38 & -- & 0.40 & 0.35 & -- \\
Bian+18 & 0.10 & -- & 0.31 & 0.42 & 0.76 & -- & 0.65 & 0.47 & -- \\
Curti+20 & 0.29 & 0.49 & 0.18 & 0.31 & 0.63 & -- & 0.64 & 0.28 & 0.52 \\
Nakajima+22 & 0.16 & 0.68 & 0.11 & 0.33 & 0.57 & -- & 0.26 & 0.38 & -- \\
Laseter+23 & -- & -- & -- & -- & -- & 0.25 & -- & -- & -- \\
Sanders+23 & 0.070 & 0.25 & 0.11 & 0.30 & -- & -- & -- & 0.32 & -- \\
Chakraborty+24 & 0.12 & 0.25 & 0.10 & 0.20 & 0.75 & -- & -- & 0.32 & -- \\
Scholte+25 & -- & -- & -- & -- & -- & 0.15 & -- & -- & -- \\
This work & 0.10 & 0.26 & 0.077 & 0.32 & 0.28 & -- & 0.22 & 0.34 & -- \\
\hline
\end{tabular}
\tablefoot{Offsets in line ratios are measured as the difference between the observed values and the values predicted by the corresponding strong-line calibration for any given galaxy metallicity in the sample. Both median offsets and dispersions (scatter) are then calculated over the entire metallicity range considered.}
\end{table*}

\begin{figure*}
    \centering
    \includegraphics[width=\textwidth]{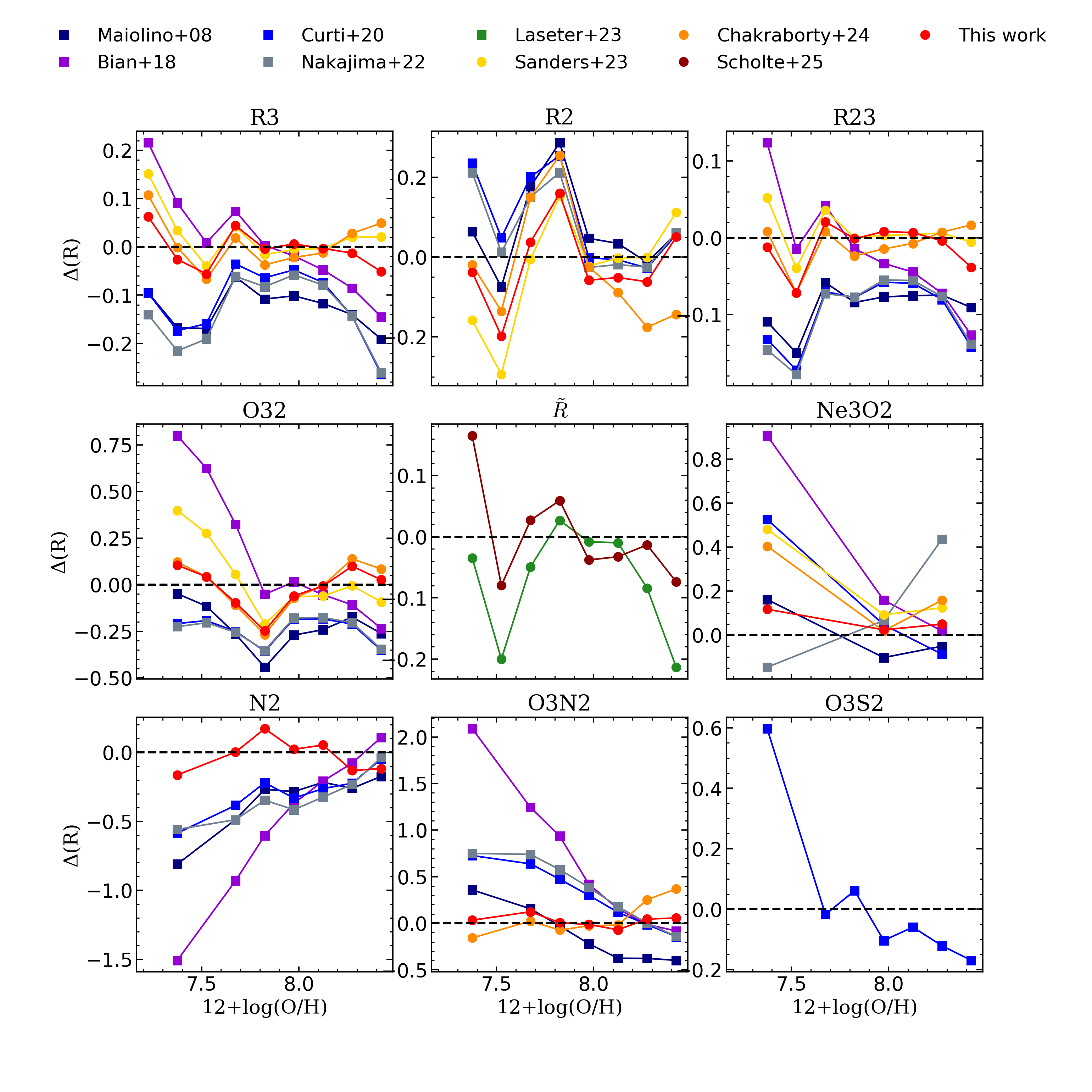}
    \caption{Median offset of the strong-line calibrations analysed in this work relative to the high-z sample (z$\sim$1-9.5), computed in bins of metallicity. The bins have a width of 0.15 dex and are included only if they contain at least five objects. The figure displays six panels, each corresponding to a different diagnostic. The color palette is the same as used in Fig. \ref{fig:diagnostics} to distinguish between low-z and high-z based calibrations, with markers also representing this division: circles for high-z and squares for z$\sim$0 based calibrations.}
\label{fig:all_offsets}

\end{figure*}

\subsection{Comparison with literature calibrations}

We analysed the behaviour of each calibration curve considered in this work by 
computing, for the combined JWST high-$z$ galaxy sample, the offset between observed and predicted line ratios at fixed 12+log(O/H).
We report the median offset and the associated dispersion for each diagnostic and calibration in Table \ref{tab:median_offsets}.
Beside the new calibrations provided in this work (for which the scatter in the calibration sample is minimised by definition), the calibrations presented in \cite{sanders_calibrations_2023} and \cite{2025ApJ...985...24C} behave well. Both were based on similar (though smaller) samples of high-$z$ galaxies.
In contrast, most of the calibrations based on local galaxy samples show larger offsets, as expected given the evolving physical conditions of the calibration samples.

To better visualise possible trends with metallicity, in 
Figure \ref{fig:all_offsets} we plot the median offset in the line ratio computed in $0.15$~dex bins of metallicity, as a function of 12+log(O/H) (where only bins containing at least five objects are considered).
Figure \ref{fig:all_offsets} highlights the discrepancy between the local ($z\sim0$) calibrations and the high-redshift galaxy sample. 
In diagnostics like R3, R23, and O32, the local calibration curves (e.g., \citealt{maiolino_amaze_2008, curti_massmetallicity_2020, nakajima_empress_2022}) tend to lie below the observed data points for high-redshift galaxies, suggesting a redshift-dependent evolution of these ratios. Such evolution must be factored into metallicity measurements for high-redshift galaxies. \cite{2025arXiv250810099S} also recently reported a clear redshift evolution in oxygen-based strong-line calibrations, with high-redshift galaxies showing offsets consistent with hotter stellar populations and harder ionising spectra relative to local samples. This reinforces the need to account for cosmic evolution when applying local calibrations to early galaxies.

However, we observe that for several calibrators 
the median offset tends to decrease with increasing metallicity, possibly suggesting a metallicity-dependent redshift evolution of the involved strong-line diagnostics. 
Indeed, this is in line with the observed increase in the scatter of line ratios at fixed O/H in the low-metallicity regime.

\begin{figure}
    \centering
    \includegraphics[width=0.98\columnwidth]{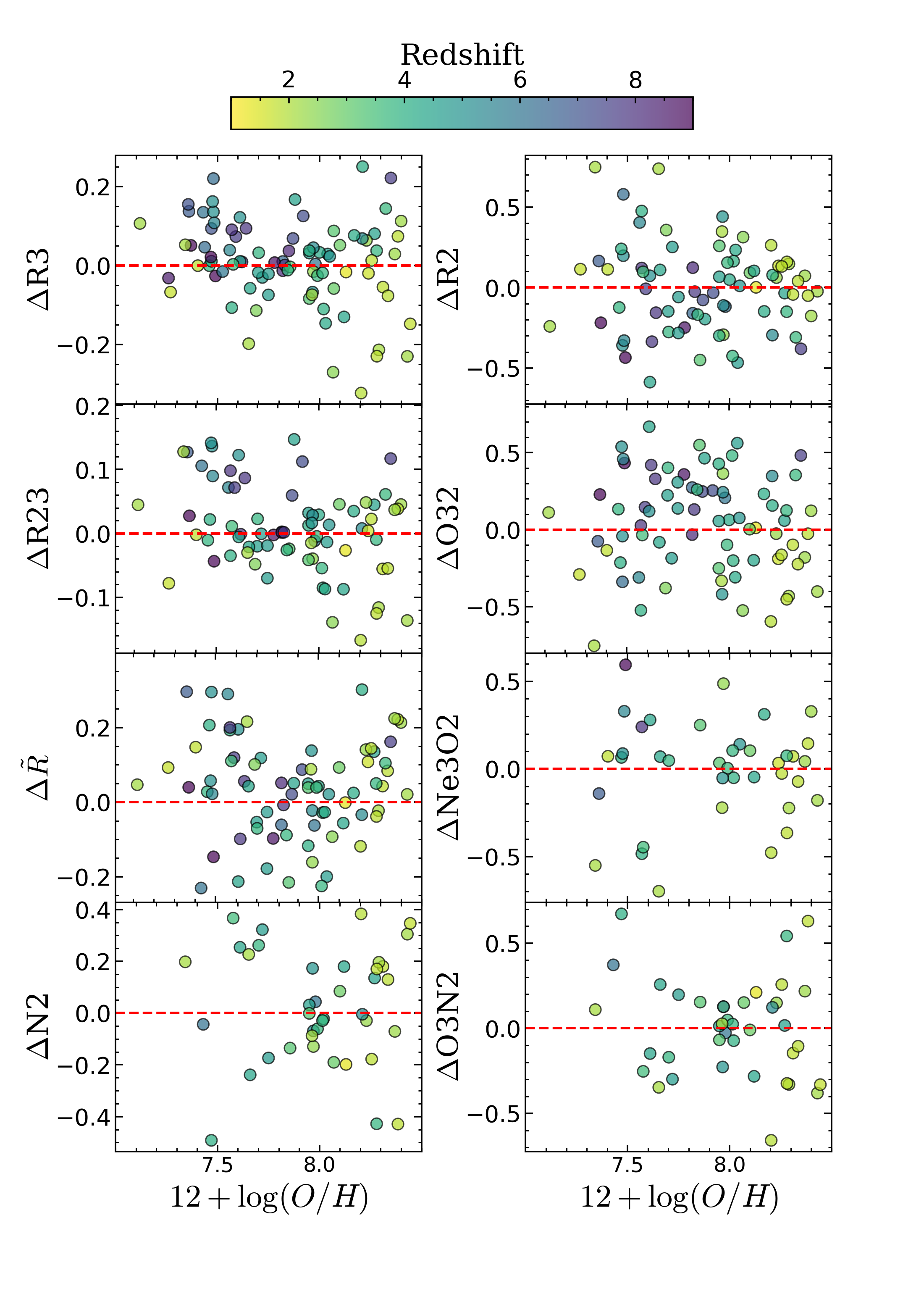}
    \caption{Offsets in the log of the chosen line ratio at fixed O/H, representing the difference between the observed ratio and the value predicted by our best-fit calibrations.
    High-z JWST sample points are color-coded by redshift to highlight potential trends. 
    }
\label{fig:offset_mycalib}

\end{figure}

As the typical ionisation conditions of the ISM in galaxies evolve with cosmic time, we would expect this behaviour to be reflected into a trend between the dispersion of each calibration and redshift. 
Although the combined JWST sample analysed in this work is still too small to perform a robust analysis of the `internal' redshift evolution (between $z\sim1$ and $z\sim10$), hints of such a correlation are already visible for some of the diagnostics in Figure~\ref{fig:diagnostics}, and are better visualised in Figure \ref{fig:offset_mycalib}, where we display the offsets in line ratios for individual high-z sources from the best-fit relationships recalibrated in this work, colour-coded by galaxy redshift. 
At low metallicity (12+log(O/H)$\lesssim 8$), galaxies at the lower-redshift boundary of the JWST sample tend to be offset in the opposite direction to the highest redshift systems from the best-fit calibration, especially in line-ratio diagnostics where the dependence on the ionisation parameter is more prominent such as R2, O32, and Ne3O2.

\section{Summary and outlook}
\label{sec:summary}

In this work, we have analysed deep JWST/NIRSpec R1000 spectra of 16 star-forming galaxies at $1.8<z<4.7$, observed in the framework of the `Measuring Abundances at high-Redshift with the Te Approach' (MARTA) programme.  
We report the detection of faint auroral lines such as \OIIIopt $\lambda$4363, \OII $\lambda\lambda$7320,7330, \SII$\lambda\lambda$4068,4076 and \SIII$\lambda$6312 (Figure~\ref{fig:all_aurorals} and \ref{fig:aurorals_silver}), derive Te-based metallicities, and discuss trends in the temperature-temperature relationships and the evolution of strong-line metallicity calibrations. 
The key results from this work are summarised below.

\begin{itemize}
    \item The temperature-temperature (T$_2$-T$_3$) relation between different oxygen ionised species, established on low-redshift samples, seems to hold at earlier cosmic epochs, offering, despite its non-negligible scatter, a reliable framework for metallicity determinations at high-z (Figure~\ref{fig:tt}).
    \item The dispersion of the T$_2$-T$_3$ relationship mainly correlate with variations in the ionisation parameter (traced by \OIIIopt/\OII) and/or in the hardening of the ionising continuum (traced by the `softness parameter' $\eta$, Figure~\ref{fig:tt_hexbin} and Table~\ref{tab:partial_corr}). Variations in the electron density mainly correlate with T$_2$ (at fixed T$_3$), but have lower impact on the scatter of the relationship.
    \item We explored the relationship between temperatures derived for both low-ionisation and high-ionisation sulphur and oxygen species, demonstrating their consistency with local trends (Figure~\ref{fig:TT_sulfur}).
    \item The MARTA auroral-line galaxies, while still subsolar on average, populate the upper end of the metallicity distribution among high-redshift galaxies with direct metallicity measurements. As such, they provide valuable insights into the applicability of strong-line metallicity calibrations at $z \sim 2$, extending beyond the low-metallicity regime typically probed by earlier studies. By combining our sample with existing T$_\mathrm{e}$-based metallicity measurements from JWST in the literature, we assess the redshift evolution in several popular strong-line diagnostics over a wide range in oxygen abundance (12+log(O/H)~$\in$~[7.0–8.4], Figure~\ref{fig:diagnostics} and \ref{fig:all_offsets}), and provide updated calibrations for some of these diagnostics in the form of standard polynomial relations (Table~\ref{tab:strong_lines}).

    \item While `strong-line methods' remain broadly applicable across cosmic time, clear systematic shifts emerge in specific line ratios, particularly R2, O32, and Ne3O2, as likely driven by evolving ionization conditions at high redshift. 
    This trend is reflected into a correlation between the offset in line ratio at fixed metallicity and redshift in these diagnostics, as tentatively observed in our combined JWST sample (Figure~\ref{fig:offset_mycalib}).
    \item We observe a systematic offset in the N2 vs O/H diagram (especially at low metallicity), in line with previous findings that high-redshift galaxies exhibit enhanced \NII/H$\alpha$ at fixed O/H compared to local counterparts \citep[e.g.][]{steidel_strong_2014, masters_physical_2014, shapley_mosdef_2015, strom_nebular_2017}.
    Furthermore, a subset of low-metallicity galaxies in the MARTA sample deviates significantly from the average relationships also in displaying elevated \OII/H$\beta$ and \SII/H$\beta$ ratios given their O/H, highlighting the variety of ionisation conditions observed in the high redshift Universe and the potential impact of selection effects on the determination of the average trends in high-z galaxy population properties.
\end{itemize}

Future research will build on the insights gained from this and other recent JWST-enabled studies to achieve a more self-consistent determination of galaxy metallicity properties and their scaling relations across cosmic epochs.
In parallel, studies of chemical abundance patterns in high-redshift galaxies, focusing on elements such as nitrogen, neon, and argon, will provide complementary insights into star formation histories at z$\sim$ 2 and beyond, helping to refine next-generation chemical evolution models. A key avenue for future exploration is the derivation of direct, Te-based nitrogen-to-oxygen (N/O) abundance ratios is. 
Variations in N/O ratios at high redshift can reveal crucial details about rapid nitrogen enrichment, differences in star formation efficiency, or localized enrichment processes such as those driven by Wolf-Rayet stars\citep{Fern'andez-Mart'in2012Ionization, Rivera_Thorsen_WR_sunburst_arc_2024, 2024ApJ...962L...6K}, highlighting the importance of exploring this diagnostic in detail.

Furthermore, we plan to conduct detailed comparisons with state-of-the-art photoionization models, such as HOMERUN (\citealt{marconi_homerun_2024}; Moreschini et al., in prep.). Unlike single-cloud models, HOMERUN incorporates contributions from multiple gas clouds with varying metallicities, ionization parameters, and densities. This approach enables a more nuanced analysis of emission line ratios, allowing us to assess the impact of high-density gas clumps on metallicity determinations and refine our understanding of biases in classical methods.

\begin{acknowledgements}
FB and FM acknowledge support from the INAF Fundamental Astrophysics programme 2022 and 2023. FM and FB acknowledge support from grant PRIN-MUR 202223XDPZM "Prometeus"
GC and EB acknowledge financial support from INAF under the Large Grant 2022 ``The metal circle: a new sharp view of the baryon cycle up to Cosmic Dawn with the latest generation IFU facilities''. A.F. acknowledges the support from project "VLT- MOONS" CRAM 1.05.03.07.
AM acknowledges support from project PRIN-MUR project “PROMETEUS”  financed by the European Union -  Next Generation EU, Mission 4 Component 1 CUP B53D23004750006.
FC acknowledges support from a UKRI Frontier Research Guarantee Grant (PI Cullen; grant reference EP/X021025/1). CK acknowledges funding from the UK Science and Technology Facility Council through grant ST/Y001443/1.
This work is based in on observations made with the NASA/ESA/CSA James Webb Space Telescope. The data were obtained from the Mikulski Archive for Space Telescopes at the Space Telescope Science Institute, which is operated by the Association of Universities for Research in Astronomy, Inc., under NASA contract NAS 5-03127 for JWST. These observations are associated with program JWST 1879.\\
The data presented in this paper were obtained from the Mikulski Archive for Space Telescopes (MAST). The specific observations analyzed can be accessed via \url{https://doi.org/10.17909/v7wt-1r79}. STScI is operated by the Association of Universities for Research in Astronomy, Inc., under NASA contract NAS5–26555.

\end{acknowledgements}

\bibliography{marta}{}
\bibliographystyle{aa}

\begin{appendix}

\section{Reliability of errors estimates }
\label{sec:AppA}

We estimate the S/N of the weak emission lines by considering both the fit residuals on one hand and the pipeline errors on the other. In the first approach, we consider the peak of each emission line and dividing it by the standard deviation of the residuals from a line-free region adjacent to the emission line. To approximate the total S/N over the line profile, this value was then multiplied by the number of spectral channels corresponding to its full width at half maximum. At a spectral resolution of $R \approx 1000$, our emission lines are unresolved, meaning their widths are dominated by the instrumental line spread function (LSF). We therefore adopt a fiducial FWHM of 3 pixels.

We compared these values with a second S/N estimate based on the ratio of the line flux obtained from the Gaussian fit to the error on this flux. Here, the flux error was multiplied by a factor derived for each galaxy of the sample from the analysis of a line-free region in its spectrum - ranging from 4125 to 4325 \AA. This factor represents the average ratio between the error spectrum produced by the JWST data reduction pipeline and the standard deviation of the residuals between the data and the pPXF best-fit within the same region. This factor is on average $\gtrsim 1.4$, which indicates that pipeline error estimates tend to be conservative. We therefore rescale the error spectra downwards.

From this comparison between S/N ratios, we found that the two were generally in good agreement. In a few cases, however, the second estimate was consistently higher than the first.

\section{Additional details on the spectral fitting}
\label{sec:AppB}
\subsection*{Possible [FeII] contamination}

Galaxies MARTA\_4195 and MARTA\_5014 both exhibit an evident emission feature at $\sim$ 4360 \AA\ — visible in the 2D and 1D extracted spectra - which is blue-shifted with respect to the \OIIIopt$\lambda$4363 line, aligning more closely with the expected position of the [Fe II]$\lambda$4360 line. This emission feature has been documented in various studies using both stacked \citep{curti_new_2017} and individual spectra of high-metallicity galaxies in the local Universe (e.g., \citealt{rogers_chaos_2021, rogers2022chaos}). Furthermore, [Fe II] contamination has recently been tentatively identified in high-redshift galaxies (see, e.g., \cite{shapley2024aurora} for a galaxy at z$\sim$7 from the AURORA survey).

\begin{figure*}
    \centering
    \includegraphics[width=0.75\linewidth]{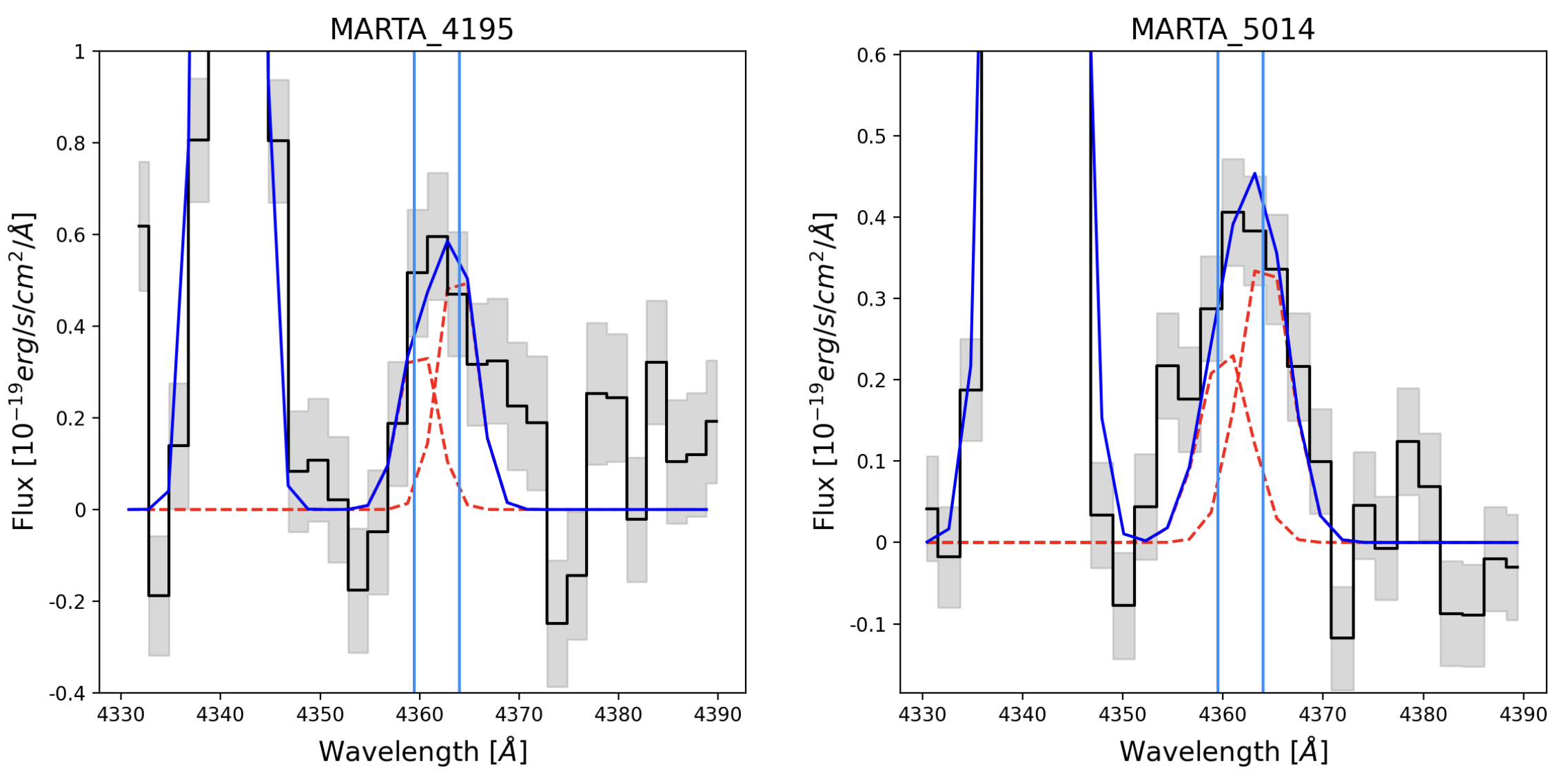}
    \caption{Spectral fits in the region around the \OIIIopt$\lambda4363$ line, containing three emission lines (H$\gamma$, [FeII]$\lambda4360$, and \OIIIopt$\lambda4363$) for galaxies 4195 (left panel) and 5014 (right panel). The overall fit is represented by the solid blue line, while the two Gaussian components centred at 4360 and 4364 $\AA$  are shown as dashed red lines. The fit demonstrates that a single Gaussian at 4364 $\AA$ alone cannot adequately reproduce the spectral features of these two galaxies. The inclusion of the [FeII] component provides a significantly better fit.}
    \label{fig:FeII_gaussians}
\end{figure*}

We therefore included [Fe II]$\lambda$4360 as an additional component in our fitting setup for these two galaxies, significantly improving the fit of \OIIIopt$\lambda$4363 (Fig. \ref{fig:FeII_gaussians}). Notably, in the case of ID MARTA\_4195, we also (tentatively) detected the [FeII]$\lambda$4288 emission line, which is expected to accompany [FeII]$\lambda$4360 emission. This approach is adopted as our fiducial fitting method in this work.
The inclusion or exclusion of [Fe II] in the fit for these two galaxies results in a 20–40\% decrease in the flux of the auroral \OIIIopt$\lambda$4363 line. This, in turn, affects the derived $T_3$ values, leading to a variation of 10–15\%, with respect to an estimate that associates the entire flux to \OIIIopt$\lambda$4363.

For the other galaxies in the sample, we performed the same analysis by fitting the \OIIIopt$\lambda$4363 line both with and without the inclusion of [Fe II]. However, in these cases, the [Fe II] flux was negligible, on the order of $\sim$1/10–1/100 of the \OIIIopt$\lambda$4363 flux, and its inclusion did not significantly affect the measured \OIIIopt$\lambda$4363 flux. Moreover, no additional evidence of [Fe II] emission is observed in the spectra of these galaxies. Therefore, to avoid introducing unnecessary complexity to the modelling, we excluded [Fe II] as a component for all other cases. The measured fluxes of the auroral lines, along with their corresponding S/N ratios, are provided in Table \ref{tab:aurorals}. 

\subsection*{The case of MARTA\_3942}

Object 3942 presented unique challenges due to its extended nature and the presence of a contaminant near the end of the slit. To address these issues, we applied a local background subtraction technique, avoiding self-subtraction and recovering part of the flux from the second shutter while excluding the contaminated region at the slit’s edge. However, the use of local background subtraction inevitably introduced some noise into the continuum, as the outer shutters were integrated for less time than the central one. This resulted in a particularly noisy continuum around the auroral line \OIIIopt4363.

A significant issue arise when fitting this region with pPXF: the algorithm attempts to model stellar features around H$\gamma$ (4340 $\AA$) but is actually fitting the noise, partially suppressing the \OIIIopt4363 line. Although this line is clearly distinguishable above the continuum, the pPXF fit underestimates its flux.

To mitigate this issue and recover a more accurate flux measurement, we subtracted the continuum modelled on the full three-nodding reduction version of the spectrum from the local background-subtracted spectrum. In this way, the model continuum was significantly smoother due to the more uniform exposure across the slit, but it was re-scaled to match the fiducial, local background-subtracted spectrum, before fitting the \OIIIoptL[4363] emission line with an individual Gaussian component.
The revised fit yields a line flux of $3.51\times 10^{-19}\,\, \rm{erg/s/cm^2}$, recovering additional flux compared to the previous fit ($3.11\times 10^{-19}\,\, \rm{erg/s/cm^2}$).

\subsection*{Different continuum modeling}
\label{sec:appB}
To assess the impact of the adopted continuum modeling on our derived quantities—particularly electron temperatures—we performed a set of tests using a variety of stellar population synthesis templates and polynomial corrections. Specifically, we repeated the spectral fits using several stellar libraries, including \texttt{eMILES} \citep{2016MNRAS.463.3409V}, \texttt{BC03} \citep{2003MNRAS.344.1000B}, \texttt{CB07} (Charlot \& Bruzual, unpublished), \texttt{MILES-HC} \citep[a high-resolution subset of MILES;][]{2018ApJ...854..139C}, and \texttt{PEGASE} \citep{2004A&A...425..881L}. 

\begin{figure}
    \centering
    \includegraphics[width=\linewidth]{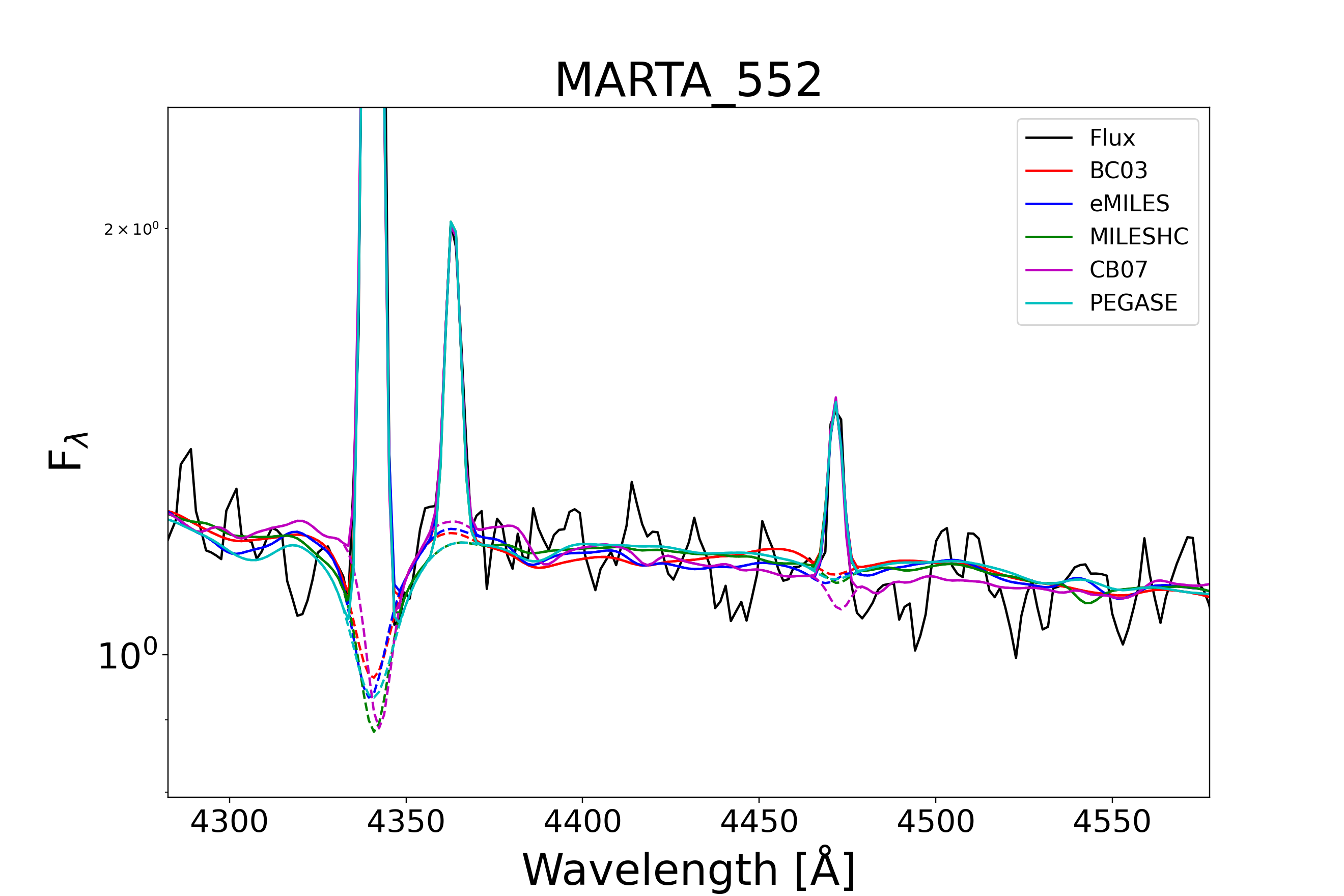}
    \caption{Example continuum fits for a representative object- MARTA\_552- using different stellar population synthesis templates. The figure illustrates the variation in the modeled absorption feature under the \Hgamma, and the corresponding stability of the \OIIIopt~$\lambda$4363 emission line flux.}
    \label{fig:templates}
\end{figure}

To directly quantify the influence of continuum modeling choices on key line measurements, we compared the fluxes of important emission lines—including \OIIIopt~$\lambda$4363—across different template fits. As shown in Figure~\ref{fig:templates}, which illustrates fits for a representative galaxy, the depth of Balmer absorption features can vary significantly depending on the chosen template \citep{2019AJ....158..160B}. Nonetheless, the effect on the flux of \OIIIopt~$\lambda$4363 remains small, typically under $\sim$10\%. The corresponding variation in the derived electron temperatures is generally within 10--15\%, although we found that in galaxies with weaker auroral lines and lower S/N—such as MARTA\_2387—the inferred $T_3$ can vary by up to $\sim$25\%. Conversely, in cases with strong lines and high S/N, variations are as low as 2\%.

For the Balmer lines used in the derivation of dust attenuation, we minimized the impact of template-dependent absorption differences by focusing on the strongest lines—\Halpha, \Hbeta, \Hgamma, and \Hdelta—whose high S/N ensures that absorption variations alter the measured fluxes by no more than $\sim$5\%.

Overall, these tests confirm that our continuum modeling approach is robust. Neither the choice of stellar template nor the polynomial order used for continuum correction introduces any systematic uncertainty that would affect the key conclusions of this work.

\section{Different parametrization of the T$_2$-T$_3$ diagram}
\label{app:yates}

We further analyse the T$_2$-T$_3$ relationship by comparing the temperatures derived for the MARTA sample with the multi-parametric calibration proposed by \cite{yates2020present}. 
This approach is motivated by two primary considerations: i) there is an anti-correlation between the electron temperature and the oxygen abundance. This means that if T(\OIIIopt) $\ll$ T(\OII), then it would be \OIIIopt $\gg$ \OII, i.e. systems where the temperature of the doubly ionized oxygen is lower than the singly ionized oxygen one, tend to have a higher proportion of doubly ionized oxygen relative to singly ionized oxygen. In these cases, the total oxygen abundance $Z(T_e)$ becomes less sensitive to changes in $T_e(\OIIIopt)$.
Conversely, systems where $T_e(\OIIIopt)>T_e(\OII)$ exhibit the opposite pattern: the singly ionized oxygen fraction dominates, and the total oxygen abundance becomes less sensitive to $T_e(\OIIIopt)$; ii) there is an empirical anticorrelation between $T_e(\OIIIopt)$ and $T_e(\OII)$ at fixed oxygen abundance. This is expected theoretically because an increase in the singly ionized oxygen fraction is balanced by a decrease in the doubly ionized oxygen fraction, and vice versa.

To model this, \cite{yates2020present} adopted a rectangular hyperbolic form centered at 0 K:

\begin{equation}
T_e(\OII) = a(Z_{Te})^2 \frac{1}{T_e(\OIIIopt)}, 
\end{equation} 
where $a(Z_{Te})$ is a function depending on metallicity.

\begin{figure}[ht]
    \centering
    \includegraphics[width=0.9\linewidth]{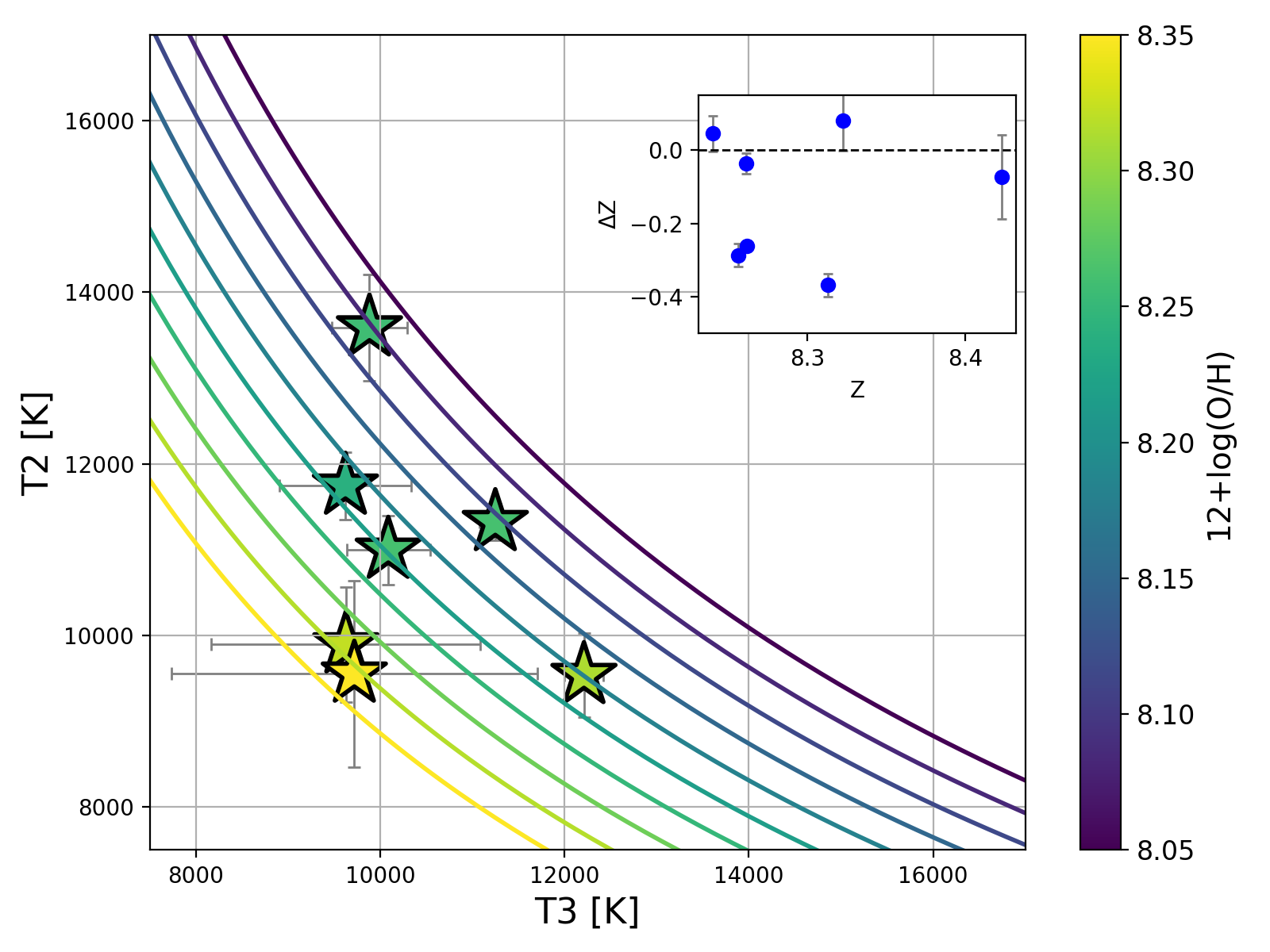}
    \caption{The parametrisation of the T2-T3 plane from \cite{yates2020present}. MARTA galaxies are overplot, colour-coded by their Te-based metallicity. The solid lines represent the T-T relation calibrated by \cite{yates2020present} for a fixed metallicity. Inset panel: The $\Delta Z$ vs. $Z$ plot, where $Z$ is the metallicity derived from the $T_e$ method for the MARTA objects and $\Delta Z$ is the difference between the `direct' metallicity and that obtained from the \cite{yates2020present} calibration for the same galaxies.}
    \label{fig:Yates}
\end{figure}

Using this parametrization, we examined how well the MARTA sample temperatures conform to the best-fit calibration of Yates et al. The results are presented in Figure \ref{fig:Yates}: the main panel shows the T2-T3 relation color coded by the metallicity, revealing that for some galaxies, the interconnected dependence of T2, T3, and metallicity is well reproduced by the Yates calibration. This alignment is particularly evident in the continuous gradients of metallicity along the T2-T3 plane.\\
The inset panel quantifies the agreement by plotting $\Delta Z$ versus Z, where $\Delta Z$ represents the difference between the observed metallicity (calculated using the $T_e$ method) and the metallicity predicted by the Yates calibration, given the observed temperatures. Although some galaxies show good agreement (with $\Delta Z$ close to zero), for others significant deviations are apparent.\\
These discrepancies might arise because of the inapplicability of such a parametrization to high-redshift galaxies, where the physical conditions of the interstellar medium may deviate substantially from those in the local universe. For instance, at higher redshifts, ionization parameters and density conditions may lead to deviations from the trends captured by the Yates calibration, necessitating further exploration of these dependencies in different galactic environments. 

\section{Emission line fluxes for MARTA galaxies}
\label{sec:appD}
Table \ref{tab:aurorals} presents line fluxes, errors, and S/N estimates for all the auroral line detections in the MARTA gold and silver sample. 

\begin{table*}[h!]
\caption{Fluxes, errors, and S/N ratios of detected Oxygen and Sulfur auroral lines for each galaxy in the sample.}
\centering
\begin{tabular}{lcccccccc}
\hline
\hline
Galaxy ID & \OIIIopt4363 Flux & S/N & \OII7320 Flux & S/N & [SII]4068 Flux & S/N & [SIII]6312 Flux & S/N \\ 
          & ($\times 10^{-19}$) &  & ($\times 10^{-19}$) &  & ($\times 10^{-19}$) &  & ($\times 10^{-19}$) &  \\ \hline
MARTA\_3942      & $3.51 \pm 0.38$ & 5.0 & $12.8 \pm 0.99$ & 6.4 & $3.3 \pm 1.1$ & 7.5 & $5.28 \pm 0.67$ & 5.0 \\
MARTA\_4195      & $1.12 \pm 0.45$ & 6.3 & $6.02 \pm 0.81$ & 6.6 & $0.634 \pm 0.12$ & 3.2 & -- & -- \\
MARTA\_5014      & $1.33 \pm 0.28$ & 6.3 & $11.6 \pm 0.56$ & 12 & -- & -- & -- & -- \\
MARTA\_552       & $7.12 \pm 0.26$ & 22 & $4.71 \pm 0.53$ & 6.9 & -- & -- & $3.04 \pm 0.58$ & 3.9 \\
MARTA\_2387      & $1.39 \pm 0.28$ & 4.2 & $2.54 \pm 0.56$ & 3.8 & -- & -- & -- & -- \\
MARTA\_4327      & $6.54 \pm 0.26$ & 15 & $14.4 \pm 0.50$ & 12 & $3.36 \pm 0.25$ & 8.2 & $1.98 \pm 0.39$ & 5.1 \\
MARTA\_4502      & $2.16 \pm 0.25$ & 13 & $6.80 \pm 0.52$ & 6.1 & $1.94 \pm 0.25$ & 3.8 & -- & -- \\ \hline

MARTA\_3408 & $1.68 \pm 0.36$ & 4.0 & -- & -- & -- & -- & -- & -- \\
MARTA\_414  & $2.59 \pm 0.30$ & 7.8 & -- & -- & -- & -- & -- & -- \\
MARTA\_4471 & $2.07 \pm 0.33$ & 4.9 & -- & -- & -- & -- & -- & -- \\
MARTA\_329  & $2.65 \pm 0.63$ & 4.3 & -- & -- & -- & -- & -- & -- \\
MARTA\_1084 & $0.51 \pm 0.4$ & 2.6$^{\dagger}$ & 2.25 $\pm$ 0.74 & 4.6 & -- & -- & -- & -- \\
MARTA\_3115 & $1.50 \pm 0.32$ & 3.3 & -- & -- & -- & -- & -- & -- \\
MARTA\_1374  & $3.12 \pm 0.12$ & 6.1 & -- & -- & -- & -- & -- & -- \\
MARTA\_3887 & $7.43 \pm 0.51$ & 11 & -- & -- & -- & -- & -- & -- \\
MARTA\_3926  & $1.10 \pm 0.43$ & 7.5 & -- & -- & -- & -- & -- & -- \\
\hline
\end{tabular}
 \tablefoot{Fluxes are given in units of 10$^{-19}$ erg/s/cm$^2$, with errors on the fluxes estimated by the \textsc{pPXF} fitting. 
 The signal-to-noise (S/N) ratios are calculated by multiplying the peak flux by $\sqrt{N}$, with N being the number of spectral channels spanned by line FWHM, and dividing it by the RMS measured in a line-free adjacent region.\\
 \tablefoottext{$\dagger$}{the S/N estimated from the \textsc{pPXF} fit is larger than that obtained from the method based on the RMS of the spectrum, hence we consider this a marginal detection.}
 }
\label{tab:aurorals}
\end{table*}

\section{On the representativeness of the auroral-line sample relative to the star-forming main sequence}
\label{sec:appE}

To investigate the potential bias of the high-redshift auroral-line sample with respect to the star-forming main sequence, we examined the location of each galaxy in the SFR–$M_\star$ plane. Stellar masses and star formation rates for the literature sample were taken from the original studies, without homogenization. We then computed the offset from the MS, defined as $\Delta \mathrm{MS} = \log(\mathrm{SFR}) - \log(\mathrm{SFR}_{\mathrm{MS}})$, where $\mathrm{SFR}_{\mathrm{MS}}$ is the expected SFR at a given redshift and stellar mass according to the parameterization by \citet{popesso_SFMS_2023}.

\begin{figure}[ht]
    \centering
    \includegraphics[width=0.99\linewidth]{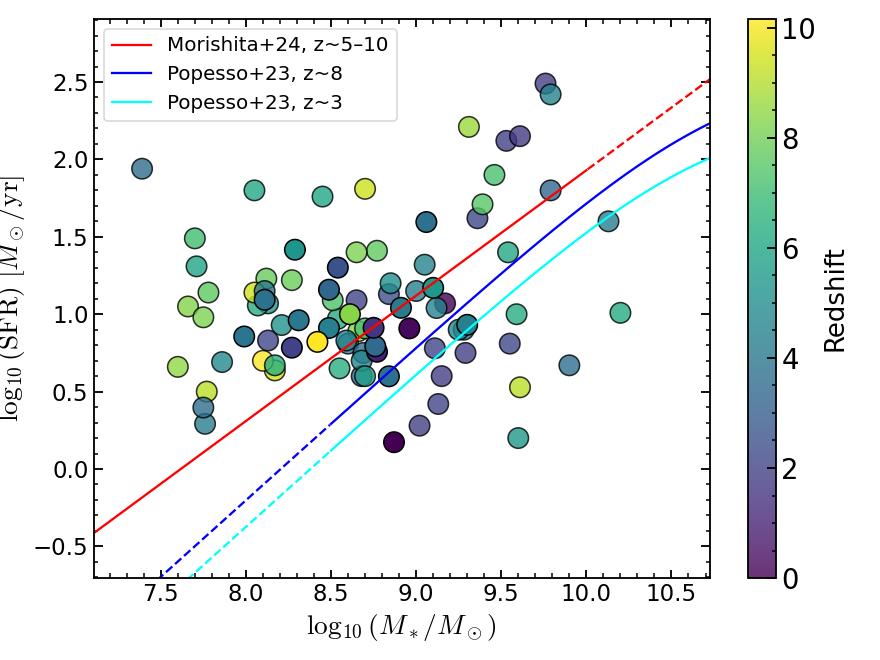}
    \caption{Location of the literature sample used for strong-line calibrations in the SFR–$M_\star$ plane. Stellar masses and SFRs are taken directly from the original studies, without re-fitting. Colored lines indicate the main sequence relations at different redshifts from \citet{popesso_SFMS_2023} and \citet{2024ApJ...963....9M}.}
    \label{fig:MS_lit}
\end{figure}

We find that the sample is, on average, significantly biased toward galaxies with elevated specific star formation rates. The median offset across the full sample is $\Delta \mathrm{MS} \approx 0.7$ dex. This is illustrated in Figure~\ref{fig:MS_lit}, where we show the SFR–$M_\star$ diagram for all galaxies used in the strong-line calibration, along with the MS relations at different redshifts from \citet{popesso_SFMS_2023} and \citet{2024ApJ...963....9M}. The offset is particularly pronounced at low stellar masses ($M_\star < 10^{8.5}\,M_\odot$) and at high redshifts ($z > 4$), where only galaxies with the highest SFRs are typically detected in emission lines, and in particular show auroral lines.

To quantify this further, we split the sample by stellar mass and redshift. Galaxies with $M_\star < 10^{8.5}\,M_\odot$ (N = 30) show a median $\Delta \mathrm{MS} = 1.19$ dex, while more massive galaxies (N = 53) have a lower median offset of 0.43 dex. Similarly, galaxies at $1 < z \leq 4$ (N = 29) have a median $\Delta \mathrm{MS} = 0.42$ dex, compared to 0.78 dex for those at $z > 4$ (N = 54). These trends reflect a combination of selection effects and the current limitations of deep spectroscopic observations at high redshift.

\begin{figure*}[ht]
    \centering
    \includegraphics[width=0.9\linewidth]{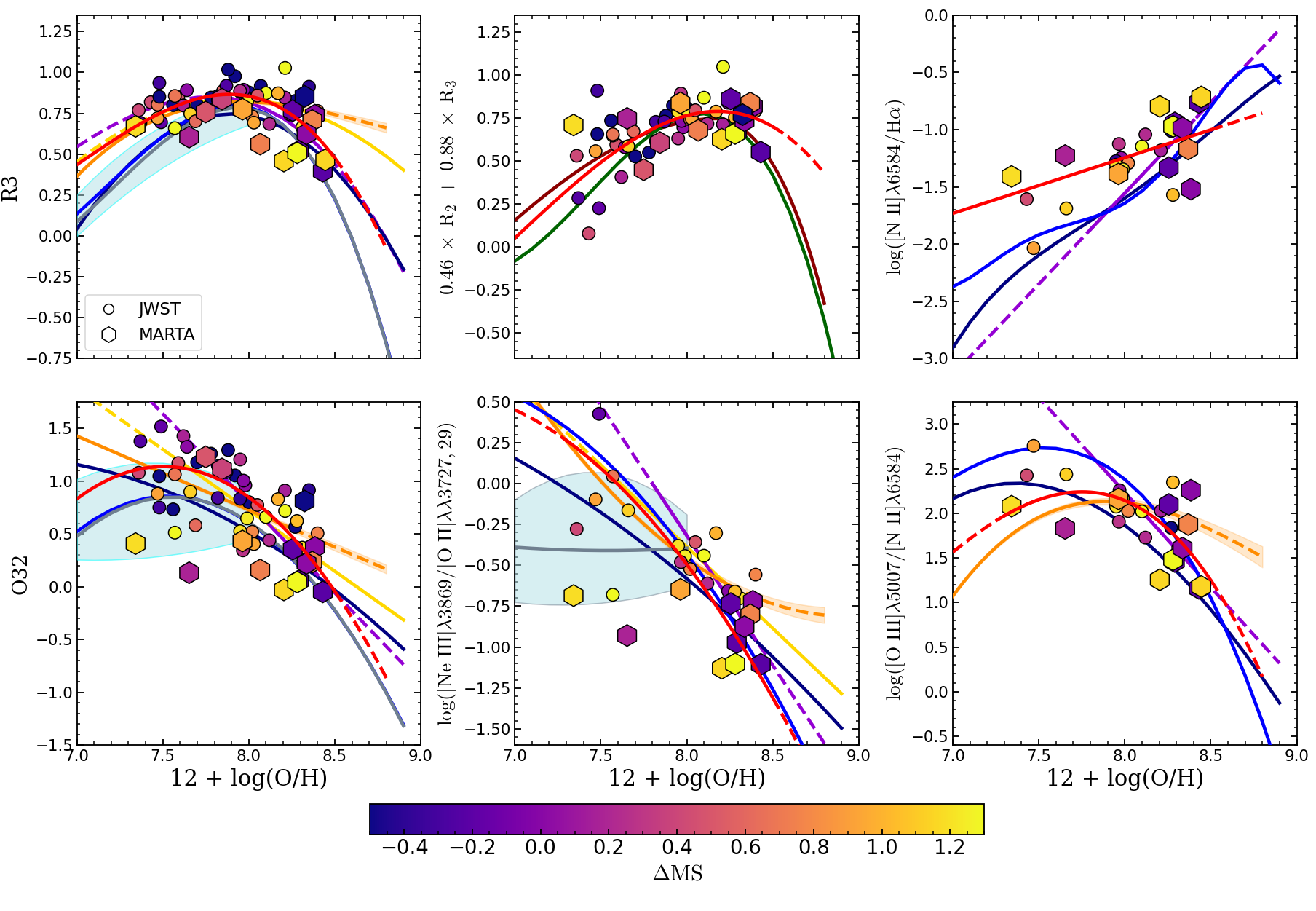}
    \caption{Strong-line calibration diagrams (see Section~\ref{sec:strong_lines}), with each galaxy color-coded by its $\Delta \mathrm{MS}$.}
    \label{fig:calib_deltaMS}
\end{figure*}

The key concern is that a bias toward high-$\mathrm{sSFR}$ systems could correlate with the physical conditions in \hii regions—particularly the ionization parameter—and thus influence the derived strong-line calibrations. To address this, we examined the calibration diagrams shown in the main text (see Section~\ref{sec:strong_lines}), this time color-coding each galaxy by its $\Delta \mathrm{MS}$. The result, shown in Figure~\ref{fig:calib_deltaMS}, reveals no clear systematic trend or shift in the location of galaxies in strong-line ratio space as a function of $\Delta \mathrm{MS}$. While our sample is biased toward elevated sSFRs, this does not appear to introduce a significant distortion in the empirical calibrations.

Nonetheless, we stress that this test is limited by sample size and the uncertainty on both the MS parameterizations and the reported SFRs. A more robust assessment would require a larger, statistically representative set of high-redshift galaxies with auroral-line detections, as well as an improved characterization of the intrinsic scatter in line ratios at fixed metallicity. We defer a more detailed investigation to future work.

\begin{figure}[ht]
    \centering
    \includegraphics[width=0.99\linewidth]{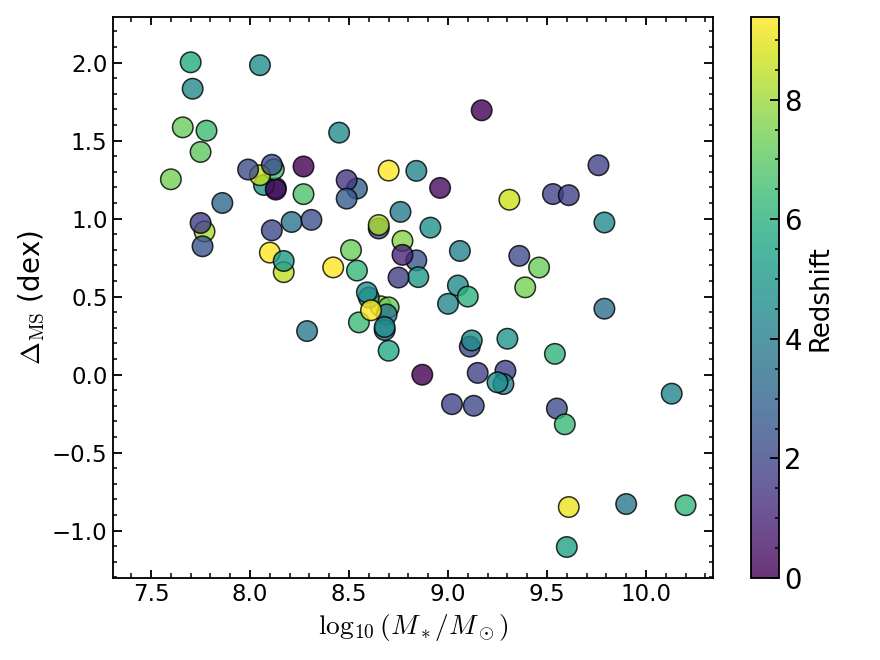}
    \caption{Offset from the main sequence of \citet{popesso_SFMS_2023} ($\Delta \mathrm{MS}$) as a function of stellar mass.}
    \label{fig:deltaMS_lit}
\end{figure}

\begin{table}[h!]
\caption{List of emission lines included in the fitting procedure.}
\centering
\begin{tabular}{l c | l c}
\hline
Line & $\lambda$ [\AA] & Line & $\lambda$ [\AA] \\
\hline
\OII & 3726.03 & \OIIIopt & 5006.84 \\
\OII & 3728.82 & [NI] & 5197.90 \\
H13 (B) & 3734.37 & [NI] & 5200.26 \\
H12 (B)& 3750.15 & [OI] & 6300.30 \\
H11 (B) & 3770.63 & [SIII] & 6312.06 \\
H10 (B) & 3797.91 & [OI] & 6363.78 \\
H9 (B) & 3835.40 & [NII] & 6548.05 \\
\NeIII & 3868.76 & \Halpha & 6562.79 \\
HeI + H8 (B) & 3889.00 & [NII] & 6583.45 \\
\NeIII & 3967.47 & HeI & 6678.15 \\
H7 (B) & 3970.08 & \SII & 6716.44 \\
\SII & 4068.60 & \SII & 6730.82 \\
\SII & 4076.35 & HeI & 7065.25 \\
\Hdelta & 4101.73 & [ArIII] & 7135.79 \\
\Hgamma & 4340.47 & HeI & 7281.35 \\
\OIIIopt & 4363.21 & \OII & 7319.40 \\
HeI & 4471.50 & \OII & 7330.35 \\
HeII & 4685.70 & [ArIII] & 7751.15 \\
{[ArIV]} & 4711.37 & H11 (P) & 8862.89 \\
HeI & 4713.17 & H10 (P) & 9015.30 \\
ArIV & 4740.17 & \SIII & 9068.60 \\
\Hbeta & 4861.35 & \SIII & 9530.60 \\
HeI & 4921.93 & HI & 9545.93 \\
\OIIIopt & 4958.91 &  HeI & 10830.20 \\

\hline
\end{tabular}
\tablefoot{Wavelengths are in air. (B) denotes Balmer and (P) Paschen lines.}
\label{tab:lines}
\end{table}

\end{appendix}

\end{document}